\documentclass[10pt]{article}
\usepackage{bbm}
\usepackage{bm}
\usepackage{mathrsfs}
\usepackage{amsmath}
\usepackage{slashed}
\usepackage{upgreek}
\usepackage{amsmath,amssymb,color}
\usepackage{amsfonts}
\usepackage[all]{xy}
\usepackage{graphics}
\usepackage{cancel}
\usepackage{ulem}

\DeclareFontFamily{OT1}{pzc}{}
\DeclareFontShape{OT1}{pzc}{m}{it}{<-> s * [1.200] pzcmi7t}{}
\DeclareMathAlphabet{\mathpzc}{OT1}{pzc}{m}{it}
\DeclareMathAlphabet{\mathbbmsl}{U}{bbm}{m}{sl}

\newcommand{\ii}{{\rm i}}
\newcommand{\ep}{{\rm e}}
\newcommand{\be}{\begin{equation}}
\newcommand{\ee}{\end{equation}}
\newcommand{\bea}{\setlength\arraycolsep{2pt} \begin{eqnarray}}
\newcommand{\eea}{\end{eqnarray}}
\newcommand{\nn}{\nonumber}

\newcommand{\rd}{{\rm d}}
\newcommand{\rD}{{\rm D}}

\newcommand{\mD}{{\mathbbm D}}

\newcommand{\mY}{{\mathbbm Y}}

\newcommand{\mA}{{\mathbbm A}}

\newcommand{\mF}{{\mathbbm F}}

\newcommand{\cA}{{\mathcal A}}
\newcommand{\cJ}{{\mathcal J}}
\newcommand{\cK}{{\mathcal K}}

\newcommand{\cP}{{\mathcal P}}

\newcommand{\cD}{{\mathcal D}}

\newcommand{\cL}{{\mathcal L}}

\newcommand{\halpha}{{\hat\alpha}}
\newcommand{\hbeta}{{\hat\beta}}
\newcommand{\hgamma}{{\hat\gamma}}

\newcommand{\hs}{{\hat s}}
\newcommand{\ha}{{\hat a}}
\newcommand{\hb}{{\hat b}}

\newcommand{\Ht}{{\hat t}}
\newcommand{\hA}{{\hat A}}

\newcommand{\hF}{{\hat F}}
\newcommand{\hrD}{{\hat\rD}}

\newcommand{\hblt}{{\hat\bullet}}

\renewcommand{\thefootnote}{\fnsymbol{footnote}}
\begin{document}

\begin{titlepage}
\vfill
\begin{flushright}

\end{flushright}
\vfill
\begin{center}
{\Large\bf Holographic $SO(2,d)$ anomaly }

\vskip 1cm

Zhao-Long Wang$^{1,2,3}$\footnote{\tt zlwang@nwu.edu.cn},
Jie Guo$^{1}$,
Yi Yan$^{4}$

\vskip 5mm

$^1${\it Institute of Modern Physics, Northwest University, Xi'an 710127, China}
\\$^2${\it Peng Huanwu Center for Fundamental Theory, Xi'an 710127, China}
\\$^3${\it Shaanxi Key Laboratory for Theoretical Physics Frontiers, Xi'an 710127, China}
\\$^4${\it College of Arts and Sciences, Shaanxi A$\&$F Technology University, Yangling, Shaanxi 712000, China}
\end{center}
\vfill

\begin{abstract}
\noindent
In the $SO(2,d)$ gauge theory formalism of AdS gravity established in \cite{Wang:2018tyq}, the dynamics of bulk gravity emerges from the vanishing of the boundary covariant anomaly for the  $SO(2,d)$ conservation law. In parallel with the known results on chiral anomalies, we establish the descendent structure of the holographic $SO(2,d)$ anomaly. The corresponding anomaly characteristic class, bulk Chern-Simons like action as well as the boundary effective action are constructed systematically. The anomalous conservation law is presented both in the covariant and consistent formalisms. Due to the existence of the ruler field, not only the Bardeen-Zumino polynomial, but also the covariant and consistent currents are explicitly constructed.

\end{abstract}

\vfill
\end{titlepage}

\tableofcontents\newpage
\renewcommand{\thefootnote}{\#\arabic{footnote}}
\setcounter{footnote}{0}

\section{Introduction}
To understand the holographic emergence of bulk dynamics in AdS/CFT \cite{Maldacena:1997re,Gubser:1998bc,Witten:1998qj}, various approaches have been suggested \cite{de Boer:1999xf,Heemskerk:2010hk,Kabat:2011rz,Swingle:2009bg,Czech:2015qta,Wang:2015qfa}.
In  \cite{Wang:2015qfa}, after including the conformal transformation of the renormalization scale, it was shown that the bulk dynamics of a scalar field is highly constrained by the $SO(2,d)$ conformal symmetry of the dual CFT scalar operators. Furthermore, the holographic emergence of the dynamics for the bulk  gravity itself is governed by the generic duality relation between the boundary global symmetry and the bulk gauge symmetry. Applying it to the $SO(2,d)$ conformal symmetry in CFT$_{d}$, the bulk AdS$_{d+1}$ gravity is reformulated as a $SO(2,d)$ gauge theory in  \cite{Wang:2018tyq}. In this formalism, the pullback of the bulk Einstein equation on a co-dimension one hypersurface $\Sigma$ can be naturally related to the covariant anomaly of the CFT $SO(2,d)$ conservation law.  Providing that the $SO(2,d)$ conservation law is not anomalous for any local renormalization scale, the full bulk Einstein equation will be automatically implied.

In this paper, we study the formal mathematical structure of the holographic $SO(2,d)$ anomaly systematically. In Section 2.1, we briefly review the bulk $SO(2,d)$ covariant action \cite{Wang:2018tyq} for AdS gravity. The covariant formula of anomalous boundary conservation law is revealed during the Noether procedure in Section 2.2. In Section 3.1.1,  the bulk Chern-Simons like action is constructed by the homotopic integration. To resolve the issue for large gauge transformation, a topological term is imposed in Section 3.1.2. In Section 3.2, the boundary effective action is also constructed via the homotopic integration method. Then in Section 3.3, the descendent structure of the holographic $SO(2,d)$ anomaly is established after obtaining the $d+2$ form anomaly characteristic class. In Section 4, we summarize our results and discuss possible generalizations. The details of our derivations are summarized in the Appendices.

\section{AdS gravity as $SO(2,d)$ gauge theory}
\subsection{Bulk covariant action}
By unifying the vielbein $e^{\ha}$ and the $SO(1,d)$ spin connection $\omega^{\ha}{}_{\hb}$ to the $SO(2,d)$ gauge field $A^{\halpha}{}_{\hbeta}$, the Palatini's 1st order formalism for the AdS$_{d+1}$ gravity was reformulated as a $SO(2,d)$ gauge theory in  \cite{Wang:2018tyq}. The $SO(2,d)$ indices $\halpha,\hbeta=\hat\bullet,\hat 0,\cdots,\hat d$ contain one more extra direction $\hat\bullet$ than the vielbein indices $\ha,\hb=0,\cdots,d$. Thus naively we have \begin{eqnarray}
A^{\halpha}{}_{\hbeta}=\left(
                                 \begin{array}{cc}
                                   A^{\hblt}{}_{\hblt} & A^{\hblt}{}_{\hb}\\
                                   A^{\ha}{}_{\hblt} & A^{\ha}{}_{\hb} \\
                                 \end{array}
                               \right)=\left(
                                 \begin{array}{cc}
                                   0 & -\ell^{-1}e_{\hb}\\
                                   \ell^{-1}e^{\ha} & \omega^{\ha}{}_{\hb} \\
                                 \end{array}
                               \right)
\end{eqnarray}
To figure out the $SO(2,d)\rightarrow SO(1,d)$ reduction in the $SO(2,d)$ covariant way, we further introduce the ruler field $Y^{\halpha}$ which satisfies the constraint
\begin{eqnarray}
\eta_{\halpha\hbeta} Y^{\halpha}Y^{\hbeta}=-\ell^2\,.
\end{eqnarray}
In terms of $Y^{\halpha}$ and the curvature 2-form $F^{\halpha}{}_{\hbeta}=\rd A^{\halpha}{}_{\hbeta}+A^{\halpha}{}_{\hgamma}\wedge A^{\hgamma}{}_{\hbeta}$, the Palatini action of the AdS$_{d+1}$ Einstein gravity is reformulated as
\begin{eqnarray}\label{S^cov}
\!\!\!\!&&\!\!\!\!S^{\rm cov}_{(d+1)}=\int_{M}\cL^{\rm cov}_{(d+1)}
\cr\!\!\!\!&=&\!\!\!\!\frac1{2\kappa^2\ell\,(d-1)!}\int_{M} \epsilon_{\halpha\hbeta\halpha_1\cdots \halpha_{d}} \left[ F^{\halpha\hbeta}-\frac{2}{(d+1)\ell^2} \rD Y^{\halpha}\wedge  \rD Y^{\hbeta}\right]\wedge  \rD Y^{\halpha_1}\wedge\cdots \wedge  \rD Y^{\halpha_{d-1}}Y^{\halpha_{d}}\,.~~~~~~~~~
\end{eqnarray}
where $\rD=\rd+A$ is the $SO(2,d)$ gauge covariant derivative. We shall refer (\ref{S^cov}) as the bulk covariant action since it is given by the summation of manifestly $SO(2,d)$ covariant terms. Obviously, $Y^{\halpha}$ is just an auxiliary field since it can be totally fixed via the $SO(2,d)$ gauge transformation. The physical degrees of freedom remain same as the original Palatini action. A more systematical discussion can be found in Appx.A.  

In this formalism, the spacetime metric is given by the gauge invariant quadratic form
\begin{eqnarray}
g_{MN}\!\!\!\!&=&\!\!\!\!\eta_{\halpha\hbeta}\rD_{M}Y^{\halpha}\rD_{N}Y^{\hbeta}\,.~~~
\end{eqnarray}
When the metric is not degenerate, $\{Y^{\halpha},\rD_{M }Y^{\halpha}\}$ forms an intrinsic basis of the $d+2$ dimensional space of the $SO(2,d)$ vector representation.  The spacetime geometric quantities can be recovered by expanding the $SO(2,d)$ covariant quantities on the intrinsic basis. For example, the $SO(2,d)$ field strength is expanded as
\begin{eqnarray}\label{flux}
(F_{M_1M_2})^{\halpha\hbeta}\!\!\!\!&=&\!\!\!\! \left(R^{N_1N_2}{}_{M_1M_2}+2\ell^{-2}\delta^{N_1}_{[M_1}\delta^{N_2}_{M_2]}\right) \rD_{N_1}Y^{[\halpha}\rD_{N_2}Y^{\hbeta]} -4\ell^{-2} t^{N}{}_{[M_1M_2]}Y^{[\halpha}\rD_{N} Y^{\hbeta]}\,,~~~~~~
\end{eqnarray}
where $t^{N}{}_{M_1M_2}$ is the torsion and $R^{N_1N_2}{}_{M_1M_2}$ is the Riemann curvature for the torsional connection. We also notice that
\begin{eqnarray}\label{ddual}
\!\!\!\!&&\!\!\!\!\frac{1}{\ell\,p!(d+1-p)!} \epsilon_{\halpha_0\cdots\halpha_{p-1}\halpha_{p}\cdots\halpha_{d+1}}\epsilon^{M_0\cdots M_{p-1}M_{p}\cdots M_{d}} \rD_{M_{p}}Y^{\halpha_{p}}\cdots\rD_{M_{d}}Y^{\halpha_{d}}Y^{\halpha_{d+1}}
=\rD^{[M_{0}}Y_{[\halpha_{0}}\cdots\rD^{M_{p-1}]}Y_{\halpha_{p-1}]}\,,~~~~
\end{eqnarray}
where the spacetime indices are raised by $g^{MN}$ and
\begin{eqnarray}
\epsilon^{01\cdots d}=\frac1{\sqrt{g}}\,,~~~~~~~~~~~g=-\det(g_{MN})\,.
\end{eqnarray}
As explained in detail in Appx.B, it can be shown directly that the action (\ref{S^cov}) is equivalent to the original Einstein-Hilbert action by using (\ref{flux}) and (\ref{ddual}).

The variation of the bulk covariant action on a manifold $M$ with boundary $\Sigma=\partial M$ is given by
\begin{eqnarray}\label{VarScov}
\!\!\!\!&&\!\!\!\!\delta S^{\rm cov}_{(d+1)}
=\int_{M} \left[\delta A^{\halpha\hbeta}\wedge (\cJ_{(d+1)})_{\halpha\hbeta}+ \delta Y^{\halpha}  (\cJ_{(d+1)})_{\halpha}\right]
+\int_{\Sigma} \left[\delta A^{\halpha\hbeta}\wedge (\cK^{\rm cov}_{(d+1)})_{\halpha\hbeta}+ \delta Y^{\halpha}  (\cK^{\rm cov}_{(d+1)})_{\halpha}\right]\,,~~~~~~~
\end{eqnarray}
where $\cJ_{(d+1)}$ are the bulk off-shell currents 
\begin{eqnarray}\label{JA}
\!\!\!\!&&\!\!\!\!(\cJ_{(d+1)})_{\halpha\hbeta}=\frac{\partial \cL^{\rm cov}_{(d+1)}}{\partial A^{\halpha\hbeta}}+\rd\left(\frac{\partial \cL^{\rm cov}_{(d+1)}}{\partial (\rd A^{\halpha\hbeta})}\right)
=\frac{(-1)^{d+1}\ell}{4\kappa^2(d-2)!}\epsilon_{\halpha\hbeta\halpha_1\cdots \halpha_{d}}F^{\halpha_1\halpha_2}\wedge  \rD Y^{\halpha_3}\wedge\cdots \wedge  \rD Y^{\halpha_{d}}\,,
\\\label{JY}
\!\!\!\!&&\!\!\!\!(\cJ_{(d+1)})_{\halpha}=\frac{\partial \cL^{\rm cov}_{(d+1)}}{\partial Y^{\halpha}}-\rd\left(\frac{\partial \cL^{\rm cov}_{(d+1)}}{\partial (\rd Y^{\halpha})}\right)
=\frac{(-1)^{d+1}}{2\kappa^2\ell\,(d-2)!}\epsilon_{\halpha\hbeta\halpha_1\cdots \halpha_{d}}Y^{\hbeta}\rD(F^{\halpha_1\halpha_2}\wedge  \rD Y^{\halpha_2}\wedge\cdots \wedge  \rD Y^{\halpha_{d}})\,,~~~~~~~~~~
\end{eqnarray}
and $\cK^{\rm cov}_{(d+1)}$ are the bulk off-shell Noether potentials \cite{Lee:1990nz}
\begin{eqnarray}\label{KA}
(\cK^{\rm cov}_{(d+1)})_{\halpha\hbeta}\!\!\!\!&=&\!\!\!\!\frac{\partial \cL^{\rm cov}_{(d+1)}}{\partial (\rd A^{\halpha\hbeta})}
=
\frac{(-1)^{d+1}}{2\kappa^2\ell\,(d-1)!}\epsilon_{\halpha\hbeta\halpha_1\cdots \halpha_{d}} Y^{\halpha_1}\rD Y^{\halpha_2}\wedge\cdots \wedge  \rD Y^{\halpha_{d}}\,,
\\\label{KY}
(\cK^{\rm cov}_{(d+1)})_{\halpha}\!\!\!\!&=&\!\!\!\!\frac{\partial \cL^{\rm cov}_{(d+1)}}{\partial (\rd Y^{\halpha})}
\cr\!\!\!\!&=&\!\!\!\!\frac{(-1)^d}{2\kappa^2\ell\,(d-2)!}\epsilon_{\halpha\hbeta\halpha_1\cdots \halpha_{d}}
Y^{\hbeta} \left[F^{\halpha_1\halpha_2}\wedge\rD Y^{\halpha_3}\wedge\cdots \wedge  \rD Y^{\halpha_{d}}
-\frac{2}{\ell^2(d-1)} \rD Y^{\halpha_1}\wedge\cdots \wedge  \rD Y^{\halpha_{d}}\right]\,.~~~~~~
\end{eqnarray}
As analysed in  \cite{Wang:2018tyq} as well as Appx.A, the bulk equations of motion for $\delta A$
\begin{eqnarray}\label{BulkEOM}
\!\!\!\!&&\!\!\!\!(\cJ_{(d+1)})_{\halpha\hbeta}=\frac{(-1)^{d+1}\ell}{4\kappa^2(d-2)!}\epsilon_{\halpha\hbeta\halpha_1\cdots \halpha_{d}}F^{\halpha_1\halpha_2}\wedge  \rD Y^{\halpha_3}\wedge\cdots \wedge  \rD Y^{\halpha_{d}}=0
\end{eqnarray}
are equivalent to the Einstein equation plus the torsion free condition. Providing (\ref{BulkEOM}), the bulk equations of motion for $\delta Y$
\begin{eqnarray}
\!\!\!\!&&\!\!\!\!(\cJ_{(d+1)})_{\halpha}
=\frac{(-1)^{d+1}}{2\kappa^2\ell\,(d-2)!}\epsilon_{\halpha\hbeta\halpha_1\cdots \halpha_{d}}Y^{\hbeta}\rD(F^{\halpha_1\halpha_2}\wedge  \rD Y^{\halpha_3}\wedge\cdots \wedge  \rD Y^{\halpha_{d}})=0~~~~~~~~~~
\end{eqnarray}
are automatically satisfied. Therefore, the introducing of $Y^{\halpha}$ does not impose any additional constraints beyond the original Palatini equations.
\subsection{The off-shell conservation laws}
Substituting the infinitesimal $SO(2,d)$ gauge transformation
\begin{eqnarray}
\delta_{u} Y^{\halpha}=u^{\halpha}{}_{\hbeta}Y^{\hbeta}\,,~~~~~~~~~
\delta_u A^{\halpha\hbeta}=-\rD u^{\halpha\hbeta}\,,
\end{eqnarray}
into the general variation (\ref{VarScov}), the gauge invariance of the action $\delta_{u} S^{\rm cov}_{(d+1)}=0$
implies the off-shell bulk conservation law
\begin{eqnarray}
\rD(\cJ_{(d+1)})_{\halpha\hbeta}+ Y_{[\hbeta}(\cJ_{(d+1)})_{\halpha]}\!\!\!\!&=&\!\!\!\!0\,.
\end{eqnarray}
For the manifold with boundary $\Sigma$
\begin{eqnarray}
\mY^{\halpha}\!\!\!\!&=&\!\!\!\!f_{\Sigma}^{*}[Y^{\halpha}]\,,~~~~~~~~~~
\mA^{\halpha\hbeta}=f_{\Sigma}^{*}[A^{\halpha\hbeta}]\,,
\end{eqnarray}
the gauge invariance also implies the boundary off-shell anomalous conservation law
\begin{eqnarray}\label{cov cons}
\mD(\cJ^{\rm cov}_{(d)})_{\halpha\hbeta}+\mY_{[\hbeta}(\cJ^{\rm cov}_{(d)})_{\halpha]}=(\cA^{\rm cov}_{(d)})_{\halpha\hbeta}\,,
\end{eqnarray}
where the boundary currents and anomaly are given by pullback of the bulk Noether potential and current on $\Sigma$ respectively
\begin{eqnarray}\label{cov currents}
(\cJ^{\rm cov}_{(d)})_{\halpha\hbeta}\!\!\!\!&=&\!\!\!\!-f_{\Sigma}^{*}[(\cK^{\rm cov}_{(d+1)})_{\halpha\hbeta}]\,,~~~~~~~~~~
(\cJ^{\rm cov}_{(d)})_{\halpha}=-f_{\Sigma}^{*}[(\cK^{\rm cov}_{(d+1)})_{\halpha}]
\,,~~~~~~~~~~~
\cr (\cA^{\rm cov}_{(d)})_{\halpha\hbeta}\!\!\!\!&=&\!\!\!\!-f_{\Sigma}^{*}[(\cJ_{(d+1)})_{\halpha\hbeta}]\,.
\end{eqnarray}
The same results were obtained in the Hamiltonian analysis in  \cite{Wang:2018tyq}. The boundary currents and anomaly are manifestly covariant under the local $SO(2,d)$ transformation. From the dual CFT point of view, (\ref{cov currents}) give rise to the covariant formalism of the CFT  currents
\begin{eqnarray}\label{S-R}
(\cJ^{\rm cov}_{(d)})_{\halpha\hbeta}=\left\langle\frac{\delta S_{\rm CFT}}{\delta \mA^{\halpha\hbeta}}\right\rangle\,, ~~~~~~~~~~~~
(\cJ^{\rm cov}_{(d)})_{\halpha}=\left\langle\frac{\delta S_{\rm CFT}}{\delta \mY^{\halpha}}\right\rangle\,,
\end{eqnarray}
as well as the corresponding covariant $SO(2,d)$ anomaly. When the pullback of the bulk Einstein equations on the hypersurface $z=\zeta(x)$ is satisfied, the covariant anomaly vanishes for the CFT with the local renormalization scale $\upmu=\zeta^{-1}$. Inversely, if we require that the CFT $SO(2,d)$ covariant conservation law is not anomalous for any local renormalization scale, the full bulk Einstein equations are automatically implied.

\section{The descendent structure of $SO(2,d)$ anomaly}
\subsection{Bulk consistent action}
\subsubsection{Chern-Simons like action}
To describe the descendent structure of $SO(2,d)$ anomaly,  we need to introduce the homotopic quantities
\begin{eqnarray}
\rD(s)Y=\rd Y+ sA Y\,,~~~~~~F(s)=\rd A(s)+A(s)\wedge A(s)
=sF-s(1-s)A\wedge A
\end{eqnarray}
by the replacement
\begin{eqnarray}
A\to A(s)=sA\,.
\end{eqnarray}
As in the case of chiral anomalies  \cite{Bertlmann:1996xk}, the bulk Chern-Simons like action can be constructed in terms of the homotopic integration over the boundary covariant anomaly. That is
\begin{eqnarray}\label{CS}
S^{\rm CS}_{(d+1)}\!\!\!\!&=&\!\!\!\!\int_{M}\cL_{(d+1)}^{\rm CS}
=-\int_{M}\int_0^1\rd s\,A^{\halpha\hbeta}\wedge (\cA^{\rm cov}_{(d)})_{\halpha\hbeta}(A(s),Y)
=\int_{M}\int_0^1\rd s\,A^{\halpha\hbeta}\wedge (\cJ_{(d+1)})_{\halpha\hbeta}(A(s),Y)
\cr\!\!\!\!&=&\!\!\!\!\frac{(-1)^{d+1}\ell}{4\kappa^2(d-2)!}\int_{M}\int_0^1\rd s\,\epsilon_{\halpha\hbeta\halpha_1\cdots\halpha_{d}}A^{\halpha\hbeta}\wedge F^{\halpha_1\halpha_2}(s)\wedge\rD(s)Y^{\halpha_3}\wedge\cdots\wedge\rD(s)Y^{\halpha_{d}}
\,.~~
\end{eqnarray}

The variation of $S^{\rm CS}_{(d+1)}$ is given by
\begin{eqnarray}\label{VarSCS}
\!\!\!\!&&\!\!\!\!\delta S^{\rm CS}_{(d+1)}=\int_{M} \left[\delta A^{\halpha\hbeta}\wedge (\cJ_{(d+1)})_{\halpha\hbeta}+ \delta Y^{\halpha}  (\cJ_{(d+1)})_{\halpha}\right]
+\int_{\Sigma} \left[\delta A^{\halpha\hbeta}\wedge (\cK^{\rm CS}_{(d+1)})_{\halpha\hbeta}+ \delta Y^{\halpha}  (\cK^{\rm CS}_{(d+1)})_{\halpha}\right]\,,~~~~~~~
\end{eqnarray}
where the Noether potentials are
\begin{eqnarray}
(\cK^{\rm CS}_{(d+1)})_{\halpha\hbeta}\!\!\!\!&=&\!\!\!\!
\frac{(-1)^{d+1}\ell}{4\kappa^2(d-2)!}\int_0^1\!\!\rd s\,\epsilon_{\halpha\hbeta\halpha_1\cdots\halpha_{d}} A^{\halpha_1\halpha_2}(s) \wedge\rD(s)Y^{\halpha_3}\wedge\cdots\wedge\rD(s)Y^{\halpha_{d}}\,,
\\
(\cK^{\rm CS}_{(d+1)})_{\halpha}\!\!\!\!&=&\!\!\!\!
\frac{(-1)^d}{2\kappa^2\ell\,(d-2)!}\epsilon_{\halpha\hbeta\halpha_1\cdots \halpha_{d}}
Y^{\hbeta} \Big[F^{\halpha_1\halpha_2}\wedge \rD Y^{\halpha_3}\wedge\cdots\wedge\rD Y^{\halpha_{d}}
\cr&&~~~~~~~~~~~~~~~~~~~~~~~~~~~~~~~~~~-\int_0^1\!\!\!\rd s\,\rD(s)\Big(A^{\halpha_1\halpha_2}\wedge \rD(s)Y^{\halpha_3}\wedge\cdots\wedge\rD(s)Y^{\halpha_{d}}\Big)\Big]\,.~~~~~~
\end{eqnarray}
We notice that $\delta S^{\rm CS}_{(d+1)}$ gives rise to the same bulk terms as $\delta S^{\rm cov}_{(d+1)}$. Thus, these two Lagrangians differ only  by a closed term, and both of them can be viewed as the $SO(2,d)$ uplifting of the bulk Einstein gravity.
For $d=2$, $S^{\rm CS}_{(d+1)}$ comes back to the $SO(2,2)=SL(2,R)\times SL(2,R)$ Chern-Simons gauge theory \cite{Witten:1988hc}.

As in the chiral anomaly case, the Chern-Simons like action (\ref{CS}) is invariant under the perturbative gauge transformations, but not invariant under the large gauge transformations. Let us consider the finite $SO(2,d)$ gauge transformation
\begin{eqnarray}
Y\to \tilde Y=UY\,,~~~~~
A\to \tilde A=UAU^{-1}-\rd U U^{-1}=U(A-\hat A)U^{-1}\,,
\end{eqnarray}
where we denote $\hA=U^{-1}\rd U$\,.
The corresponding homotopic quantities are
\begin{eqnarray}
\hA(\hat s)\!\!\!\!&=&\!\!\!\!\hat s\hA=\hat sU^{-1}\rd U\,,~~~~~~~\hrD(\hat s)Y=\rd Y+\hat s\hat AY=\rd Y+sU^{-1}\rd UY\,,
\cr \hF(\hat s)\!\!\!\!&=&\!\!\!\!\rd \hA(\hat s)+\hA(\hat s)\wedge \hA(\hat s)=-\hat s(1-\hat s)\hA\wedge \hA=\hF(1-\hat s)\,,
\cr A(s,\hat s)\!\!\!\!&=&\!\!\!\!sA+\hat s\hA\,,~~~~~~~\rD(s,\hat s)Y=\rd Y+sA Y+\hs\hat AY\,,
\cr
F(s,\hs)\!\!\!\!&=&\!\!\!\!\rd A(s,\hs)+A(s,\hs)\wedge A(s,\hs)
=F(s) +\hF(\hs)+s\hs(A \wedge\hA+\hA\wedge A)\,.
\end{eqnarray}
In terms of these homotopic quantities, the finite gauge transformation of the $\cL^{\rm CS}_{(d+1)}[A,Y]$ can be expressed as
\begin{eqnarray}\label{LGT-CS}
\!\!\!\!&&\!\!\!\!\frac{4\kappa^2(d-2)!}{(-1)^{d+1}\ell}\left(\cL^{\rm CS}_{(d+1)}[\tilde A,\tilde Y]-\cL^{\rm CS}_{(d+1)}[A,Y]\right)
\cr\!\!\!\!&=&\!\!\!\!
-\int_0^1\rd \hs\,\epsilon_{\halpha\hbeta\halpha_1\cdots\halpha_{d}}\hA^{\halpha\hbeta}\wedge \hF^{\halpha_1\halpha_2}(\hs)\wedge\hrD(\hs)Y^{\halpha_3}\wedge\cdots\wedge\hrD(\hs)Y^{\halpha_{d}}
\cr\!\!\!\!&&\!\!\!\!+\rd\left[\int_0^1\rd s\int_0^{1-s}\rd \hs\,\epsilon_{\halpha\hbeta\halpha_1\cdots\halpha_{d}} \hA^{\halpha\hbeta}\wedge A^{\halpha_1\halpha_2}
\wedge\rD(s,\hs)Y^{\halpha_3}\wedge\cdots\wedge \rD(s,\hs)Y^{\halpha_{d}}\right]\,.
\end{eqnarray}
Obviously,  the bulk term vanishes for infinitesimal gauge transformations.  It means that the perturbative gauge invariance is unbroken up to the boundary term. However, for large gauge transformations, the bulk term is no longer zero and the full $SO(2,d)$ invariance is broken.
\subsubsection{Topological term and bulk consistent action}
Due to the existence of the ruler field, one can always introduce a $A$-independent term as
\begin{eqnarray}\label{Stop}
S^{\rm top}_{(d+1)}
\!\!\!\!&=&\!\!\!\!\int_{M}\cL^{\rm top}_{(d+1)}=\frac{d}{\kappa^2(d+1)!\ell^3}\int_{M} \epsilon_{\halpha_0\cdots\halpha_{d+1}}\rd Y^{\halpha_0}\wedge\cdots\wedge\rd Y^{\halpha_{d}}Y^{\halpha_{d+1}}
\,.~~~~~~~
\end{eqnarray}
It is a topological term since $\cL^{\rm top}_{(d+1)}$ is closed
\begin{eqnarray}
\rd(\epsilon_{\halpha_0\cdots\halpha_{d+1}}\rd Y^{\halpha_0}\wedge\cdots\wedge\rd Y^{\halpha_{d}}Y^{\halpha_{d+1}})=(-1)^{d+1}\epsilon_{\halpha_0\cdots\halpha_{d+1}}\rd Y^{\halpha_0}\wedge\cdots\wedge\rd Y^{\halpha_{d+1}}=0
\,.~~~~~~~
\end{eqnarray}
Due to the topological nature, the variation of (\ref{Stop}) is just a boundary term
\begin{eqnarray}\label{variS_top}
\delta S^{\rm top}_{(d+1)}
\!\!\!\!&=&\!\!\!\!\frac{(-1)^d}{{\kappa^2\ell^3(d-1)!}}\int_{\Sigma}\epsilon_{\halpha_1\cdots\halpha_{d+2}}\delta  Y^{\halpha_1} Y^{\halpha_{2}}\rd  Y^{\halpha_3}\wedge\cdots\wedge\rd  Y^{\halpha_{d+2}}
\,.~~~~~~~
\end{eqnarray}
Especially, $S^{\rm top}_{(d+1)}$ is invariant under the bulk perturbative $SO(2,d)$ gauge transformations.

Under the finite $SO(2,d)$ gauge transformations, $\cL^{\rm top}_{(d+1)}$ gives rise to
\begin{eqnarray}\label{LGT-TP}
\!\!\!\!&&\!\!\!\!\frac{4\kappa^2(d-2)!}{(-1)^{d+1}\ell}\left(\cL^{\rm top}_{(d+1)}[\tilde Y]-\cL^{\rm top}_{(d+1)}[Y]\right)
\cr\!\!\!\!&=&\!\!\!\!-\int_0^{1}\rd \hs\,\epsilon_{\halpha_0\cdots\halpha_{d+1}}\hA^{\halpha_0\halpha_{1}}\wedge \hF^{\halpha_{2}\halpha_{3}}(\hs)\wedge\hrD(\hs) Y^{\halpha_{4}}\wedge\cdots\wedge\hrD(\hs)Y^{\halpha_{d+1}}
\cr\!\!\!\!&&\!\!\!\!+\frac{2(-1)^d}{(d-1)\ell^2}\rd\Big[\int_0^{1}\rd \hs\,\epsilon_{\halpha_0\cdots\halpha_{d+1}}\hA^{\halpha_0\halpha_{1}}\wedge \hrD(\hs) Y^{\halpha_{2}}\wedge\cdots\wedge\hrD(\hs)Y^{\halpha_{d}} Y^{\halpha_{d+1}}\Big]
\,.~~~~~~~
\end{eqnarray}
The bulk term in (\ref{LGT-TP}) is exactly same as the one appeared in (\ref{LGT-CS}). Therefore, the large gauge transformation of the Chern-Simons like action can be compensated by adding the topological term
\begin{eqnarray}
S^{\rm con}_{(d+1)}\!\!\!\!&=&\!\!\!\!\int_{M}\cL^{\rm con}_{(d+1)}=\int_{M}[\cL^{\rm CS}_{(d+1)}-\cL^{\rm top}_{(d+1)}]\,.~~~~~~~
\end{eqnarray}
Up to the boundary terms,
this bulk consistent action $S^{\rm con}_{(d+1)}$
is invariant under both perturbative and large gauge transformations.

Since $\delta S^{\rm top}_{(d+1)}$ is just a boundary term, the variation of the consistent action gives rise to the same bulk term as in $\delta S^{\rm CS}_{(d+1)}$ and  $\delta S^{\rm cov}_{(d+1)}$. We have
\begin{eqnarray}\label{VarScon}
\!\!\!\!&&\!\!\!\!\delta S^{\rm con}_{(d+1)}=\int_{M} \left[\delta A^{\halpha\hbeta}\wedge (\cJ_{(d+1)})_{\halpha\hbeta}+ \delta Y^{\halpha}  (\cJ_{(d+1)})_{\halpha}\right]
+\int_{\Sigma} \left[\delta A^{\halpha\hbeta}\wedge (\cK^{\rm con}_{(d+1)})_{\halpha\hbeta}+ \delta Y^{\halpha}  (\cK^{\rm con}_{(d+1)})_{\halpha}\right]\,,~~~~~~~
\end{eqnarray}
where $(\cK^{\rm con}_{(d+1)})_{\halpha\hbeta}=(\cK^{\rm CS}_{(d+1)})_{\halpha\hbeta}$ since $S^{\rm top}_{(d+1)}$ is $A$-independent and only $(\cK_{(d+1)})_{\halpha}$ is modified by the topological term
\begin{eqnarray}
(\cK^{\rm con}_{(d+1)})_{\halpha}\!\!\!\!&=&\!\!\!\!(\cK^{\rm CS}_{(d+1)})_{\halpha}-\frac{(-1)^d}{{\kappa^2\ell^3(d-1)!}}\epsilon_{\halpha\hbeta\halpha_1\cdots\halpha_{d}}Y^{\hbeta}\rd  Y^{\halpha_1}\wedge\cdots\wedge\rd  Y^{\halpha_{d}}
\,.~~~~~~
\end{eqnarray}
The pullback of the consistent Noether potentials gives rise to the Bardeen-Zumino polynomial \cite{Bardeen:1984pm}
\begin{eqnarray}\label{con currents}
(\cP_{(d)})_{\halpha\hbeta}\!\!\!\!&=&\!\!\!\!-f_{\Sigma}^{*}[(\cK^{\rm con}_{(d+1)})_{\halpha\hbeta}]\,,~~~~~~~~~~
(\cP_{(d)})_{\halpha}=-f_{\Sigma}^{*}[(\cK^{\rm con}_{(d+1)})_{\halpha}]\,,
\end{eqnarray}
which are basically the differences between the boundary covariant and  consistent currents.
\subsection{Boundary effective action}
Similar to (\ref{CS}), one can also construct the boundary relative effective action by the homotopic integration over  the boundary current \cite{Fujikawa:2004cx}. That is,
\begin{eqnarray}
W_{(d)}\!\!\!\!&=&\!\!\!\!\int_{\Sigma}\cL^{\rm eff}_{(d)}=\int_{\Sigma}\int_{0}^1\rd s\, A^{\halpha\hbeta }\wedge(\cJ_{(d)}^{\rm cov})_{\halpha\hbeta}(A(s),Y)
\cr\!\!\!\!&=&\!\!\!\!-\frac1{2\kappa^2\ell(d-1)!}\int_{\Sigma}\int_{0}^1\rd s\,\epsilon_{\halpha\hbeta\halpha_1\cdots\halpha_{d}} A^{\halpha\hbeta }\wedge \rD(s)Y^{\halpha_1}\wedge\cdots\wedge \rD(s)Y^{\halpha_{d-1}}Y^{\halpha_{d}}
\,.
\end{eqnarray}
Under the finite $SO(2,d)$ gauge transformations, $W_{(d)}$ transforms as
\begin{eqnarray}\label{LGT-eff}
\!\!\!\!&&\!\!\!\!W_{(d)}[\tilde A,\tilde Y]-W_{(d)}[A,Y]
\cr\!\!\!\!&=&\!\!\!\!\frac{(-1)^{d+1}\ell}{4\kappa^2(d-2)!}\int_{\Sigma}\int_0^1\rd \hs\,\epsilon_{\halpha\hbeta\halpha_1\cdots\halpha_{d}} \Big[\int_0^{1-\hs}\rd s\,\hA^{\halpha\hbeta}\wedge A^{\halpha_1\halpha_2}
\wedge\rD(s,\hs)Y^{\halpha_3}\wedge\cdots\wedge \rD(s,\hs)Y^{\halpha_{d}}
\cr&&~~~~~~~~~~~~~~~~~~~~~~~~~~~~~
+\frac{2}{(d-1)\ell^2}\hA^{\halpha\hbeta}Y^{\halpha_{1}}\wedge \hrD(\hs) Y^{\halpha_{2}}\wedge\cdots\wedge\hrD(\hs)Y^{\halpha_{d}} \Big]\,.
\end{eqnarray}
Comparing with (\ref{LGT-CS}) and (\ref{LGT-TP}), we find (\ref{LGT-eff}) is explicitly the boundary term produced by the gauge transformation of the bulk consistent action $S^{\rm con}_{(d+1)}$. Thus we have the expected finite descendent relation
\begin{eqnarray}\label{fDR}
\Delta_{U}\cL^{\rm con}_{(d+1)}[A,Y]=\rd (\Delta_{U}\cL^{\rm eff}_{(d)}[A,Y])\,.
\end{eqnarray}
At the infinitesimal limit, (\ref{LGT-eff}) gives rise to the consistent anomaly
\begin{eqnarray}
\!\!\!\!&&\!\!\!\!\int_{\Sigma}u^{\halpha\hbeta}(\cA^{\rm con}_{(d)})_{\halpha\hbeta}=\delta W_{(d)}[A,Y]
\cr\!\!\!\!&=&\!\!\!\!\frac{(-1)^{d}\ell}{4\kappa^2\,(d-2)!}\int_{\Sigma}u^{\halpha\hbeta} \epsilon_{\halpha\hbeta\halpha_1\cdots\halpha_{d}}\Big[
\int_0^1\rd s\, (1-s)\,\rd\! \left( A^{\halpha_1\halpha_2}\wedge\rD(s)Y^{\halpha_3}\wedge\cdots\wedge \rD(s)Y^{\halpha_{d}}\right)
\cr&&~~~~~~~~~~~~~~~~~~~~~~~~~~~~~~~~~~~~~+\frac{2}{(d-1)\ell^2}\rd Y^{\halpha_1}\wedge\cdots\wedge \rd Y^{\halpha_{d}}\Big]\,,
\end{eqnarray}
where the total derivative terms have been subtracted since $\partial\Sigma=\partial\partial M=0$.
The boundary consistent anomaly satisfies that the infinitesimal descendent relation
\begin{eqnarray}
\delta_{u} I^{\rm con}_{(d+1)}\!\!\!\!&=&\!\!\!\!\rd I^{\rm con}_{(d)}\,,~~~~~
\end{eqnarray}
where we denote that
\begin{eqnarray}
I^{\rm con}_{(d)}\!\!\!\!&=&\!\!\!\!u^{\halpha\hbeta}(\cA^{\rm con}_{(d)})_{\halpha\hbeta}\,,~~~~~~~~~~~~~~I^{\rm con}_{(d+1)}=\cL^{\rm con}_{(d+1)}\,.
\end{eqnarray}

The corresponding consistent currents are
\begin{eqnarray}
(\cJ^{\rm con}_{(d)})_{\halpha\hbeta}\!\!\!\!&=&\!\!\!\!\frac{\delta W_{(d)}}{\delta A^{\halpha\hbeta}}
\cr\!\!\!\!&=&\!\!\!\!\frac{(-1)^d}{2\kappa^2\ell(d-1)!}\epsilon_{\halpha\hbeta\halpha_1\cdots\halpha_{d}}\Big[ Y^{\halpha_1}\rD Y^{\halpha_2}\wedge\cdots\wedge \rD Y^{\halpha_{d}}
\cr&&~~~~~~~~~~~~~~~~~~~~~~~~~~~~~-\frac{\ell^2(d-1)}2 \int_{0}^1\rd s\,A^{\halpha_1\halpha_2}(s)\wedge \rD(s)Y^{\halpha_3}\wedge\cdots\wedge \rD(s)Y^{\halpha_{d}}\Big]\,,~~~~~~~~
\\ (\cJ^{\rm con}_{(d)})_{\halpha}\!\!\!\!&=&\!\!\!\!\frac{\delta W_{(d)}}{\delta Y^{\halpha}}
\cr\!\!\!\!&=&\!\!\!\!\frac{(-1)^d}{2\kappa^2\ell(d-1)!}\epsilon_{\halpha\hbeta\halpha_1\cdots\halpha_{d}}
 Y^{\hbeta}\Big[\frac{2}{\ell^2}\,(\rD Y^{\halpha_2}\wedge\cdots\wedge  \rD Y^{\halpha_{d}}-\rd Y^{\halpha_2}\wedge\cdots\wedge  \rd Y^{\halpha_{d}})
\cr&&~~~~~~~~~~~~~~~~~~~~~~~~~~~~~~~-(d-1) \int_{0}^1\rd s\,\rD(s)[A^{\halpha_1\halpha_2}  \wedge \rD(s)Y^{\halpha_3}\wedge\cdots\wedge \rD(s)Y^{\halpha_{d}}]\Big]\,.
\end{eqnarray}
One can further verify that the consistent currents satisfy the off-shell consistent conservation law
\begin{eqnarray}\label{cov cons}
\rD(\cJ^{\rm con}_{(d)})_{\halpha\hbeta}+Y_{[\hbeta}(\cJ^{\rm con}_{(d)})_{\halpha]}=(\cA^{\rm con}_{(d)})_{\halpha\hbeta}\,,
\end{eqnarray}
as well as the relations  \cite{Bardeen:1984pm}
\begin{eqnarray}
(\cJ^{\rm con}_{(d)})_{\halpha\hbeta}\!\!\!\!&=&\!\!\!\!(\cJ^{\rm cov}_{(d)})_{\halpha\hbeta}-(\cP_{(d)})_{\halpha\hbeta}\,,~~~~~~~~~~~~~~~
(\cJ^{\rm con}_{(d)})_{\halpha}=(\cJ^{\rm cov}_{(d)})_{\halpha}-(\cP_{(d)})_{\halpha}\,.
\end{eqnarray}
Since the Bardeen-Zumino polynomial does not contribute to the homotopic integration
\begin{eqnarray}
\!\!\!\!&&\!\!\!\!A^{\halpha\hbeta}\wedge (\cP_{(d)})_{\halpha\hbeta}(sA,Y)\propto \int_0^1\rd \tilde s\,\epsilon_{\halpha\hbeta\halpha_1\cdots\halpha_{d}} A^{\halpha\hbeta}\wedge A^{\halpha_1\halpha_2}(\tilde s)\wedge\rD(s \tilde s)Y^{\halpha_3}\wedge\cdots\wedge\rD(s \tilde s)Y^{\halpha_{d}}=0\,,~~~~~~
\end{eqnarray}
the effective action can also be constructed in terms of the homotopic integration over the consistent current \cite{Fujikawa:2004cx}
\begin{eqnarray}
W_{(d)}\!\!\!\!&=&\!\!\!\!\int_{\Sigma}\int_{0}^1\rd s\, A^{\halpha\hbeta }\wedge(\cJ_{(d)}^{\rm cov})_{\halpha\hbeta}(A(s),Y)
=\int_{\Sigma}\int_{0}^1\rd s\, A^{\halpha\hbeta }\wedge(\cJ_{(d)}^{\rm con})_{\halpha\hbeta}(A(s),Y)
\,.
\end{eqnarray}

\subsection{Characteristic class}
The characteristic class of the $SO(2,d)$ anomaly is given by the exterior derivative of the bulk consistent  Lagrangian
\begin{eqnarray}
I_{(d+2)}\!\!\!\!&=&\!\!\!\!\rd \cL_{(d+1)}^{\rm con} 
=\frac{(-1)^{d+1}\ell}{8\kappa^2(d-2)!}\epsilon_{\halpha\hbeta\halpha_1\cdots\halpha_{d}}F^{\halpha\hbeta}\wedge F^{\halpha_1\halpha_2}\wedge\rD Y^{\halpha_3}\wedge\cdots\wedge\rD Y^{\halpha_{d}}
\,.
\end{eqnarray}
As shown in \cite{Wang:2018tyq} as well as Appx.C, the exterior derivative of the bulk covariant Lagrangian gives rise to the same result
\begin{eqnarray}
I_{(d+2)}\!\!\!\!&=&\!\!\!\!\rd\cL_{(d+1)}^{\rm cov}\,.
\end{eqnarray}
In fact, the bulk consistent action is just the summation of the bulk covariant action and the boundary effective action
\begin{eqnarray}
\cL_{(d+1)}^{\rm cov}\!\!\!\!&=&\!\!\!\!\cL_{(d+1)}^{\rm con}-\rd \cL_{(d)}^{\rm eff}=\cL_{(d+1)}^{\rm CS}-\cL_{(d+1)}^{\rm top}-\rd\cL_{(d)}^{\rm eff}\,.
\end{eqnarray}
Under the finite gauge transformations, the bulk and boundary terms of $\Delta_{U}S_{(d+1)}^{\rm CS}$ are compensated respectively by the transformations of topological term and the boundary effective action. This is consistent with the fact that $\cL_{(d+1)}^{\rm cov}$ is manifestly $SO(2,d)$ gauge invariant.

As in the usual treatment of chiral anomalies, one can also establish the descendent structure
\begin{eqnarray}
I_{(d+2)}\!\!\!\!&=&\!\!\!\!\rd\cL_{(d+1)}^{\rm CS}
\cr\delta_{u} I^{\rm CS}_{(d+1)}\!\!\!\!&=&\!\!\!\!\rd I^{\rm CS}_{(d)}
\end{eqnarray}
directly by the bulk Chern-Simons term
\begin{eqnarray}
I^{\rm CS}_{(d+1)}=\cL^{\rm CS}_{(d+1)}\,,
~~~~~~~~~~
I^{\rm CS}_{(d)}=u^{\halpha\hbeta}(\cA^{\rm CS}_{(d)})_{\halpha\hbeta}\,.
\end{eqnarray}
The corresponding anomaly
\begin{eqnarray}
(\cA^{\rm CS}_{(d)})_{\halpha\hbeta}=\frac{(-1)^{d}\ell}{4\kappa^2\,(d-2)!} \epsilon_{\halpha\hbeta\halpha_1\cdots\halpha_{d}}
\int_0^1\rd s\, (1-s)\,\rd\! \left( A^{\halpha_1\halpha_2}\wedge\rD(s)Y^{\halpha_3}\wedge\cdots\wedge \rD(s)Y^{\halpha_{d}}\right)
\end{eqnarray}
appears from the infinitesimal gauge transformation of the topologically improved effective action
\begin{eqnarray}
\hat W=W_{(d)}+S_{(d+1)}^{\rm top} \,.
\end{eqnarray}
Due to the existence of the bulk term in $\Delta_{U}S_{(d+1)}^{\rm CS}$ and $\Delta_{U}S_{(d+1)}^{\rm top}$, the Chern-Simons descendent relation is correct only for the perturbative gauge transformations. The finite descendent relation like (\ref{fDR}) is absent in this approach.
\section{Summary}
In this paper, we establish the descendent structure of the holographic $SO(2,d)$ anomaly. Due to the existence of the ruler field, one can write down the explicit form of the boundary covariant and consistent currents. The bulk Chern-Simons like action and the boundary effective action are constructed in terms of the homotopic integration method. To compensate the large gauge transformation of the Chern-Simons like action, we introduce a gauge field independent topological term which is constructed only by the ruler field. It is curious to discuss the effect of the topological term in the story of cobordism invariance \cite{Atiyah:1975jf,Witten:2019bou}.

In the Einstein gravity, the $SO(2,d)$ anomaly is governed by the characteristic class
\begin{eqnarray}
I_{(d+2)}\!\!\!\!&=&\!\!\!\!\frac{(-1)^{d+1}\ell}{8\kappa^2(d-2)!}\epsilon_{\halpha_0\cdots\halpha_{d+1}}F^{\halpha_0\halpha_1}\wedge F^{\halpha_2\halpha_3}\wedge\rD Y^{\halpha_4}\wedge\cdots\wedge\rD Y^{\halpha_{d+1}}
\,.
\end{eqnarray}
It is natural to consider the generalization of this formula in various cases. For example, one can investigate its supersymmetric generalization based on the local $OSp(1|4;{\rm R})$ covariant formalism \cite{Han:2024txq} of $N=1$ AdS$_4$ supergravity.
On the other hand, the most general formula of the characteristic class in the pure bosonic case could be 
\begin{eqnarray}
I_{(d+2)}\!\!\!\!&=&\!\!\!\!\sum_{i=1}^{\lfloor \frac{d}2\rfloor+1}a_i\epsilon_{\halpha_0\cdots\halpha_{d+1}}F^{\halpha_0\halpha_1}\wedge\cdots\wedge F^{\halpha_{2i-2}\halpha_{2i-1}}\wedge\rD Y^{\halpha_{2i}}\wedge\cdots\wedge\rD Y^{\halpha_{d+1}}
\,.
\end{eqnarray}
Now the corresponding bulk theory will be the $SO(2,d)$ gauge theory uplifting of the Lovelock gravity \cite{Lovelock:1971yv} for which the bulk covariant action is in the form
\begin{eqnarray}\label{lovelock}
S^{\rm cov}_{(d+1)}\!\!\!\!&=&\!\!\!\!\int_{M}\sum_{i=0}^{\lfloor \frac{d+1}2\rfloor}\lambda_i\epsilon_{\halpha_0\cdots\halpha_{d+1}}F^{\halpha_0\halpha_1}\wedge\cdots\wedge F^{\halpha_{2i-2}\halpha_{2i-1}}\wedge\rD Y^{\halpha_{2i}}\wedge\cdots\wedge\rD Y^{\halpha_{d}}Y^{\halpha_{d+1}}
\,.
\end{eqnarray}
In \cite{Wang:2020eln}, it is suggested that the string low energy effective action could always be recast as Lovelock type of theory to all orders in $\alpha'$. Therefore, by considering the gauge field and ruler field associated with the stringy gauge symmetries, it is possible to recast the string low energy effective theory as a gauge theory in the form of (\ref{lovelock}).

\vskip 1cm
\centerline{\bf\large Acknowledgments}
\vskip 5mm\noindent
The authors thank Chen-Xu Han, Xing Huang, Pu-Jian Mao, Seung-Joo Lee, Bo-Han Li, Hong L\"u, Jian-Xin Lu, Jun-Jin Peng, Jun-Bao Wu, Hai-Tang Yang and Piljin Yi for useful conversations.
This work is supported by National Natural Science Foundation of China(Grants No. 11305125, No. 12047502, No. 12275217, No. 12247103), the Basic Research Program of Natural Science of Shaanxi Province(Grants No. 2019JM-026), and the Double First-class University Construction Project of Northwest University.

\appendix
\setcounter{equation}{0}
\setcounter{subsection}{0}
\renewcommand{\theequation}{A.\arabic{equation}}
\section{$SO(2,d)$ uplifting of Einstein gravity}
\subsection{Vielbein formalism of Einstein gravity}
In terms of the vielbein formalism, the $D=d+1$-dimensional Einstein-Hilbert action with negative cosmological constant
\begin{eqnarray}
\Lambda\!\!\!\!&=&\!\!\!\!-\frac{(D-1)(D-2)}{2 \ell^2}
\end{eqnarray}
can be re-written as
\begin{eqnarray}\label{2ndGR}
S[e^{a}]\!\!\!\!&=&\!\!\!\!\frac{1}{2\kappa^2\,(D-2)!}\int\epsilon_{a_1\cdots a_D} \left[\Theta^{a_1a_2}+\frac{(D-2)}{D\,\ell^2} e^{a_1}\wedge e^{a_2}\right]\wedge e^{a_3}\wedge\cdots \wedge e^{a_D}\,.
\end{eqnarray}
Our convention for the unit total antisymmetric tensor $\epsilon_{a_1\cdots a_D}$ of the $SO(1,d)$ vielbein indices $a_i=0,\cdots,d$ is  
\begin{eqnarray} 
\epsilon_{0\cdots d}=1.
\end{eqnarray}
In the usual 2nd order understanding of Einstein gravity, the fundamental dynamical field is just the vielbein 1-form $e^a=e^{a}{}_{M}\rd x^{M}$ where $\{x^{M}\}$ are the bulk spacetime coordinates.
The curvature 2-form
\begin{eqnarray}
\Theta^{a}{}_{b}\!\!\!\!&=&\!\!\!\!\rd\omega^{a}{}_{b}+\omega^{a}{}_{c}\wedge\omega^{c}{}_{b}=\tfrac12R^{a}{}_{bMN}\rd x^{M}\wedge\rd x^{N}\,,
\end{eqnarray}
is the field strength of the spin connection 1-form $\omega^{ab}=-\omega^{ba}$ which is just the gauge field of the local $SO(1,d)$ group. Providing the torsion free condition
\begin{eqnarray}\label{TF2}
\rD^{(\omega)} e^{a}=\rd e^a+\omega^{a}{}_{b}\wedge e^b=0\,,
\end{eqnarray}
the spin connection is totally fixed by the vielbein.
Taking the variation of the action (\ref{2ndGR}), we get the Einstein equation
\begin{eqnarray}
\left(\Theta^{[a_1a_2}+\ell^{-2}e^{[a_1}\wedge e^{a_2}\right)
\wedge e^{a_3}\wedge\cdots\wedge e^{a_{D-1}]}=0\,.
\end{eqnarray}

In the Palatini's 1st order understanding of gravity, both the vielbein and the spin connection are regarded  as independent variables
\begin{eqnarray}\label{1stGR}
S[e^{a},\omega^{a}{}_{b}]\!\!\!\!&=&\!\!\!\!\frac{1}{2\kappa^2\,(D-2)!}\int\epsilon_{a_1\cdots a_D} \left[\Theta^{a_1a_2}+\frac{(D-2)}{D\,\ell^2} e^{a_1}\wedge e^{a_2}\right]\wedge e^{a_3}\wedge\cdots \wedge e^{a_D}\,.~~~
\end{eqnarray}
The corresponding EOM's from the variation of $e^a$ and $\omega^{ab}$ are respectively
\begin{eqnarray}\label{Ein1}
\left(\Theta^{[a_1a_2}+\ell^{-2}e^{[a_1}\wedge e^{a_2}\right)
\wedge e^{a_3}\wedge\cdots\wedge e^{a_{D-1}]}=0\,,
\\\label{TF1}
\rD^{(\omega)} e^{[a_1}\wedge e^{a_{2}}\wedge\cdots \wedge e^{a_{D-2}]}=0\,.
\end{eqnarray}
Providing that the vielbein $e^{a}$ is not degenerate, the second equation (\ref{TF1}) is equivalent to the torsion free condition (\ref{TF2}). Therefore, the Palatini action is equivalent to the original AdS gravity classically.
\subsection{From $SO(1,d)$ to $SO(2,d)$}
The Palatini's only have the $SO(1,d)$ gauge symmetry which comes from the ambiguities‌ in choosing the vielbein. As explained in the main text, there is a natural way to unify the vielbein $e^a$ and the $SO(1,d)$  spin connection $\omega^{ab}$ as different components of the $SO(2,d)$ gauge field ${A}^{\halpha\hbeta}$. That is
\begin{eqnarray}\label{initi}
{A}^{\hat a\hat b}=\omega^{ab}\,,~~~{A}^{\hat a\hat\bullet}=\ell^{-1} e^{a}=-{A}^{\hat\bullet\hat a}\,,
\end{eqnarray}
where we use $\hat\bullet$ to denote the additional indices of $SO(2,d)$ vector space and $\halpha,\hbeta$ are the vector indices of $SO(2,d)$ group.
Then the corresponding field strength is related to the curvature as well as the torsion
\begin{eqnarray}
{F}^{ab}\!\!\!\!&=&\!\!\!\!
\rd{A}^{ab}+{A}^{a}{}_{\hgamma}\wedge{A}^{\hgamma b}
=\rd\omega^{ab}+\omega^{a}{}_c\wedge\omega^{cb}+\ell^{-2}e^{a}\wedge e^{b}
=\Theta^{ab}+\ell^{-2}e^{a}\wedge e^{b}\,,~~~
\cr
{F}^{a\hat\bullet}\!\!\!\!&=&\!\!\!\!
\rd{A}^{a\hat\bullet}+{A}^{a}{}_{\hgamma}\wedge{A}^{\hgamma\hat\bullet}
=\ell^{-1}(\rd e^a+\omega^{a}{}_c\wedge e^{c})=\ell^{-1}\,\rD^{(\omega)} e^a\,.
\end{eqnarray}
From the $SO(2,d)$ point of view, the choice of the $\hat\bullet$ direction itself is only invariant under the $SO(1,d)$ subgroup perpendicular to the $\hat\bullet$ direction. By a generic $SO(2,d)$ rotation, the choice of the $\hat\bullet$ direction shall also be changed covariant. To exhibit the choice of the $\hat\bullet$ direction in a $SO(2,d)$ covariant way, we introduce an auxiliary field $Y^{\halpha}$ which is in the vector representation of $SO(2,d)$. Since the $Y^{\halpha}$ field is imposed only for pointing out the $\hat\bullet$ direction covariantly, its norm is not relevant in the $SO(1,d)\to SO(2,d)$ uplifting. Thus we also need to impose the $SO(2,d)$ invariant constraint 
\begin{eqnarray}
Y^{\halpha}Y_{\halpha}=-\ell^2\,.
\end{eqnarray}
which fixes the norm of $Y^{\halpha}$. Such an auxiliary field is named as ruler field in \cite{Wang:2018tyq}.  
Obviously, the formula of the ruler field $Y^{\halpha}$ can be totally fixed by the $SO(2,d)$ gauge choice, e.g., the Einstein gauge  
\begin{eqnarray}\label{Egauge}
Y_{\hat a}=0\,,~~~~Y_{\hat\bullet}=\ell\,,
\end{eqnarray}
which recovers our initial idea of $SO(2,d)$ uplifting (\ref{initi}).
After the gauge choice,  $Y^{\halpha}$ is totally fixed and the residual gauge symmetry becomes $SO(1,d)$. Thus the introducing of the ruler field  does not introduce any additional physical degree of freedom. In the language of the fiber bundle, the ruler field is just a section $y^{\halpha}=Y^{\halpha}(x)$ of the  AdS bundle $E$
\begin{eqnarray}\begin{array}{ccccc}
{\rm AdS}_{d+1} &\longrightarrow& E &\overset{\pi}{\longrightarrow}&  M_{D} \\
\{y^{\halpha}\} &\longrightarrow& \{y^{\halpha},x^{M}\} &\overset{\pi}{\longrightarrow}& \{x^{M}\}
                \end{array}
\end{eqnarray}
upon the $D=d+1$ dimensional base manifold $M_{D}$ with coordinates $\{x^{M}\}$.
The fiber space is the pure AdS space
\begin{eqnarray}
{\rm AdS}_{d+1}=\frac{SO(2,d)}{SO(1,d)}
\end{eqnarray}
which is described by the redundant embedding coordinates $y^{\halpha}$ with the constraint
\begin{eqnarray}
\eta_{\halpha\hbeta}y^{\halpha}y^{\hbeta}=-\ell^2\,.
\end{eqnarray}

In the Einstein gauge (\ref{Egauge}), we have
\begin{eqnarray}
\rD Y^{\hat a}\!\!\!\!&=&\!\!\!\!\rd Y^{\hat a}+{A}^{\hat a}{}_{\hbeta}Y^{\hbeta}=\ell{A}^{\hat a\hat\bullet}=e^a\,,~~~~~~
\rD Y^{\hat\bullet}=\rd Y^{\hat\bullet}+{A}^{\hat\bullet}{}_{\hbeta}Y^{\hbeta}=0\,,
\cr \rD\rD Y^{\hat a}\!\!\!\!&=&\!\!\!\!{F}^{\hat a}{}_{\hbeta}Y^{\hbeta}
=\ell{F}^{\hat a\hat\bullet}=\rD^{(\omega)} e^a\,,~~~~~~~~~~
\rD\rD Y^{\hat\bullet}={F}^{\hat\bullet}{}_{\hat b}Y^{\hat b}=0\,,
\end{eqnarray}
where $\rD=\rd+A$ is the $SO(2,d)$ covariant derivative.
Now the Palatini EOMs (\ref{Ein1}) and (\ref{TF1})
can be nicely unified in a $SO(2,d)$ covariant way
\begin{eqnarray}\label{UPala}
 F^{[\halpha_1\halpha_2}\wedge\rD Y^{\halpha_3}\wedge\cdots\wedge \rD Y^{\halpha_{d}]}=0\,.
\end{eqnarray}
One can further realize the uplifting at the action level. In the Einstein gauge (\ref{Egauge}), we have
\begin{eqnarray}
\!\!\!\!&&\!\!\!\!\left(\Theta^{[a_1a_2}+\ell^{-2}e^{[a_1}\wedge e^{a_2}\right)
\wedge e^{a_3}\wedge\cdots\wedge e^{a_{D}]}
\cr \!\!\!\!&=&\!\!\!\!(D+1)\,\ell^{-1} F^{[\halpha_1\halpha_2}\wedge \rD Y^{\halpha_3}\wedge\cdots\wedge \rD Y^{\halpha_{D}}Y^{\halpha_{D+1}]}\,,
\cr\!\!\!\!&&\!\!\!\!e^{[a_1}\wedge e^{a_2}
\wedge e^{a_3}\wedge\cdots\wedge e^{a_{D}]}
\cr \!\!\!\!&=&\!\!\!\!(D+1)\ell^{-1}\rD Y^{[\halpha_1}\wedge \rD Y^{\halpha_2}\wedge\cdots\wedge \rD Y^{\halpha_{D}}Y^{\halpha_{D+1}]}\,.
\end{eqnarray}
It suggests that the following manifestly $SO(2,d)$ gauge invariant action is 
\begin{eqnarray}\label{GRGT}
S^{\rm cov}[Y,A]
\!\!\!\!&=&\!\!\!\!\frac{1}{2\kappa^2\ell\,(D-2)!}\int_{M} \epsilon_{\halpha_1\cdots \halpha_{D+1}} \left[ F^{\halpha_1\halpha_2}-\frac{2}{D \ell^2} \rD Y^{\halpha_1}\wedge  \rD Y^{\halpha_2}\right]\wedge  \rD Y^{\halpha_3}\wedge\cdots \wedge  \rD Y^{\halpha_D}Y^{\halpha_{D+1}}
\,.~~~~~~~~~~
\end{eqnarray}
where $\epsilon_{\halpha_1\cdots \halpha_{D+1}}$ is the unit total antisymmetric tensor of the $SO(2,d)$ indices $\halpha_i=\hat\bullet,\hat 0,\cdots,\hat d.$ In the Einstein gauge (\ref{Egauge}), it recovers the 1st order formalism (\ref{1stGR}) of Einstein gravity. 
\subsection{Equations of motion}
Now let us check that the equations of motion for (\ref{GRGT}) is just the unified Palatini equations of motion (\ref{UPala}). We notice that
\begin{eqnarray}
\!\!\!\!&&\!\!\!\!\delta \int_{M} \epsilon_{\halpha_1\cdots \halpha_{D+1}}
F^{\halpha_1\halpha_2}\wedge  \rD Y^{\halpha_3}\wedge\cdots \wedge  \rD Y^{\halpha_D}Y^{\halpha_{D+1}}
\cr\!\!\!\!&=&\!\!\!\!\int_{M} \Big\{\epsilon_{\halpha_1\cdots \halpha_{D+1}}
\rD\delta A^{\halpha_1\halpha_2}\wedge  \rD Y^{\halpha_3}\wedge\cdots \wedge  \rD Y^{\halpha_D}Y^{\halpha_{D+1}}
\cr&&~~~~+(D-2)\epsilon_{\halpha_1\cdots \halpha_{D+1}}
F^{\halpha_1\halpha_2}\wedge (\rD \delta Y^{\halpha_3}+  \delta A^{\halpha_3}{}_{\hbeta}Y^{\hbeta})\wedge\rD Y^{\halpha_4}\wedge\cdots \wedge  \rD Y^{\halpha_D}Y^{\halpha_{D+1}}
\cr&&~~~~+\epsilon_{\halpha_1\cdots \halpha_{D+1}}
F^{\halpha_1\halpha_2} \wedge  \rD Y^{\halpha_3}\wedge\cdots \wedge  \rD Y^{\halpha_D}\delta Y^{\halpha_{D+1}}\Big\}
\cr\!\!\!\!&=&\!\!\!\!\int_{M} \Big\{(D-2)\epsilon_{\halpha_1\cdots \halpha_{D+1}}
\delta A^{\halpha_1\halpha_2}\wedge F^{\halpha_3}{}_{\hbeta} Y^{\hbeta}\wedge \rD Y^{\halpha_4}\wedge\cdots \wedge  \rD Y^{\halpha_D}Y^{\halpha_{D+1}}
\cr&&~~~~~~+(-1)^{D}\epsilon_{\halpha_1\cdots \halpha_{D+1}}
\delta A^{\halpha_1\halpha_2}\wedge\rD Y^{\halpha_3}\wedge\cdots \wedge  \rD Y^{\halpha_D} \wedge  \rD Y^{\halpha_{D+1}}
\cr&&~~~~~~-(D-2)(D-3)\epsilon_{\halpha_1\cdots \halpha_{D+1}}
\delta Y^{\halpha_1} F^{\halpha_2\halpha_3}\wedge F^{\halpha_4}{}_{\hbeta} Y^{\hbeta} \wedge\rD Y^{\halpha_5}\wedge\cdots \wedge  \rD Y^{\halpha_D}Y^{\halpha_{D+1}}
\cr&&~~~~~~+(-1)^D(D-2)\epsilon_{\halpha_1\cdots \halpha_{D+1}}\delta Y^{\halpha_1}
 F^{\halpha_2\halpha_3}\wedge\rD Y^{\halpha_4}\wedge\cdots \wedge  \rD Y^{\halpha_D} \wedge \rD Y^{\halpha_{D+1}}
\cr&&~~~~~~+(D-2)\epsilon_{\halpha_1\cdots \halpha_{D+1}}
\delta A^{\halpha_1}{}_{\hbeta}Y^{\hbeta}\wedge F^{\halpha_2\halpha_3}\wedge\rD Y^{\halpha_4}\wedge\cdots \wedge  \rD Y^{\halpha_D}Y^{\halpha_{D+1}}
\cr&&~~~~~~+\epsilon_{\halpha_1\cdots \halpha_{D+1}}
F^{\halpha_1\halpha_2} \wedge  \rD Y^{\halpha_3}\wedge\cdots \wedge  \rD Y^{\halpha_D}\delta Y^{\halpha_{D+1}}\Big\}
\,,\\
\!\!\!\!&&\!\!\!\!\delta\int_{M} \epsilon_{\halpha_1\cdots \halpha_{D+1}} \rD Y^{\halpha_1} \wedge\cdots \wedge  \rD Y^{\halpha_D}Y^{\halpha_{D+1}}
\cr\!\!\!\!&=&\!\!\!\!\int_{M}\Big\{ D\epsilon_{\halpha_1\cdots \halpha_{D+1}} (\rD \delta Y^{\halpha_1}+\delta A^{\halpha_1}{}_{\hbeta}Y^{\hbeta})\wedge  \rD Y^{\halpha_2}\wedge \cdots \wedge  \rD Y^{\halpha_D}Y^{\halpha_{D+1}}
\cr&&~~~~~~+\epsilon_{\halpha_1\cdots \halpha_{D+1}}
\rD Y^{\halpha_1}\wedge\cdots \wedge  \rD Y^{\halpha_D}\delta Y^{\halpha_{D+1}}
\Big\}
\cr\!\!\!\!&=&\!\!\!\!\int_{M}\Big\{-D(D-1)\epsilon_{\halpha_1\cdots \halpha_{D+1}} \delta Y^{\halpha_1} F^{\halpha_2}{}_{\hbeta}Y^{\hbeta}\wedge  \rD Y^{\halpha_3}\wedge \cdots \wedge  \rD Y^{\halpha_D}Y^{\halpha_{D+1}}
\cr&&~~~~~~+(-1)^DD\,\epsilon_{\halpha_1\cdots \halpha_{D+1}} \delta Y^{\halpha_1} \wedge  \rD Y^{\halpha_2}\wedge  \rD Y^{\halpha_3}\wedge \cdots \wedge  \rD Y^{\halpha_D}\wedge  \rD Y^{\halpha_{D+1}}
\cr&&~~~~~~
+D\epsilon_{\halpha_1\cdots \halpha_{D+1}} \delta A^{\halpha_1}{}_{\hbeta}Y^{\hbeta}\wedge  \rD Y^{\halpha_2}\wedge \cdots \wedge  \rD Y^{\halpha_D}Y^{\halpha_{D+1}}
\cr&&~~~~~~+\epsilon_{\halpha_1\cdots \halpha_{D+1}}
\rD Y^{\halpha_1}\wedge\cdots \wedge  \rD Y^{\halpha_D}\delta Y^{\halpha_{D+1}}
\Big\}
\,.~~~~~
\end{eqnarray}
At the 3rd steps, we have used the integration by parts as well as 
\begin{eqnarray}
\rD\rD Y=FY\,,~~~~~~~\rD F=0\,.~~~~~~~~~~
\end{eqnarray}
By using the constraint $Y^{\hbeta}Y_{\hbeta}=-\ell^2$ as well as the facts
\begin{eqnarray}
\rD Y^{\hbeta}Y_{\hbeta}=0\,,~~~~~~~\delta Y^{\hbeta}Y_{\hbeta}=0\,,
\end{eqnarray}
which are implied by the constraint, we can get the resummation identities{\footnote{The illustration for the proof of the resummation identities is presented in the Appendix A.4}}
\begin{eqnarray}
\!\!\!\!&&\!\!\!\!\epsilon_{\halpha_1\cdots \halpha_{D+1}}\delta A^{\halpha_1}{}_{\hbeta}Y^{\hbeta}\wedge F^{\halpha_2\halpha_3}\wedge\rD Y^{\halpha_4}\wedge\cdots \wedge  \rD Y^{\halpha_D}Y^{\halpha_{D+1}}
\cr\!\!\!\!&=&\!\!\!\!-\epsilon_{\halpha_1\cdots \halpha_{D+1}}\delta A^{\halpha_1\halpha_2}\wedge F^{\halpha_3}{}_{\hbeta}Y^{\hbeta}\wedge\rD Y^{\halpha_4}\wedge\cdots \wedge  \rD Y^{\halpha_D}Y^{\halpha_{D+1}} 
\cr\!\!\!\!&&\!\!\!\!
+\frac{(-1)^D\ell^2}2\epsilon_{\halpha_1\cdots \halpha_{D+1}}\delta A^{\halpha_1\halpha_{2}}\wedge F^{\halpha_3\halpha_4}\wedge\rD Y^{\halpha_5}\wedge\cdots \wedge  \rD Y^{\halpha_{D+1}}\,,
\\\cr\!\!\!\!&&\!\!\!\!\label{resum2}
\epsilon_{\halpha_1\cdots \halpha_{D+1}}
\delta Y^{\halpha_1} F^{\halpha_2\halpha_3}\wedge F^{\halpha_4}{}_{\hbeta} Y^{\hbeta} \wedge\rD Y^{\halpha_5}\wedge\cdots \wedge  \rD Y^{\halpha_D}Y^{\halpha_{D+1}}
\cr\!\!\!\!&=&\!\!\!\!\frac{(-1)^{D+1}\ell^2}4\epsilon_{\halpha_1\cdots \halpha_{D+1}}
\delta Y^{\halpha_1} F^{\halpha_2\halpha_3}\wedge F^{\halpha_4\halpha_{5}}  \wedge\rD Y^{\halpha_6}\wedge\cdots \wedge  \rD Y^{\halpha_{D+1}}\,,
\\\cr
\!\!\!\!&&\!\!\!\!\epsilon_{\halpha_1\cdots \halpha_{D+1}}
 \delta A^{\halpha_1}{}_{\hbeta}Y^{\hbeta}\wedge  \rD Y^{\halpha_2}\wedge \cdots \wedge  \rD Y^{\halpha_D}Y^{\halpha_{D+1}}
\cr\!\!\!\!&=&\!\!\!\!\frac{(-1)^D\ell^2}2\epsilon_{\halpha_1\cdots \halpha_{D+1}}\delta A^{\halpha_1\halpha_{2}}\wedge F^{\halpha_3\halpha_4}\wedge\rD Y^{\halpha_5}\wedge\cdots \wedge  \rD Y^{\halpha_{D+1}}
\,,
\\\cr
\!\!\!\!&&\!\!\!\!\epsilon_{\halpha_1\cdots \halpha_{D+1}}
\delta Y^{\halpha_1}F^{\halpha_2}{}_{\hbeta}Y^{\hbeta}\wedge  \rD Y^{\halpha_3}\wedge \cdots \wedge  \rD Y^{\halpha_D}Y^{\halpha_{D+1}}
\cr\!\!\!\!&=&\!\!\!\!\frac{(-1)^{D+1}\ell^2}2\epsilon_{\halpha_1\cdots \halpha_{D+1}}\delta Y^{\halpha_1}  F^{\halpha_2\halpha_{3}}\wedge\rD Y^{\halpha_4}\wedge\cdots \wedge  \rD Y^{\halpha_{D+1}}\,.
\end{eqnarray} 
Thus
\begin{eqnarray}
\!\!\!\!&&\!\!\!\!\delta \int_{M} \epsilon_{\halpha_1\cdots \halpha_{D+1}}
F^{\halpha_1\halpha_2}\wedge  \rD Y^{\halpha_3}\wedge\cdots \wedge  \rD Y^{\halpha_D}Y^{\halpha_{D+1}}
\cr\!\!\!\!&=&\!\!\!\!\int_{M} \Big\{(-1)^{D}\epsilon_{\halpha_1\cdots \halpha_{D+1}}
\delta A^{\halpha_1\halpha_2}\wedge \left[\frac{\ell^2(D-2)}2F^{\halpha_3\halpha_4}+\rD Y^{\halpha_3}\wedge\rD Y^{\halpha_4}\right]\wedge\rD Y^{\halpha_5}\wedge\cdots \wedge  \rD Y^{\halpha_{D+1}}
\cr&&~~~~~~+(-1)^D\epsilon_{\halpha_1\cdots \halpha_{D+1}}\delta Y^{\halpha_1}\Big[\frac{(D-2)(D-3)\ell^2}4
F^{\halpha_2\halpha_3}\wedge F^{\halpha_4\halpha_{5}}  \wedge\rD Y^{\halpha_6}\wedge\cdots \wedge  \rD Y^{\halpha_{D+1}}
\cr&&~~~~~~~~~~~~~~~~~~~~~~~~~~~~~~~~~~~~~~+(D-1) F^{\halpha_2\halpha_3}\wedge
\rD Y^{\halpha_4}\wedge\cdots \wedge  \rD Y^{\halpha_{D+1}} \Big]\Big\}
\,,~~~~~~~~\\\cr
\!\!\!\!&&\!\!\!\!\delta\int_{M} \epsilon_{\halpha_1\cdots \halpha_{D+1}} \rD Y^{\halpha_1} \wedge\cdots \wedge  \rD Y^{\halpha_D}Y^{\halpha_{D+1}}
\cr\!\!\!\!&=&\!\!\!\!\int_{M}\Big\{
\frac{(-1)^D\ell^2D}2\epsilon_{\halpha_1\cdots \halpha_{D+1}}\delta A^{\halpha_1\halpha_{2}}\wedge F^{\halpha_3\halpha_4}\wedge\rD Y^{\halpha_5}\wedge\cdots \wedge  \rD Y^{\halpha_{D+1}}
\cr&&~~~~~~+(-1)^D\,\epsilon_{\halpha_1\cdots \halpha_{D+1}}\delta Y^{\halpha_1}\Big[
\frac{D(D-1)\ell^2}2 F^{\halpha_2\halpha_{3}}\wedge\rD Y^{\halpha_4}\wedge\cdots \wedge  \rD Y^{\halpha_{D+1}}
+ (D+1)\rD Y^{\halpha_2}\wedge \cdots \wedge    \rD Y^{\halpha_{D+1}}\Big]
\Big\}
\cr\!\!\!\!&=&\!\!\!\!\int_{M}\Big\{
\frac{(-1)^D\ell^2D}2\epsilon_{\halpha_1\cdots \halpha_{D+1}}\delta A^{\halpha_1\halpha_{2}}\wedge F^{\halpha_3\halpha_4}\wedge\rD Y^{\halpha_5}\wedge\cdots \wedge  \rD Y^{\halpha_{D+1}}
\cr&&~~~~~~+\frac{(-1)^D D(D-1)\ell^2}2\,\epsilon_{\halpha_1\cdots \halpha_{D+1}}\delta Y^{\halpha_1} 
F^{\halpha_2\halpha_{3}}\wedge\rD Y^{\halpha_4}\wedge\cdots \wedge  \rD Y^{\halpha_{D+1}}\Big\}
\,,~~~~~
\end{eqnarray}
where we have used that $\delta Y^{[\halpha_1}\rD Y^{\halpha_2}\wedge \cdots \wedge    \rD Y^{\halpha_{D+1}]}=0$ since there are only $D$ linear independent directions which are orthogonal with  $Y^{\halpha}$.
In total, we have 
\begin{eqnarray}
\!\!\!\!&&\!\!\!\!2\kappa^2\ell\,\,(D-2)!\,\delta S^{\rm cov}[Y,A]
\cr\!\!\!\!&=&\!\!\!\!\delta\int_{M} \epsilon_{\halpha_1\cdots \halpha_{D+1}} \left[ F^{\halpha_1\halpha_2}-\frac{2}{D \ell^2} \rD Y^{\halpha_1}\wedge  \rD Y^{\halpha_2}\right]\wedge  \rD Y^{\halpha_3}\wedge\cdots \wedge  \rD Y^{\halpha_D}Y^{\halpha_{D+1}}
\cr\!\!\!\!&=&\!\!\!\!\int_{M} \Big\{\frac{(-1)^{D}\ell^2(D-2)}2\epsilon_{\halpha_1\cdots \halpha_{D+1}}
\delta A^{\halpha_1\halpha_2}\wedge F^{\halpha_3\halpha_4}\wedge\rD Y^{\halpha_5}\wedge\cdots \wedge  \rD Y^{\halpha_{D+1}}
\cr&&~~~~~~+\frac{(-1)^D(D-2)(D-3)\ell^2}4\epsilon_{\halpha_1\cdots \halpha_{D+1}}\delta Y^{\halpha_1}
F^{\halpha_2\halpha_3}\wedge F^{\halpha_4\halpha_{5}}  \wedge\rD Y^{\halpha_6}\wedge\cdots \wedge  \rD Y^{\halpha_{D+1}}\Big\}
\cr\!\!\!\!&=&\!\!\!\!\int_{M} \Big\{\frac{(-1)^{D}\ell^2(D-2)}2\epsilon_{\halpha_1\cdots \halpha_{D+1}}
\delta A^{\halpha_1\halpha_2}\wedge F^{\halpha_3\halpha_4}\wedge\rD Y^{\halpha_5}\wedge\cdots \wedge  \rD Y^{\halpha_{D+1}}
\cr&&~~~~~~-(D-2)(D-3)
\epsilon_{\halpha_1\cdots \halpha_{D+1}}
\delta Y^{\halpha_1}F^{\halpha_2\halpha_3}\wedge F^{\halpha_4}{}_{\hbeta} Y^{\hbeta} \wedge\rD Y^{\halpha_5}\wedge\cdots \wedge  \rD Y^{\halpha_D}Y^{\halpha_{D+1}}\Big\}
\,.~~~~~~~~~~
\end{eqnarray}
At the last step, by using the resummation identity (\ref{resum2}) backwardly, we have reformulated the $\delta Y^{\halpha}V_{\halpha}$ term  such that
\begin{eqnarray}
V_{\halpha} Y^{\halpha}=0~~\Leftrightarrow~~~~~ V^{\halpha}=(\delta^{\halpha}_{\hbeta}+\ell^{-2}Y^{\halpha}Y_{\hbeta}) V^{\hbeta}\,.\nn
\end{eqnarray}

In summary, the bulk equations of motion for $\delta A$ are
\begin{eqnarray}
\!\!\!\!&&\!\!\!\!\epsilon_{\halpha\hbeta\halpha_1\cdots \halpha_{d}}F^{\halpha_1\halpha_2}\wedge  \rD Y^{\halpha_3}\wedge\cdots \wedge  \rD Y^{\halpha_{d}}=0
\end{eqnarray}
which are exactly the unified Palatini equations (\ref{UPala}). On the other hand, since the components of $\delta Y^{\halpha}$ are constrained by the linear relation  
\begin{eqnarray}
Y_{\halpha}\delta Y^{\halpha}=0\,,
\end{eqnarray} 
the correct EOM of $\delta Y^{\halpha}$ should be given by the $V_{\halpha}$ satisfying $V_{\halpha} Y^{\halpha}=0$. That is
\begin{eqnarray}
\epsilon_{\halpha\halpha_1\cdots \halpha_{d+1}}
F^{\halpha_1\halpha_2}\wedge F^{\halpha_3}{}_{\hbeta} Y^{\hbeta} \wedge\rD Y^{\halpha_4}\wedge\cdots \wedge  \rD Y^{\halpha_d}Y^{\halpha_{d+1}}=0\,.
\end{eqnarray}
Since
\begin{eqnarray}
\!\!\!\!&&\!\!\!\!\epsilon_{\halpha\hbeta\halpha_1\cdots \halpha_{d}}Y^{\hbeta}\rD(F^{\halpha_1\halpha_2}\wedge  \rD Y^{\halpha_3}\wedge\cdots \wedge  \rD Y^{\halpha_{d}})
\cr\!\!\!\!&=&\!\!\!\!(-1)^d(d-2)\epsilon_{\halpha\halpha_1\cdots \halpha_{d+1}}F^{\halpha_1\halpha_2}\wedge F^{\halpha_3}{}_{\hbeta} Y^{\hbeta} \wedge\rD Y^{\halpha_4}\wedge\cdots \wedge  \rD Y^{\halpha_d}Y^{\halpha_{d+1}}\,,
\end{eqnarray}
the $\delta Y^{\halpha}$ EOM is automatically satisfied if the EOM for $\delta A$ is satisfied. Therefore, the introducing of $Y^{\halpha}$ does not impose any additional constraints beyond the original Palatini equations of motion.
\subsection{Proof of resummation identities} 
Let us take 
\begin{eqnarray}
\!\!\!\!&&\!\!\!\!\epsilon_{\halpha_1\cdots \halpha_{D+1}}\delta A^{\halpha_1}{}_{\hbeta}Y^{\hbeta}\wedge F^{\halpha_2\halpha_3}\wedge\rD Y^{\halpha_4}\wedge\cdots \wedge  \rD Y^{\halpha_D}Y^{\halpha_{D+1}}
\cr\!\!\!\!&=&\!\!\!\!-\epsilon_{\halpha_1\cdots \halpha_{D+1}}\delta A^{\halpha_1\halpha_2}\wedge F^{\halpha_3}{}_{\hbeta}Y^{\hbeta}\wedge\rD Y^{\halpha_4}\wedge\cdots \wedge  \rD Y^{\halpha_D}Y^{\halpha_{D+1}} 
\cr\!\!\!\!&&\!\!\!\!
+\frac{(-1)^D\ell^2}2\epsilon_{\halpha_1\cdots \halpha_{D+1}}\delta A^{\halpha_1\halpha_{2}}\wedge F^{\halpha_3\halpha_4}\wedge\rD Y^{\halpha_5}\wedge\cdots \wedge  \rD Y^{\halpha_{D+1}}
\end{eqnarray}
as a example to illustrate the proof of the resummation identities. To avoid confusion of the notations in the middle steps, let us recover the notation of summations which is ignored in the Einstein summation convention. We find    
\begin{eqnarray}
\!\!\!\!&&\!\!\!\!\sum_{\halpha_1,\cdots ,\halpha_{D+1},\hbeta}\epsilon_{\halpha_1\cdots \halpha_{D+1}}\delta A^{\halpha_1}{}_{\hbeta}Y^{\hbeta}\wedge F^{\halpha_2\halpha_3}\wedge\rD Y^{\halpha_4}\wedge\cdots \wedge  \rD Y^{\halpha_D}Y^{\halpha_{D+1}}
\cr\!\!\!\!&=&\!\!\!\!\sum_{\halpha_1,\cdots ,\halpha_{D+1}}\epsilon_{\halpha_1\cdots \halpha_{D+1}}\sum_{\hbeta}\delta A^{\halpha_1}{}_{\hbeta}Y^{\hbeta}\wedge F^{\halpha_2\halpha_3}\wedge\rD Y^{\halpha_4}\wedge\cdots \wedge  \rD Y^{\halpha_D}Y^{\halpha_{D+1}}
\cr\!\!\!\!&=&\!\!\!\!\sum_{\halpha_1,\cdots ,\halpha_{D+1}}\epsilon_{\halpha_1\cdots \halpha_{D+1}}\delta A^{\halpha_1}{}_{\halpha_1}Y^{\halpha_1}\wedge F^{\halpha_2\halpha_3}\wedge\rD Y^{\halpha_4}\wedge\cdots \wedge  \rD Y^{\halpha_D}Y^{\halpha_{D+1}}
\cr\!\!\!\!&&\!\!\!\!+\sum_{\halpha_1,\cdots ,\halpha_{D+1}}\epsilon_{\halpha_1\cdots \halpha_{D+1}}\delta A^{\halpha_1}{}_{\halpha_2}Y^{\halpha_2}\wedge F^{\halpha_2\halpha_3}\wedge\rD Y^{\halpha_4}\wedge\cdots \wedge  \rD Y^{\halpha_D}Y^{\halpha_{D+1}}
\cr\!\!\!\!&&\!\!\!\!+\sum_{\halpha_1,\cdots ,\halpha_{D+1}}\epsilon_{\halpha_1\cdots \halpha_{D+1}}\delta A^{\halpha_1}{}_{\halpha_3}Y^{\halpha_3}\wedge F^{\halpha_2\halpha_3}\wedge\rD Y^{\halpha_4}\wedge\cdots \wedge  \rD Y^{\halpha_D}Y^{\halpha_{D+1}}
\cr\!\!\!\!&&\!\!\!\!+\sum_{\halpha_1,\cdots ,\halpha_{D+1}}\epsilon_{\halpha_1\cdots \halpha_{D+1}}\delta A^{\halpha_1}{}_{\halpha_4}Y^{\halpha_4}\wedge F^{\halpha_2\halpha_3}\wedge\rD Y^{\halpha_4}\wedge\cdots \wedge  \rD Y^{\halpha_D}Y^{\halpha_{D+1}}
\cr\!\!\!\!&&\!\!\!\!+\cdots+\sum_{\halpha_1,\cdots ,\halpha_{D+1}}\epsilon_{\halpha_1\cdots \halpha_{D+1}}\delta A^{\halpha_1}{}_{\halpha_D}Y^{\halpha_D}\wedge F^{\halpha_2\halpha_3}\wedge\rD Y^{\halpha_4}\wedge\cdots \wedge  \rD Y^{\halpha_D}Y^{\halpha_{D+1}}
\cr\!\!\!\!&&\!\!\!\!+\sum_{\halpha_1,\cdots ,\halpha_{D+1}}\epsilon_{\halpha_1\cdots \halpha_{D+1}}\delta A^{\halpha_1}{}_{\halpha_{D+1}}Y^{\halpha_{D+1}}\wedge F^{\halpha_2\halpha_3}\wedge\rD Y^{\halpha_4}\wedge\cdots \wedge  \rD Y^{\halpha_D}Y^{\halpha_{D+1}}
\cr\!\!\!\!&=&\!\!\!\!0
+2\sum_{\halpha_1,\cdots ,\halpha_{D+1}}\epsilon_{\halpha_1\cdots \halpha_{D+1}}\delta A^{\halpha_1}{}_{\halpha_2}Y^{\halpha_2}\wedge F^{\halpha_2\halpha_3}\wedge\rD Y^{\halpha_4}\wedge\cdots \wedge  \rD Y^{\halpha_D}Y^{\halpha_{D+1}}
\cr\!\!\!\!&&\!\!\!\!+(D-3)\sum_{\halpha_1,\cdots ,\halpha_{D+1}}\epsilon_{\halpha_1\cdots \halpha_{D+1}}\delta A^{\halpha_1}{}_{\halpha_4}Y^{\halpha_4}\wedge F^{\halpha_2\halpha_3}\wedge\rD Y^{\halpha_4}\wedge\cdots \wedge  \rD Y^{\halpha_D}Y^{\halpha_{D+1}}
\cr\!\!\!\!&&\!\!\!\!+\sum_{\halpha_1,\cdots ,\halpha_{D+1}}\epsilon_{\halpha_1\cdots \halpha_{D+1}}\delta A^{\halpha_1}{}_{\halpha_{D+1}}Y^{\halpha_{D+1}}\wedge F^{\halpha_2\halpha_3}\wedge\rD Y^{\halpha_4}\wedge\cdots \wedge  \rD Y^{\halpha_D}Y^{\halpha_{D+1}}
\,.~~~~~
\end{eqnarray}
At the 3rd step, we have used the fact that in our present computation
\begin{eqnarray}
\epsilon_{\halpha_1\cdots \halpha_{D+1}}\sum_{\hbeta=\hblt,\hat0,\cdots,\hat d}
\!\!\!\!&\Leftrightarrow&\!\!\!\!\epsilon_{\halpha_1\cdots \halpha_{D+1}}\sum_{\hbeta=\halpha_1,\cdots,\halpha_{D+1}}
\,.~~~~~
\end{eqnarray}
since $\epsilon_{\halpha_1\cdots \halpha_{D+1}}$ is non-zero only when $\halpha_i\neq\halpha_{j}~(\forall i\neq j)$. At the 4th step, the terms with same structures are collected together.  

On the other hand, we similarly notice that  
\begin{eqnarray}
\!\!\!\!&&\!\!\!\!2\sum_{\halpha_1,\cdots ,\halpha_{D+1}}\epsilon_{\halpha_1\cdots \halpha_{D+1}}\delta A^{\halpha_1\halpha_2}Y_{\halpha_2}\wedge F^{\halpha_2\halpha_3}\wedge\rD Y^{\halpha_4}\wedge\cdots \wedge  \rD Y^{\halpha_D}Y^{\halpha_{D+1}}
\cr\!\!\!\!&=&\!\!\!\!\sum_{\halpha_1,\cdots ,\halpha_{D+1}}\epsilon_{\halpha_1\cdots \halpha_{D+1}}\sum_{\hbeta}\delta A^{\halpha_1\halpha_2}Y_{\hbeta}\wedge F^{\hbeta\halpha_3}\wedge\rD Y^{\halpha_4}\wedge\cdots \wedge  \rD Y^{\halpha_D}Y^{\halpha_{D+1}}
\cr\!\!\!\!&&\!\!\!\!-\sum_{\halpha_1,\cdots ,\halpha_{D+1}}\epsilon_{\halpha_1\cdots \halpha_{D+1}}\delta A^{\halpha_1\halpha_2}Y_{\halpha_3}\wedge F^{\halpha_3\halpha_3}\wedge\rD Y^{\halpha_4}\wedge\cdots \wedge  \rD Y^{\halpha_D}Y^{\halpha_{D+1}}
\cr\!\!\!\!&&\!\!\!\!-(D-3)\sum_{\halpha_1,\cdots ,\halpha_{D+1}}\epsilon_{\halpha_1\cdots \halpha_{D+1}}\delta A^{\halpha_1\halpha_2}Y_{\halpha_4}\wedge F^{\halpha_4\halpha_3}\wedge\rD Y^{\halpha_4}\wedge\cdots \wedge  \rD Y^{\halpha_D}Y^{\halpha_{D+1}}
\cr\!\!\!\!&&\!\!\!\!-\sum_{\halpha_1,\cdots ,\halpha_{D+1}}\epsilon_{\halpha_1\cdots \halpha_{D+1}}\delta A^{\halpha_1\halpha_{2}}Y_{\halpha_{D+1}}\wedge F^{\halpha_{D+1}\halpha_3}\wedge\rD Y^{\halpha_4}\wedge\cdots \wedge  \rD Y^{\halpha_D}Y^{\halpha_{D+1}}
\cr\!\!\!\!&=&\!\!\!\!\sum_{\halpha_1,\cdots ,\halpha_{D+1}}\epsilon_{\halpha_1\cdots \halpha_{D+1}}\sum_{\hbeta}\delta A^{\halpha_1\halpha_2}Y_{\hbeta}\wedge F^{\hbeta\halpha_3}\wedge\rD Y^{\halpha_4}\wedge\cdots \wedge  \rD Y^{\halpha_D}Y^{\halpha_{D+1}}-0
\cr\!\!\!\!&&\!\!\!\!-(D-3)\sum_{\halpha_1,\cdots ,\halpha_{D+1}}\epsilon_{\halpha_1\cdots \halpha_{D+1}}\delta A^{\halpha_1\halpha_2}Y_{\halpha_4}\wedge F^{\halpha_4\halpha_3}\wedge\rD Y^{\halpha_4}\wedge\cdots \wedge  \rD Y^{\halpha_D}Y^{\halpha_{D+1}}
\cr\!\!\!\!&&\!\!\!\!-\sum_{\halpha_1,\cdots ,\halpha_{D+1}}\epsilon_{\halpha_1\cdots \halpha_{D+1}}\delta A^{\halpha_1\halpha_{2}}Y_{\halpha_{D+1}}\wedge F^{\halpha_{D+1}\halpha_3}\wedge\rD Y^{\halpha_4}\wedge\cdots \wedge  \rD Y^{\halpha_D}Y^{\halpha_{D+1}}
\,,~~~~~
\\\cr
\!\!\!\!&&\!\!\!\!2\sum_{\halpha_1,\cdots ,\halpha_{D+1}}\epsilon_{\halpha_1\cdots \halpha_{D+1}}\delta A^{\halpha_1\halpha_4}Y_{\halpha_4}\wedge F^{\halpha_2\halpha_3}\wedge\rD Y^{\halpha_4}\wedge\rD Y^{\halpha_5}\wedge\cdots \wedge  \rD Y^{\halpha_D}Y^{\halpha_{D+1}}
\cr\!\!\!\!&=&\!\!\!\!\sum_{\halpha_1,\cdots ,\halpha_{D+1}}\epsilon_{\halpha_1\cdots \halpha_{D+1}}\sum_{\hbeta}\delta A^{\halpha_1\halpha_4}Y_{\hbeta}\wedge F^{\halpha_2\halpha_3}\wedge\rD Y^{\hbeta}\wedge\rD Y^{\halpha_5}\wedge\cdots \wedge  \rD Y^{\halpha_D}Y^{\halpha_{D+1}}
\cr\!\!\!\!&&\!\!\!\!-2\sum_{\halpha_1,\cdots ,\halpha_{D+1}}\epsilon_{\halpha_1\cdots \halpha_{D+1}}\delta A^{\halpha_1\halpha_4}Y_{\halpha_2}\wedge F^{\halpha_2\halpha_3}\wedge\rD Y^{\halpha_2}\wedge\rD Y^{\halpha_5}\wedge\cdots \wedge  \rD Y^{\halpha_D}Y^{\halpha_{D+1}}
\cr\!\!\!\!&&\!\!\!\!-(D-4)\sum_{\halpha_1,\cdots ,\halpha_{D+1}}\epsilon_{\halpha_1\cdots \halpha_{D+1}}\delta A^{\halpha_1\halpha_4}Y_{\halpha_5}\wedge F^{\halpha_2\halpha_3}\wedge\rD Y^{\halpha_5}\wedge\rD Y^{\halpha_5}\wedge\cdots \wedge  \rD Y^{\halpha_D}Y^{\halpha_{D+1}}
\cr\!\!\!\!&&\!\!\!\!-\sum_{\halpha_1,\cdots ,\halpha_{D+1}}\epsilon_{\halpha_1\cdots \halpha_{D+1}}\delta A^{\halpha_1\halpha_4}Y_{\halpha_{D+1}}\wedge F^{\halpha_2\halpha_3}\wedge\rD Y^{\halpha_{D+1}}\wedge\rD Y^{\halpha_5}\wedge\cdots \wedge  \rD Y^{\halpha_D}Y^{\halpha_{D+1}}
\cr\!\!\!\!&=&\!\!\!\!\sum_{\halpha_1,\cdots ,\halpha_{D+1}}\epsilon_{\halpha_1\cdots \halpha_{D+1}}\sum_{\hbeta}\delta A^{\halpha_1\halpha_4}Y_{\hbeta}\wedge F^{\halpha_2\halpha_3}\wedge\rD Y^{\hbeta}\wedge\rD Y^{\halpha_5}\wedge\cdots \wedge  \rD Y^{\halpha_D}Y^{\halpha_{D+1}}
\cr\!\!\!\!&&\!\!\!\!-2\sum_{\halpha_1,\cdots ,\halpha_{D+1}}\epsilon_{\halpha_1\cdots \halpha_{D+1}}\delta A^{\halpha_1\halpha_4}Y_{\halpha_2}\wedge F^{\halpha_2\halpha_3}\wedge\rD Y^{\halpha_2}\wedge\rD Y^{\halpha_5}\wedge\cdots \wedge  \rD Y^{\halpha_D}Y^{\halpha_{D+1}}-0
\cr\!\!\!\!&&\!\!\!\!-\sum_{\halpha_1,\cdots ,\halpha_{D+1}}\epsilon_{\halpha_1\cdots \halpha_{D+1}}\delta A^{\halpha_1\halpha_4}Y_{\halpha_{D+1}}\wedge F^{\halpha_2\halpha_3}\wedge\rD Y^{\halpha_{D+1}}\wedge\rD Y^{\halpha_5}\wedge\cdots \wedge  \rD Y^{\halpha_D}Y^{\halpha_{D+1}}\,,
\\
\cr\!\!\!\!&&\!\!\!\!
2\sum_{\halpha_1,\cdots ,\halpha_{D+1}}\epsilon_{\halpha_1\cdots \halpha_{D+1}}\delta A^{\halpha_1\halpha_{D+1}}Y_{\halpha_{D+1}}\wedge F^{\halpha_2\halpha_3}\wedge\rD Y^{\halpha_4}\wedge\cdots \wedge  \rD Y^{\halpha_D}Y^{\halpha_{D+1}}
\cr\!\!\!\!&=&\!\!\!\!
\sum_{\halpha_1,\cdots ,\halpha_{D+1}}\epsilon_{\halpha_1\cdots \halpha_{D+1}}\sum_{\hbeta}\delta A^{\halpha_1\halpha_{D+1}}Y_{\hbeta}\wedge F^{\halpha_2\halpha_3}\wedge\rD Y^{\halpha_4}\wedge\cdots \wedge  \rD Y^{\halpha_D}Y^{\hbeta}
\cr\!\!\!\!&&\!\!\!\!
-2\sum_{\halpha_1,\cdots ,\halpha_{D+1}}\epsilon_{\halpha_1\cdots \halpha_{D+1}}\delta A^{\halpha_1\halpha_{D+1}}Y_{\halpha_2}\wedge F^{\halpha_2\halpha_3}\wedge\rD Y^{\halpha_4}\wedge\cdots \wedge  \rD Y^{\halpha_D}Y^{\halpha_2}
\cr\!\!\!\!&&\!\!\!\!
-(D-3)\sum_{\halpha_1,\cdots ,\halpha_{D+1}}\epsilon_{\halpha_1\cdots \halpha_{D+1}}\delta A^{\halpha_1\halpha_{D+1}}Y_{\halpha_4}\wedge F^{\halpha_2\halpha_3}\wedge\rD Y^{\halpha_4}\wedge\cdots \wedge  \rD Y^{\halpha_D}Y^{\halpha_4}\,.
\end{eqnarray}
It follows that 
\begin{eqnarray}
\!\!\!\!&&\!\!\!\!\sum_{\halpha_1,\cdots ,\halpha_{D+1},\hbeta}\epsilon_{\halpha_1\cdots \halpha_{D+1}}\delta A^{\halpha_1}{}_{\hbeta}Y^{\hbeta}\wedge F^{\halpha_2\halpha_3}\wedge\rD Y^{\halpha_4}\wedge\cdots \wedge  \rD Y^{\halpha_D}Y^{\halpha_{D+1}}
\cr\!\!\!\!&=&\!\!\!\!
2\sum_{\halpha_1,\cdots ,\halpha_{D+1}}\epsilon_{\halpha_1\cdots \halpha_{D+1}}\delta A^{\halpha_1}{}_{\halpha_2}Y^{\halpha_2}\wedge F^{\halpha_2\halpha_3}\wedge\rD Y^{\halpha_4}\wedge\cdots \wedge  \rD Y^{\halpha_D}Y^{\halpha_{D+1}}
\cr\!\!\!\!&&\!\!\!\!+(D-3)\sum_{\halpha_1,\cdots ,\halpha_{D+1}}\epsilon_{\halpha_1\cdots \halpha_{D+1}}\delta A^{\halpha_1}{}_{\halpha_4}Y^{\halpha_4}\wedge F^{\halpha_2\halpha_3}\wedge\rD Y^{\halpha_4}\wedge\cdots \wedge  \rD Y^{\halpha_D}Y^{\halpha_{D+1}}
\cr\!\!\!\!&&\!\!\!\!+\sum_{\halpha_1,\cdots ,\halpha_{D+1}}\epsilon_{\halpha_1\cdots \halpha_{D+1}}\delta A^{\halpha_1}{}_{\halpha_{D+1}}Y^{\halpha_{D+1}}\wedge F^{\halpha_2\halpha_3}\wedge\rD Y^{\halpha_4}\wedge\cdots \wedge  \rD Y^{\halpha_D}Y^{\halpha_{D+1}}
\cr\!\!\!\!&=&\!\!\!\!
\sum_{\halpha_1,\cdots ,\halpha_{D+1}}\epsilon_{\halpha_1\cdots \halpha_{D+1}}\sum_{\hbeta}\delta A^{\halpha_1\halpha_2}Y_{\hbeta}\wedge F^{\hbeta\halpha_3}\wedge\rD Y^{\halpha_4}\wedge\cdots \wedge  \rD Y^{\halpha_D}Y^{\halpha_{D+1}} 
\cr\!\!\!\!&&\!\!\!\!-(D-3)\sum_{\halpha_1,\cdots ,\halpha_{D+1}}\epsilon_{\halpha_1\cdots \halpha_{D+1}}\delta A^{\halpha_1\halpha_2}Y_{\halpha_4}\wedge F^{\halpha_4\halpha_3}\wedge\rD Y^{\halpha_4}\wedge\cdots \wedge  \rD Y^{\halpha_D}Y^{\halpha_{D+1}}
\cr\!\!\!\!&&\!\!\!\!-\sum_{\halpha_1,\cdots ,\halpha_{D+1}}\epsilon_{\halpha_1\cdots \halpha_{D+1}}\delta A^{\halpha_1\halpha_{2}}Y_{\halpha_{D+1}}\wedge F^{\halpha_{D+1}\halpha_3}\wedge\rD Y^{\halpha_4}\wedge\cdots \wedge  \rD Y^{\halpha_D}Y^{\halpha_{D+1}}
\cr\!\!\!\!&&\!\!\!\!+\frac{D-3}2\sum_{\halpha_1,\cdots ,\halpha_{D+1}}\epsilon_{\halpha_1\cdots \halpha_{D+1}}\sum_{\hbeta}\delta A^{\halpha_1\halpha_4}Y_{\hbeta}\wedge F^{\halpha_2\halpha_3}\wedge\rD Y^{\hbeta}\wedge\rD Y^{\halpha_5}\wedge\cdots \wedge  \rD Y^{\halpha_D}Y^{\halpha_{D+1}}
\cr\!\!\!\!&&\!\!\!\!-(D-3)\sum_{\halpha_1,\cdots ,\halpha_{D+1}}\epsilon_{\halpha_1\cdots \halpha_{D+1}}\delta A^{\halpha_1\halpha_4}Y_{\halpha_2}\wedge F^{\halpha_2\halpha_3}\wedge\rD Y^{\halpha_2}\wedge\rD Y^{\halpha_5}\wedge\cdots \wedge  \rD Y^{\halpha_D}Y^{\halpha_{D+1}}
\cr\!\!\!\!&&\!\!\!\!-\frac{D-3}2\sum_{\halpha_1,\cdots ,\halpha_{D+1}}\epsilon_{\halpha_1\cdots \halpha_{D+1}}\delta A^{\halpha_1\halpha_4}Y_{\halpha_{D+1}}\wedge F^{\halpha_2\halpha_3}\wedge\rD Y^{\halpha_{D+1}}\wedge\rD Y^{\halpha_5}\wedge\cdots \wedge  \rD Y^{\halpha_D}Y^{\halpha_{D+1}}
\cr\!\!\!\!&&\!\!\!\!
+\frac12\sum_{\halpha_1,\cdots ,\halpha_{D+1}}\epsilon_{\halpha_1\cdots \halpha_{D+1}}\sum_{\hbeta}\delta A^{\halpha_1\halpha_{D+1}}Y_{\hbeta}\wedge F^{\halpha_2\halpha_3}\wedge\rD Y^{\halpha_4}\wedge\cdots \wedge  \rD Y^{\halpha_D}Y^{\hbeta}
\cr\!\!\!\!&&\!\!\!\!
-\sum_{\halpha_1,\cdots ,\halpha_{D+1}}\epsilon_{\halpha_1\cdots \halpha_{D+1}}\delta A^{\halpha_1\halpha_{D+1}}Y_{\halpha_2}\wedge F^{\halpha_2\halpha_3}\wedge\rD Y^{\halpha_4}\wedge\cdots \wedge  \rD Y^{\halpha_D}Y^{\halpha_2}
\cr\!\!\!\!&&\!\!\!\!
-\frac{D-3}2\sum_{\halpha_1,\cdots ,\halpha_{D+1}}\epsilon_{\halpha_1\cdots \halpha_{D+1}}\delta A^{\halpha_1\halpha_{D+1}}Y_{\halpha_4}\wedge F^{\halpha_2\halpha_3}\wedge\rD Y^{\halpha_4}\wedge\cdots \wedge  \rD Y^{\halpha_D}Y^{\halpha_4}
\cr\!\!\!\!&=&\!\!\!\!
\sum_{\halpha_1,\cdots ,\halpha_{D+1}}\epsilon_{\halpha_1\cdots \halpha_{D+1}}\sum_{\hbeta}\delta A^{\halpha_1\halpha_2}Y_{\hbeta}\wedge F^{\hbeta\halpha_3}\wedge\rD Y^{\halpha_4}\wedge\cdots \wedge  \rD Y^{\halpha_D}Y^{\halpha_{D+1}} 
\cr\!\!\!\!&&\!\!\!\!+\frac{D-3}2\sum_{\halpha_1,\cdots ,\halpha_{D+1}}\epsilon_{\halpha_1\cdots \halpha_{D+1}}\sum_{\hbeta}\delta A^{\halpha_1\halpha_4}Y_{\hbeta}\wedge F^{\halpha_2\halpha_3}\wedge\rD Y^{\hbeta}\wedge\rD Y^{\halpha_5}\wedge\cdots \wedge  \rD Y^{\halpha_D}Y^{\halpha_{D+1}}
\cr\!\!\!\!&&\!\!\!\!
+\frac12\sum_{\halpha_1,\cdots ,\halpha_{D+1}}\epsilon_{\halpha_1\cdots \halpha_{D+1}}\sum_{\hbeta}\delta A^{\halpha_1\halpha_{D+1}}Y_{\hbeta}\wedge F^{\halpha_2\halpha_3}\wedge\rD Y^{\halpha_4}\wedge\cdots \wedge  \rD Y^{\halpha_D}Y^{\hbeta}
\cr\!\!\!\!&&\!\!\!\!-(D-3)\sum_{\halpha_1,\cdots ,\halpha_{D+1}}\epsilon_{\halpha_1\cdots \halpha_{D+1}} \left[\delta A^{\halpha_1\halpha_4}Y_{\halpha_2}\wedge F^{\halpha_2\halpha_3}\wedge\rD Y^{\halpha_2}\wedge\rD Y^{\halpha_5}\wedge\cdots \wedge  \rD Y^{\halpha_D}Y^{\halpha_{D+1}}+(\halpha_4\leftrightarrow\halpha_2)\right]
\cr\!\!\!\!&&\!\!\!\!-\frac{D-3}2\sum_{\halpha_1,\cdots ,\halpha_{D+1}}\epsilon_{\halpha_1\cdots \halpha_{D+1}}\left[\delta A^{\halpha_1\halpha_4}Y_{\halpha_{D+1}}\wedge F^{\halpha_2\halpha_3}\wedge\rD Y^{\halpha_{D+1}}\wedge\rD Y^{\halpha_5}\wedge\cdots \wedge  \rD Y^{\halpha_D}Y^{\halpha_{D+1}}
+(\halpha_4\leftrightarrow\halpha_{d+1})\right]
\cr\!\!\!\!&&\!\!\!\!
-\sum_{\halpha_1,\cdots ,\halpha_{D+1}}\epsilon_{\halpha_1\cdots \halpha_{D+1}}\left[\delta A^{\halpha_1\halpha_{D+1}}Y_{\halpha_2}\wedge F^{\halpha_2\halpha_3}\wedge\rD Y^{\halpha_4}\wedge\cdots \wedge  \rD Y^{\halpha_D}Y^{\halpha_2}+(\halpha_2\leftrightarrow\halpha_{d+1})\right]
\cr\!\!\!\!&=&\!\!\!\!
\sum_{\halpha_1,\cdots ,\halpha_{D+1}}\epsilon_{\halpha_1\cdots \halpha_{D+1}}\sum_{\hbeta}\delta A^{\halpha_1\halpha_2}Y_{\hbeta}\wedge F^{\hbeta\halpha_3}\wedge\rD Y^{\halpha_4}\wedge\cdots \wedge  \rD Y^{\halpha_D}Y^{\halpha_{D+1}} 
\cr\!\!\!\!&&\!\!\!\!+\frac{D-3}2\sum_{\halpha_1,\cdots ,\halpha_{D+1}}\epsilon_{\halpha_1\cdots \halpha_{D+1}}\sum_{\hbeta}\delta A^{\halpha_1\halpha_4}Y_{\hbeta}\wedge F^{\halpha_2\halpha_3}\wedge\rD Y^{\hbeta}\wedge\rD Y^{\halpha_5}\wedge\cdots \wedge  \rD Y^{\halpha_D}Y^{\halpha_{D+1}}
\cr\!\!\!\!&&\!\!\!\!
+\frac12\sum_{\halpha_1,\cdots ,\halpha_{D+1}}\epsilon_{\halpha_1\cdots \halpha_{D+1}}\sum_{\hbeta}\delta A^{\halpha_1\halpha_{D+1}}Y_{\hbeta}\wedge F^{\halpha_2\halpha_3}\wedge\rD Y^{\halpha_4}\wedge\cdots \wedge  \rD Y^{\halpha_D}Y^{\hbeta}\,,
\end{eqnarray}
where all the terms containing $[\cdots]$ vanish since $\epsilon_{\halpha_1\cdots \halpha_{D+1}}$ is totally antisymmetric. In summary, we have 
\begin{eqnarray}
\!\!\!\!&&\!\!\!\!\delta A^{\halpha_1}{}_{\hbeta}Y^{\hbeta}\wedge F^{\halpha_2\halpha_3}\wedge\rD Y^{\halpha_4}\wedge\cdots \wedge  \rD Y^{\halpha_D}Y^{\halpha_{D+1}}
\cr\!\!\!\!&=&\!\!\!\!\epsilon_{\halpha_1\cdots \halpha_{D+1}}\delta A^{\halpha_1\halpha_2}Y_{\hbeta}\wedge F^{\hbeta\halpha_3}\wedge\rD Y^{\halpha_4}\wedge\cdots \wedge  \rD Y^{\halpha_D}Y^{\halpha_{D+1}} 
\cr\!\!\!\!&&\!\!\!\!+\frac{D-3}2\epsilon_{\halpha_1\cdots \halpha_{D+1}}\delta A^{\halpha_1\halpha_4}Y_{\hbeta}\wedge F^{\halpha_2\halpha_3}\wedge\rD Y^{\hbeta}\wedge\rD Y^{\halpha_5}\wedge\cdots \wedge  \rD Y^{\halpha_D}Y^{\halpha_{D+1}}
\cr\!\!\!\!&&\!\!\!\!
+\frac12\epsilon_{\halpha_1\cdots \halpha_{D+1}}\delta A^{\halpha_1\halpha_{D+1}}Y_{\hbeta}\wedge F^{\halpha_2\halpha_3}\wedge\rD Y^{\halpha_4}\wedge\cdots \wedge  \rD Y^{\halpha_D}Y^{\hbeta}
\cr\!\!\!\!&=&\!\!\!\!\epsilon_{\halpha_1\cdots \halpha_{D+1}}\delta A^{\halpha_1\halpha_2}Y_{\hbeta}\wedge F^{\hbeta\halpha_3}\wedge\rD Y^{\halpha_4}\wedge\cdots \wedge  \rD Y^{\halpha_D}Y^{\halpha_{D+1}} 
+0
\cr\!\!\!\!&&\!\!\!\!
-\frac{\ell^2}2\epsilon_{\halpha_1\cdots \halpha_{D+1}}\delta A^{\halpha_1\halpha_{D+1}}\wedge F^{\halpha_2\halpha_3}\wedge\rD Y^{\halpha_4}\wedge\cdots \wedge  \rD Y^{\halpha_D}
\cr\!\!\!\!&=&\!\!\!\!-\epsilon_{\halpha_1\cdots \halpha_{D+1}}\delta A^{\halpha_1\halpha_2}\wedge F^{\halpha_3}{}_{\hbeta}Y^{\hbeta}\wedge\rD Y^{\halpha_4}\wedge\cdots \wedge  \rD Y^{\halpha_D}Y^{\halpha_{D+1}} 
\cr\!\!\!\!&&\!\!\!\!
+\frac{(-1)^D\ell^2}2\epsilon_{\halpha_1\cdots \halpha_{D+1}}\delta A^{\halpha_1\halpha_{2}}\wedge \wedge F^{\halpha_3\halpha_4}\wedge\rD Y^{\halpha_5}\wedge\cdots \wedge  \rD Y^{\halpha_{D+1}}
\,,~~~~~
\end{eqnarray}
where we have used that $Y^{\hbeta}Y_{\hbeta}=-\ell^2$ and
$\rD Y^{\hbeta}Y_{\hbeta}=0$. 
\section{Intrinsic expansion}
In the $SO(2,d)$ gauge theory formalism, the spacetime metric is given by 
\begin{eqnarray}
g_{MN}=\rD_{M}Y^{\halpha}\rD_{N}Y_{\halpha}\,.~~~
\end{eqnarray}
When the metric $g_{MN}$ is not degenerate, together with
\begin{eqnarray}
Y^{\halpha}Y_{\halpha}\!\!\!\!&=&\!\!\!\!-\ell^2\,,~~~~~~~~
 Y^{\halpha}\rD_{M } Y_{\halpha}=0\,.~~~
\end{eqnarray}
it is noticed that the $d+2$ vectors $\{Y^{\halpha},\rD_{M }Y^{\halpha}\}$ naturally forms an intrinsic basis of the $d+2$ dimensional space of the $SO(2,d)$ vector representation.   The completion relation  is obviously
\begin{eqnarray}
\rD_{M}Y^{\halpha}\rD^{M}Y_{\hbeta}-\ell^{-2} Y^{\halpha} Y_{\hbeta}=\delta^{\halpha}_{\hbeta}\,,
\end{eqnarray}
where the spacetime indices are rising by $g^{MN}$.

To clarify the relation between the $SO(2,d)$ gauge theory notations and the usual $SO(2,d)$ invariant notations of gravity, let us expand the quantities with $SO(2,d)$ indices by the intrinsic basis. We notice that
\begin{eqnarray}
 Y^{\halpha}\rD_{N}\rD_{M } Y_{\halpha}\!\!\!\!&=&\!\!\!\!-g_{M N}\,,
\cr
\rD_{N}Y^{\halpha}\rD_{(M_2}\rD_{M_1)} Y_{\halpha}
\!\!\!\!&=&\!\!\!\!\frac12\left[\rD_{M_2}(\rD_{N}Y^{\halpha}\rD_{M_1} Y_{\halpha})+\rD_{M_1}(\rD_{N}Y^{\halpha}\rD_{M_2} Y_{\halpha})
-\rD_{N}(\rD_{M_1}Y^{\halpha}\rD_{M_2} Y_{\halpha})\right]
\cr\!\!\!\!&&\!\!\!\!+\rD_{[N}\rD_{M_1]}Y^{\halpha}\rD_{M_2} Y_{\halpha}
+\rD_{M_1}Y^{\halpha}\rD_{[N}\rD_{M_2]} Y_{\halpha}
\cr\!\!\!\!&=&\!\!\!\!\frac12\left(\partial_{M_2}g_{N M_1}+\partial_{M_1}g_{NM_2}
-\partial_{N}g_{M_1M_2}\right)+(F_{N(M_1})^{\halpha\hbeta}\rD_{M_2)}Y_{\halpha}\,Y_{\hbeta}
\,,~~~
\cr
\rD_{N}Y^{\halpha}\rD_{[M_2}\rD_{M_1]} Y_{\halpha}
\!\!\!\!&=&\!\!\!\!\frac12(F_{M_2M_1})^{\halpha\hbeta}Y_{\hbeta}\rD_{N}Y_{\halpha}\,.
\end{eqnarray}
Thus we have the expansion of $\rD\rD Y$ as
\begin{eqnarray}\label{bulkdev}
\rD_{M_2}\rD_{M_1} Y^{\halpha}
\!\!\!\!&=&\!\!\!\!\ell^{-2} g_{M_1 M_2}Y^{\halpha}+\Gamma^{N}{}_{M_1M_2}\rD_{N}Y^{\halpha}\,,~~~~~~~~
\end{eqnarray}
which suggests the following spacetime connection $\Gamma^{M}{}_{NP}$ with torsion $t^{M}{}_{NP}$
\begin{eqnarray}
\Gamma^{M}{}_{NP}\!\!\!\!&=&\!\!\!\!\hat\Gamma^{M}{}_{NP}+{t}^M{}_{NP}\,,
\cr
\hat\Gamma_{MM_1M_2}\!\!\!\!&=&\!\!\!\!\frac12\left(\partial_{M_2}g_{N M_1}+\partial_{M_1}g_{NM_2}-\partial_{N}g_{M_1M_2}\right)\,,
\cr
{t}_{NM_1M_2}
\!\!\!\!&=&\!\!\!\!\frac{1}{2}\left[(F_{NM_1})^{\hbeta_1\hbeta_2}\rD_{M_2}Y_{\hbeta_1}
+(F_{NM_2})^{\hbeta_1\hbeta_2}\rD_{M_1}Y_{\hbeta_1} -(F_{M_1M_2})^{\hbeta_1\hbeta_2}\rD_{N}Y_{\hbeta_1}
\right]Y_{\hbeta_2}
\,.~~~~~~~~
\end{eqnarray}
The torsion free condition can be expressed covariantly as
\begin{eqnarray}
\rD_{[M_1}\rD_{M_2]} Y^{\halpha}
\!\!\!\!&=&\!\!\!\!\frac12(F_{M_1 M_2}){}^{\halpha}{}_{\hbeta}Y^{\hbeta}=0\,.~~~~~~~~
\end{eqnarray}
Now we can define the torsional spacetime covariant derivative $\nabla=\partial+\Gamma$, as well as the torsional total covariant derivative $\cD=\partial+\Gamma+A=\nabla+A$.  Similarly, the spacetime covariant derivative without torsion is  $\hat\nabla=\partial+\hat\Gamma$, and  the total covariant derivatives without torsion is $\hat\cD=\partial+\hat\Gamma+A=\hat\nabla+A$.  In terms of $\hat\cD$ and $\cD$, we have
\begin{eqnarray}\label{intrin1}
\hat\cD_{M_2}\hat\cD_{M_1} Y^{\halpha}
\!\!\!\!&=&\!\!\!\!\ell^{-2} g_{M_1 M_2}Y^{\halpha}+t^{N}{}_{M_1M_2}\rD_{N}Y^{\halpha}\,,
\\\label{intrin2}
\cD_{M_2}\cD_{M_1} Y^{\halpha}
\!\!\!\!&=&\!\!\!\!\ell^{-2} g_{M_1 M_2}Y^{\halpha}\,.~~~~~~~~
\end{eqnarray}

Now, by using (\ref{bulkdev}),  one can further get the intrinsic expansion of higher derivative terms such as $\rD_{M_1}\rD_{M_2}\rD_{M_3}Y^{\halpha}$.  In terms of the results for $\rD_{[M_1}\rD_{M_2]}\rD_{M_3}Y^{\halpha}$ and $\rD_{[M_1}\rD_{M_2]}Y^{\halpha}$, we find that the intrinsic expansion of the $SO(2,d)$ field strength is
\begin{eqnarray}\label{fluxA}
\!\!\!\!&&\!\!\!\!(F_{M_1M_2})^{\halpha\hbeta}
\cr\!\!\!\!&=&\!\!\!\! \left(R^{N_1N_2}{}_{M_1M_2}+2\ell^{-2}\delta^{N_1}_{[M_1}\delta^{N_2}_{M_2]}\right) \rD_{N_1}Y^{[\halpha}\rD_{N_2}Y^{\hbeta]} -4\ell^{-2} t^{N}{}_{[M_1M_2]}Y^{[\halpha}\rD_{N} Y^{\hbeta]}
\cr\!\!\!\!&=&\!\!\!\!\left(\hat R^{N_1N_2}{}_{M_1M_2}+2\hat\nabla_{[M_1}t^{N_1N_2}{}_{M_2]}
+2t^{N_1}{}_{N[M_1}t^{N N_2}{}_{ M_2]} +2\ell^{-2}\delta^{N_1}_{[M_1}\delta^{N_2}_{M_2]}\right)\rD_{N_1}Y^{[\halpha}\rD_{N_2}Y^{\hbeta]} \cr\!\!\!\!&&\!\!\!\!-4\ell^{-1}t^{N}{}_{[M_1M_2]}Y^{[\halpha}\rD_{N}Y^{\hbeta]}
\,.~~~~~~~~
\end{eqnarray}
where $R^{N_1N_2}{}_{M_1M_2}$ and $\hat R^{N_1N_2}{}_{M_1M_2}$ are respectively the curvature tensors for $\nabla$ and $\hat\nabla$.  
Furthermore, expanding the $SO(2,d)$ Bianchi identity on the intrinsic basis, we get
\begin{eqnarray}
0\!\!\!\!&=&\!\!\!\!\rD_{[M_3}(F_{M_1M_2]})^{\halpha\hbeta}
\cr\!\!\!\!&=&\!\!\!\!
\left(\hat\nabla_{[M_3} \hat R^{N_1N_2}{}_{M_1M_2]}-\hat R^N{}_{[M_2M_3M_1]} t^{N_1N_2}{}_{N} \right)\rD_{N_1}Y^{[\halpha}\rD_{N_2}Y^{\hbeta]}
-2\ell^{-2}\hat R^{N}{}_{[M_3M_1M_2]}Y^{[\halpha}\rD_{N}Y^{\hbeta]}\,.~~~~~~~~~
\end{eqnarray}
It is equivalent to the two Bianchi identities for the usual Riemann curvature.
\subsection{Recovering Einstein-Hilbert action}
By using the intrinsic expansion, one can directly recover the Einstein-Hilbert action without taking any gauge choice.

We notice that
\begin{eqnarray}
\epsilon_{\halpha_1\cdots\halpha_{D+1}}
\rD_{M_1} Y^{\halpha_1}\cdots \rD_{M_D} Y^{\halpha_{D}} Y^{\halpha_{D+1}} \!\!\!\!&=&\!\!\!\!\ell g^{\frac12}\varepsilon_{M_1\cdots M_D}
\end{eqnarray}
where
\begin{eqnarray}
\varepsilon_{01\cdots d}=-1\,,~~~~g=-\det(g_{MN})\,.
\end{eqnarray}
By using the resummation technique, one can further show that 
\begin{eqnarray}\label{ddualA}
\!\!\!\!&&\!\!\!\!\frac{1}{\ell\,p!(d+1-p)!} \epsilon_{\halpha_0\cdots\halpha_{p-1}\halpha_{p}\cdots\halpha_{d+1}}\epsilon^{M_0\cdots M_{p-1}M_{p}\cdots M_{d}} \rD_{M_{p}}Y^{\halpha_{p}}\cdots\rD_{M_{d}}Y^{\halpha_{d}}Y^{\halpha_{d+1}}
\cr\!\!\!\!&=&\!\!\!\!\rD^{[M_{0}}Y_{[\halpha_{0}}\cdots\rD^{M_{p-1}]}Y_{\halpha_{p-1}]}\,,~~~~
\end{eqnarray} 
where
\begin{eqnarray}
\epsilon^{01\cdots d}=\frac1{\sqrt{g}}\,.
\end{eqnarray} 
By using (\ref{ddualA}) as well as the expansion (\ref{fluxA}), we find that
\begin{eqnarray}
\!\!\!\!&&\!\!\!\!\int\epsilon_{\halpha_1\cdots\halpha_{D+1}} F^{\halpha_1\halpha_2}\wedge \rD Y^{\halpha_3}\wedge\cdots\wedge \rD Y^{\halpha_{D}}Y^{\halpha_{D+1}}\,,
\cr\!\!\!\!&=&\!\!\!\!\frac12\,\ell\,2!\,(D-2)!\int\rd^Dx\,\sqrt{g}(F_{M_1M_2})^{\halpha_1\halpha_2} \rD^{[M_{1}}Y_{[\halpha_{1}}\rD^{M_{2}]}Y_{\halpha_{2}]}
\cr\!\!\!\!&=&\!\!\!\!\ell\,(D-2)!\int\rd^Dx\,\sqrt{g}\left[R+\ell^{-2}D(D-1)\right]\,,
\\
\cr \!\!\!\!&&\!\!\!\!\int\epsilon_{\halpha_1\cdots\halpha_{D+1}} \rD Y^{\halpha_1}\wedge \rD Y^{\halpha_2}\wedge\cdots\wedge \rD Y^{\halpha_{D}}Y^{\halpha_{D+1}}
=\ell\,D!\int\rd^Dx\,\sqrt{g}\,.
\end{eqnarray}
It follows that 
\begin{eqnarray}
S^{\rm cov}[Y,A]
\!\!\!\!&=&\!\!\!\!\frac{1}{2\kappa^2\ell\,(D-2)!}\int_{M} \epsilon_{\halpha_1\cdots \halpha_{D+1}} \left[ F^{\halpha_1\halpha_2}-\frac{2}{D \ell^2} \rD Y^{\halpha_1}\wedge  \rD Y^{\halpha_2}\right]\wedge  \rD Y^{\halpha_3}\wedge\cdots \wedge  \rD Y^{\halpha_D}Y^{\halpha_{D+1}}
\cr\!\!\!\!&=&\!\!\!\!\frac{1}{2\kappa^2}\int\rd^Dx\,\sqrt{g}\left[R+\ell^{-2}(D-2)(D-1)\right]
\,,~~~~~~~~~~
\end{eqnarray}
which is exactly the Einstein-Hilbert action with $\Lambda=-\frac{(D-1)(D-2)}{2 \ell^2}$.

\section{Structure of $SO(2,d)$ anomaly}
\subsection{Generic structure of the anomaly}
In the field theory, a classical symmetry which keeps the action invariant might be anomalous at the quantum level. The reason is that the path integral measure might not be invariant under the symmetry transformations \cite{Fujikawa:2004cx}.  If the symmetry is described by a Lie group $\mathcal G$ with background gauge field $A$, the classical conservation law 
\begin{eqnarray}
\rD J =0
\end{eqnarray}
will get modified in presence of the anomaly. The anomalous conservation law 
\begin{eqnarray}
\langle\rD J \rangle=\cA
\end{eqnarray}  
has an extra source term $\cA$ which comes from the Jacobian of the matter field path integral measure under the infinitesimal gauge transformation.  

The explicit forms of the anomaly depend on the regularization procedures \cite{Bertlmann:1996xk,Bardeen:1984pm,Fujikawa:2004cx}. By integrating out the matter fields, one will get the effective action $W[A]$ of the theory 
\begin{eqnarray}
\ep^{\ii W[A]}=\int[\mathfrak{D}\phi]\ep^{\ii S[\phi;A]}\,. 
\end{eqnarray}
In presence of the anomaly, the effective action $W[A]$ is not gauge invariant and its gauge transformation 
\begin{eqnarray}
\cA^{\rm con}\sim\delta_{\mathcal G}W[A] 
\end{eqnarray}  
will gives rise to the so-called the consistent anomaly $\cA^{\rm con}$. By construction, we have
\begin{eqnarray}
\rD J^{\rm con}=\cA^{\rm con}
\end{eqnarray}
where the consistent current is simply
\begin{eqnarray}
J^{\rm con}=\frac{\delta W[A]}{\delta A}\,. 
\end{eqnarray} 
When the Lie group $\mathcal G$ is non-abelian, the consistent anomaly itself is not a covariant tensor under the gauge transformation. Alternatively, one can adjust the regularization procedures such that $\langle J \rangle$ and thus  $\langle \rD J \rangle$  are gauge covariant. This procedure gives rise to the covariant current $J^{\rm cov}$ and covariant anomaly $\cA^{\rm cov}$.

By using the BRST formalism, one can  establish a generic descendent description\cite{Bertlmann:1996xk,Bardeen:1984pm,Fujikawa:2004cx}  of the non-abelian anomaly 
\begin{eqnarray}
I_{(d+2)}\!\!\!\!&=&\!\!\!\!\rd I_{(d+1)}
\cr\delta_{u} I_{(d+1)}\!\!\!\!&=&\!\!\!\!\rd I_{(d)}
\cr&\vdots&
\end{eqnarray}
where $I_{(d)}=\cA^{\rm con}$ is the consistent anomaly. 
Now the anomaly in the $d$-dimensional QFT is uniquely governed by a gauge invariant closed $(d+2)$-form, i.e. the characteristic class $I_{(d+2)}$. Given $I_{(d+2)}$, one can locally have a $(d+1)$-form $I_{(d+1)}$ for which the exterior derivative gives rise to $I_{(d+2)}$. Under the infinitesimal gauge transformation, $I_{(d+1)}$ is invariant up to a boundary term which is exactly the anomaly $I_{(d)}=\cA^{\rm con}$. 

In terms of the descendent formula, one can easily decide the explicit form of the consistent anomaly $\cA^{\rm con}$.  However,  in many cases, the explicit form of the current $J$ would be non-local and is usually unknown.  In the generic discussion of the anomaly \cite{Bertlmann:1996xk,Bardeen:1984pm,Fujikawa:2004cx}, it is proved that the difference between the consistent and covariant currents is a local polynomial of the background gauge field, and it can be decided systematically. This polynomial is called the Bardeen-Zumino polynomial $\cP^{\rm BZ}$. Therefore, one can always get the explicit form of the covariant anomaly by adding $\rD\cP^{\rm BZ}$ to $\cA^{\rm con}$.       
Given the covariant anomaly, the term $I_{(d+1)}$ can be obtained by the homotopic integration
\begin{eqnarray}
I_{(d+1)}\!\!\!\!&=&\!\!\!\!\int_0^1\rd s\,A\wedge \cA^{\rm cov}(A(s))
\,.~~
\end{eqnarray} 
Similarly, the regularized effective action can be expressed as the homotopic integration  
\begin{eqnarray}
W_{(d)}\!\!\!\!&=&\!\!\!\!\int_{\Sigma}\int_{0}^1\rd s\, A\wedge\cJ(A(s))
\,.
\end{eqnarray}
where one can use either the consistent current or covariant current since the homotopic integration of the Bardeen-Zumino polynomial $\cP^{\rm BZ}$ is always zero.
  
The simplest example is the $d=2$  chiral anomaly.  In this case, the characteristic class is
\begin{eqnarray}
I_4\!\!\!\!&=&\!\!\!\!{\rm tr}(F\wedge F)
\end{eqnarray}
and $I_{(3)}$ is just the $d=3$ Chern-Simons term
\begin{eqnarray}
I_3\!\!\!\!&=&\!\!\!\!{\rm tr}\left(A\wedge \rd A +\frac23A\wedge A\wedge A\right)=2\int_0^1\rd s\,{\rm tr}\left[A\wedge F(s)\right]\,.
\end{eqnarray}
The consistent and covariant anomalies are respectively 
\begin{eqnarray}
\cA^{\rm con}\!\!\!\!&=&\!\!\!\!\rd A\,,~~~~~~~~~~~\cA^{\rm cov}=2F\,,
\end{eqnarray}
which are consistent with
\begin{eqnarray}
\cP^{\rm BZ}\!\!\!\!&=&\!\!\!\!A\,.
\end{eqnarray}

\subsection{The bulk covariant action}
The bulk covariant action on $D=d+1$ dimensional manifold $M$ is given by
\begin{eqnarray}
\!\!\!\!&&\!\!\!\!2\kappa^2\ell\,(d-1)!S^{\rm cov}_{(d+1)}=2\kappa^2\ell\,(d-1)!\int_{M}\cL^{\rm cov}_{(d+1)}
\cr\!\!\!\!&=&\!\!\!\!\int_{M} \epsilon_{\halpha\hbeta\halpha_1\cdots \halpha_{d}} \left[ F^{\halpha\hbeta}-\frac{2}{(d+1)\ell^2} \rD Y^{\halpha}\wedge  \rD Y^{\hbeta}\right]\wedge  \rD Y^{\halpha_1}\wedge\cdots \wedge  \rD Y^{\halpha_{d-1}}Y^{\halpha_{d}}
\,.~~~~~~~~~
\end{eqnarray}

\subsubsection{$\rd\cL_{(d+1)}^{\rm cov}=I_{(d+2)}$}
By using the resummations
\begin{eqnarray}
\!\!\!\!&&\!\!\!\!\epsilon_{\halpha\hbeta\halpha_1\cdots \halpha_{d}}F^{\halpha\hbeta}\wedge F^{\halpha_1\hgamma} Y_{\hgamma}\wedge\rD Y^{\halpha_2}\wedge\cdots\wedge\rD Y^{\halpha_{d-1}}Y^{\halpha_{d}}
\cr\!\!\!\!&=&\!\!\!\!-\frac{(-1)^d\ell^2}4\epsilon_{\halpha\hbeta\halpha_1\cdots \halpha_{d}}F^{\halpha\hbeta}\wedge F^{\halpha_1\halpha_2} \wedge\rD Y^{\halpha_3}\wedge\cdots\wedge\rD Y^{\halpha_{d}}\,,
\\\cr\!\!\!\!&&\!\!\!\!
\epsilon_{\halpha\hbeta\halpha_1\cdots \halpha_{d}}F^{\halpha\hgamma} Y_{\hgamma}\wedge\rD Y^{\hbeta}\wedge\rD Y^{\halpha_1}\wedge\cdots \wedge  \rD Y^{\halpha_{d-1}} Y^{\halpha_{d}}
\cr\!\!\!\!&=&\!\!\!\!-\frac{(-1)^d\ell^2}2\epsilon_{\halpha\hbeta\halpha_1\cdots \halpha_{d}}F^{\halpha\hbeta} \wedge\rD Y^{\halpha_1}\wedge\rD Y^{\halpha_2}\wedge\cdots \wedge  \rD Y^{\halpha_{d}}\,,
\end{eqnarray}
we find that the exterior derivative of $\cL_{(d+1)}^{\rm cov}$ is given by
\begin{eqnarray}
\!\!\!\!&&\!\!\!\!2\kappa^2\ell\,(d-1)!\rd \cL^{\rm cov}_{(d+1)}
\cr\!\!\!\!&=&\!\!\!\!\rd\left\{\epsilon_{\halpha\hbeta\halpha_1\cdots \halpha_{d}}\left[ F^{\halpha\hbeta}-\tfrac{2}{(d+1)\ell^{2}} \rD Y^{\halpha}\wedge  \rD Y^{\hbeta}\right]\wedge  \rD Y^{\halpha_1}\wedge\cdots \wedge  \rD Y^{\halpha_{d-1}}Y^{\halpha_{d}}\right\}
\cr\!\!\!\!&=&\!\!\!\!(d-1)\epsilon_{\halpha\hbeta\halpha_1\cdots \halpha_{d}}F^{\halpha\hbeta}\wedge F^{\halpha_1}{}_{\hgamma} Y^{\hgamma}\wedge\rD Y^{\halpha_2}\wedge\cdots \wedge  \rD Y^{\halpha_{d-1}} Y^{\halpha_{d}}
\cr\!\!\!\!&&\!\!\!\!-(-1)^{d}\epsilon_{\halpha\hbeta\halpha_1\cdots \halpha_{d}}F^{\halpha\hbeta}\wedge\rD Y^{\halpha_1}\wedge\cdots \wedge  \rD Y^{\halpha_{d-1}}\wedge\rD Y^{\halpha_{d}}
\cr\!\!\!\!&&\!\!\!\!-2\ell^{-2} \epsilon_{\halpha\hbeta\halpha_1\cdots \halpha_{d}}F^{\halpha}{}_{\hgamma} Y^{\hgamma}\wedge\rD Y^{\hbeta}\wedge\rD Y^{\halpha_1}\wedge\cdots \wedge  \rD Y^{\halpha_{d-1}} Y^{\halpha_{d}}
\cr\!\!\!\!&=&\!\!\!\!-\frac{(-1)^{d}(d-1)\ell^2}4\epsilon_{\halpha\hbeta\halpha_1\cdots\cdots \halpha_{d}}F^{\halpha\hbeta}\wedge F^{\halpha_1\halpha_2} \wedge\rD Y^{\halpha_3}\wedge\cdots \wedge  \rD Y^{\halpha_{d}}
\cr\!\!\!\!&&\!\!\!\!-(-1)^{d}\epsilon_{\halpha\hbeta\halpha_1\cdots \halpha_{d}}F^{\halpha\hbeta}\wedge\rD Y^{\halpha_1}\wedge\cdots \wedge  \rD Y^{\halpha_{d-1}}\wedge\rD Y ^{\halpha_{d}}
\cr\!\!\!\!&&\!\!\!\!+(-1)^{d}\epsilon_{\halpha\hbeta\halpha_1\cdots \halpha_{d}}F^{\halpha\hbeta}\wedge\rD Y^{\halpha_1}\wedge\cdots \wedge  \rD Y^{\halpha_{d-1}}\wedge\rD Y ^{\halpha_{d}}
\cr\!\!\!\!&=&\!\!\!\!\frac{(d-1)\ell^2}{4(-1)^{d+1}}\epsilon_{\halpha\hbeta\halpha_1\cdots\cdots \halpha_{d}}F^{\halpha\hbeta}\wedge F^{\halpha_1\halpha_2} \wedge\rD Y^{\halpha_3}\wedge\cdots \wedge  \rD Y^{\halpha_{d}}\,.
\end{eqnarray}
Thus the characteristic class of the anomaly is 
\begin{eqnarray}
I_{(d+2)}\!\!\!\!&=&\!\!\!\!\frac{(-1)^{d+1}\ell}{8\kappa^2\,(d-2)!}\epsilon_{\halpha\hbeta\halpha_1\cdots\cdots \halpha_{d}}F^{\halpha\hbeta}\wedge F^{\halpha_1\halpha_2} \wedge\rD Y^{\halpha_3}\wedge\cdots \wedge  \rD Y^{\halpha_{d}}\,.
\end{eqnarray}

\subsubsection{Bulk off-shell currents and Noether potentials}
In computing $\delta S^{\rm cov}_{(d+1)}$ in Appx.A.3, the boundary terms are omitted. To get the Noether potentials, we need to included the boundary terms on $\Sigma=\partial M$. By using the resummations
\begin{eqnarray}
\!\!\!\!&&\!\!\!\!\epsilon_{\halpha\hbeta\halpha_1\cdots \halpha_{d}}\delta A^{\halpha\hgamma}Y_{\hgamma}Y^{\hbeta}\wedge F^{\halpha_1\halpha_2}\wedge  \rD Y^{\halpha_3}\wedge\cdots \wedge  \rD Y^{\halpha_{d}}
\cr\!\!\!\!&=&\!\!\!\!-\frac{\ell^2}2\epsilon_{\halpha\hbeta\halpha_1\cdots \halpha_{d}}\delta A^{\halpha\hbeta}\wedge F^{\halpha_1\halpha_2}\wedge  \rD Y^{\halpha_3}\wedge\cdots \wedge  \rD Y^{\halpha_{d}}
\cr\!\!\!\!&&\!\!\!\!-\epsilon_{\halpha\hbeta\halpha_1\cdots \halpha_{d}}\delta A^{\halpha\hbeta}\wedge (FY)^{\halpha_1}Y^{\halpha_2}\wedge  \rD Y^{\halpha_3}\wedge\cdots \wedge  \rD Y^{\halpha_{d}}\,,
\\\cr
\!\!\!\!&&\!\!\!\!\epsilon_{\halpha\hbeta\halpha_1\cdots \halpha_{d}}\delta A^{\halpha\hgamma}Y_{\hgamma}Y^{\hbeta}\wedge  \rD Y^{\halpha_1}\wedge\cdots \wedge  \rD Y^{\halpha_{d}}
\cr\!\!\!\!&=&\!\!\!\!-\frac{\ell^2}2\epsilon_{\halpha\hbeta\halpha_1\cdots \halpha_{d}}\delta A^{\halpha\hbeta}\wedge  \rD Y^{\halpha_1}\wedge\cdots \wedge  \rD Y^{\halpha_{d}}\,,
\\\cr
\!\!\!\!&&\!\!\!\!-\ell^2\epsilon_{\halpha\hbeta\halpha_1\cdots \halpha_{d}}\delta Y^{\halpha}F^{\hbeta\halpha_1}\wedge  \rD Y^{\halpha_2}\wedge\cdots \wedge  \rD Y^{\halpha_{d}}
\cr\!\!\!\!&=&\!\!\!\!-2\epsilon_{\halpha\hbeta\halpha_1\cdots \halpha_{d}}\delta Y^{\halpha}Y^{\hbeta}(FY)^{\halpha_1}\wedge  \rD Y^{\halpha_2}\wedge\cdots \wedge  \rD Y^{\halpha_{d}}\,,
\\\cr
\!\!\!\!&&\!\!\!\!-\ell^2\epsilon_{\halpha\hbeta\halpha_1\cdots \halpha_{d}} \delta Y^{\halpha} \rD Y^{\hbeta}\wedge  \rD Y^{\halpha_1}\wedge\cdots \wedge  \rD Y^{\halpha_{d}}
\cr\!\!\!\!&=&\!\!\!\epsilon_{\halpha\hbeta\halpha_1\cdots \halpha_{d}} \delta Y^{\halpha} \rD Y^{\hbeta}\wedge  \rD Y^{\halpha_1}\wedge\cdots \wedge  \rD Y^{\halpha_{d}}Y^{\hgamma}Y_{\hgamma}=0\,,
\end{eqnarray}
we get
\begin{eqnarray}
\!\!\!\!&&\!\!\!\!\delta\int_{M} \epsilon_{\halpha\hbeta\halpha_1\cdots \halpha_{d}} \left[ F^{\halpha\hbeta}-\tfrac{2}{(d+1)\ell^2} \rD Y^{\halpha}\wedge  \rD Y^{\hbeta}\right]\wedge  \rD Y^{\halpha_1}\wedge\cdots \wedge  \rD Y^{\halpha_{d-1}}Y^{\halpha_{d}}
\cr\!\!\!\!&=&\!\!\!\!\int_{M} \epsilon_{\halpha\hbeta\halpha_1\cdots \halpha_{d}} \Big\{ \rD\delta A^{\halpha\hbeta}\wedge  \rD Y^{\halpha_1}\wedge\cdots \wedge  \rD Y^{\halpha_{d-1}}Y^{\halpha_{d}}
\cr&&~~~~~~~~~~~~~~~~~+(d-1)F^{\halpha\hbeta}\wedge  (\rD \delta Y+\delta AY)^{\halpha_1}\wedge\rD Y^{\halpha_2}\wedge\cdots \wedge  \rD Y^{\halpha_{d-1}}Y^{\halpha_{d}}
\cr&&~~~~~~~~~~~~~~~~~+F^{\halpha\hbeta}\wedge  \rD Y^{\halpha_1}\wedge\cdots \wedge  \rD Y^{\halpha_{d-1}}\delta Y^{\halpha_{d}}
\cr&&~~~~~~~~~~~~~~~~~-\tfrac{2}{\ell^2} (\rD \delta Y+\delta AY)^{\halpha}\wedge  \rD Y^{\hbeta}\wedge  \rD Y^{\halpha_1}\wedge\cdots \wedge  \rD Y^{\halpha_{d-1}}Y^{\halpha_{d}}
\cr&&~~~~~~~~~~~~~~~~~-\tfrac{2}{(d+1)\ell^2} \rD Y^{\halpha}\wedge  \rD Y^{\hbeta}\wedge  \rD Y^{\halpha_1}\wedge\cdots \wedge  \rD Y^{\halpha_{d-1}}\delta Y^{\halpha_{d}}\Big\}
\cr\!\!\!\!&=&\!\!\!\!\int_{M} \epsilon_{\halpha\hbeta\halpha_1\cdots \halpha_{d}} \Big\{\delta A^{\halpha\hbeta}\wedge  \left[\rD (\rD Y^{\halpha_1}\wedge\cdots \wedge  \rD Y^{\halpha_{d-1}})Y^{\halpha_{d}}-(-1)^d\rD Y^{\halpha_1} \wedge\cdots \wedge  \rD Y^{\halpha_{d}}\right]
\cr&&~~~~~~~~~~~~~~~~~~+(d-1)(\delta AY)^{\halpha_1}\wedge F^{\halpha\hbeta}\wedge  \rD Y^{\halpha_2}\wedge\cdots \wedge  \rD Y^{\halpha_{d-1}}Y^{\halpha_{d}}
\cr&&~~~~~~~~~~~~~~~~~~-\tfrac{2}{\ell^2} (\delta AY)^{\halpha}\wedge  \rD Y^{\hbeta}\wedge  \rD Y^{\halpha_1}\wedge\cdots \wedge  \rD Y^{\halpha_{d-1}}Y^{\halpha_{d}}
\cr&&~~~~~~~~~~~~~~~~~~-(d-1)\delta Y^{\halpha_1} \left[\rD(F^{\halpha\hbeta}\wedge  \rD Y^{\halpha_2}\wedge\cdots \wedge  \rD Y^{\halpha_{d-1}})Y^{\halpha_{d}}+(-1)^dF^{\halpha\hbeta}\wedge  \rD Y^{\halpha_2}\wedge\cdots \wedge  \rD Y^{\halpha_{d}}\right]
\cr&&~~~~~~~~~~~~~~~~~~+\tfrac{2}{\ell^2} \delta Y^{\halpha} \left[\rD(\rD Y^{\hbeta}\wedge  \rD Y^{\halpha_1}\wedge\cdots \wedge  \rD Y^{\halpha_{d-1}})Y^{\halpha_{d}}+(-1)^d\rD Y^{\hbeta}\wedge  \rD Y^{\halpha_1}\wedge\cdots \wedge  \rD Y^{\halpha_{d}}\right]
\cr&&~~~~~~~~~~~~~~~~~~+\delta Y^{\halpha_{d}}F^{\halpha\hbeta}\wedge  \rD Y^{\halpha_1}\wedge\cdots \wedge  \rD Y^{\halpha_{d-1}}
\cr&&~~~~~~~~~~~~~~~~~~-\tfrac{2}{(d+1)\ell^2} \delta Y^{\halpha_{d}}\rD Y^{\halpha}\wedge  \rD Y^{\hbeta}\wedge  \rD Y^{\halpha_1}\wedge\cdots \wedge  \rD Y^{\halpha_{d-1}}\Big\}
\cr\!\!\!\!&&\!\!\!\!+\int_{\Sigma} \epsilon_{\halpha\hbeta\halpha_1\cdots \halpha_{d}} \Big\{ \delta A^{\halpha\hbeta}\wedge  \rD Y^{\halpha_1}\wedge\cdots \wedge  \rD Y^{\halpha_{d-1}}Y^{\halpha_{d}}
\cr&&~~~~~~~~~~~~~~~~~~~+(d-1)\delta Y^{\halpha_1} F^{\halpha\hbeta}\wedge\rD Y^{\halpha_2}\wedge\cdots \wedge  \rD Y^{\halpha_{d-1}}Y^{\halpha_{d}}
\cr&&~~~~~~~~~~~~~~~~~~~-\tfrac{2}{\ell^2} \delta Y^{\halpha}\rD Y^{\hbeta}\wedge  \rD Y^{\halpha_1}\wedge\cdots \wedge  \rD Y^{\halpha_{d-1}}Y^{\halpha_{d}}\Big\}
\cr\!\!\!\!&=&\!\!\!\!(-1)^{d+1}(d-1)\int_{M} \epsilon_{\halpha\hbeta\halpha_1\cdots \halpha_{d}}  \Big\{\tfrac{\ell^2}2\delta A^{\halpha\hbeta}\wedge F^{\halpha_1\halpha_2}\wedge  \rD Y^{\halpha_3}\wedge\cdots \wedge  \rD Y^{\halpha_{d}}
\cr&&~~~~~~~~~~~~~~~~~~~~~~~~~~~~~~~~~~~~~~+\delta Y^{\halpha}Y^{\hbeta}\rD(F^{\halpha_1\halpha_2}\wedge  \rD Y^{\halpha_2}\wedge\cdots \wedge  \rD Y^{\halpha_{d}})
\Big\}
\cr\!\!\!\!&&\!\!\!\!+\int_{\Sigma} \epsilon_{\halpha\hbeta\halpha_1\cdots \halpha_{d}} \Big\{ \delta A^{\halpha\hbeta}\wedge  \rD Y^{\halpha_1}\wedge\cdots \wedge  \rD Y^{\halpha_{d-1}}Y^{\halpha_{d}}
\cr&&~~~~~~~~~~~~~~~~~~~+(-1)^d\delta Y^{\halpha}Y^{\hbeta} \left[(d-1)F^{\halpha_1\halpha_2}\wedge\rD Y^{\halpha_3}\wedge\cdots \wedge  \rD Y^{\halpha_{d}}
-\tfrac{2}{\ell^2} \rD Y^{\halpha_1}\wedge\cdots \wedge  \rD Y^{\halpha_{d}}\right]\Big\}
\,.
\end{eqnarray}
It implies the expressions (\ref{JA},\ref{JY},\ref{KA},\ref{KY}) for $(\cJ_{(d+1)})_{\halpha\hbeta},(\cJ_{(d+1)})_{\halpha},(\cK^{\rm cov}_{(d+1)})_{\halpha\hbeta},(\cK^{\rm cov}_{(d+1)})_{\halpha}$ in the main text. Especially, we notice that
\begin{eqnarray}
(\cJ_{(d+1)})_{\halpha}Y^{\halpha}=0\,,~~~~~~~(\cK^{\rm cov}_{(d+1)})_{\halpha}Y^{\halpha}=0\,,
\end{eqnarray}
which are the correct formula under the constraint $Y_{\halpha}\delta Y^{\halpha}=0$. 
\subsubsection{$SO(2,d)$ conservation law from $S^{\rm cov}_{(d+1)}$}
Under the $SO(2,d)$ gauge transformation
\begin{eqnarray}
Y^{\halpha}\to \tilde Y^{\halpha}=U^{\halpha}{}_{\hbeta}Y^{\hbeta}\,,~~~~~~~~~
A\to \tilde A=UAU^{-1}-\rd U U^{-1}\,,
\end{eqnarray}
this bulk covariant action is manifestly invariant without any boundary tails
\begin{eqnarray}
\Delta_{U}\cL^{\rm cov}_{(d+1)}(A,Y)=\cL^{\rm cov}_{(d+1)}(\tilde A,\tilde Y)-\cL^{\rm cov}_{(d+1)}(A,Y)=0
\,.
\end{eqnarray}

Substituting the infinitesimal gauge transformation
\begin{eqnarray}
\delta_{u} Y^{\halpha}=u^{\halpha}{}_{\hbeta}Y^{\hbeta}\,,~~~~~~~~~
\delta_u A^{\halpha\hbeta}=-\rD u^{\halpha\hbeta}\,,
\end{eqnarray}
to the variation of $S^{\rm cov}_{(d+1)}$, we get
\begin{eqnarray}
0\!\!\!\!&=&\!\!\!\!\delta_{u} S^{\rm cov}_{(d+1)}(A,Y)=\delta_{u}\int_{M}\cL^{\rm cov}_{(d+1)}(A,Y,F,\rD Y)
\cr\!\!\!\!&=&\!\!\!\!\int_{M} \left[-\rD u^{\halpha\hbeta}\wedge (\cJ^{\rm cov}_{(d+1)})_{\halpha\hbeta}+ (uY)^{\halpha} (\cJ^{\rm cov}_{(d+1)})_{\halpha}\right]
+\int_{\Sigma} \left[-\rD u^{\halpha\hbeta}\wedge (\cK^{\rm cov}_{(d+1)})_{\halpha\hbeta}+ (u Y)^{\halpha} (\cK^{\rm cov}_{(d+1)})_{\halpha}\right]
\cr\!\!\!\!&=&\!\!\!\!\int_{M} u^{\halpha\hbeta}\left[\rD  (\cJ^{\rm cov}_{(d+1)})_{\halpha\hbeta}+ Y_{[\hbeta}(\cJ^{\rm cov}_{(d+1)})_{\halpha]}\right]
\cr&&+\int_{\Sigma} u^{\halpha\hbeta}\left[-(\cJ^{\rm cov}_{(d+1)})_{\halpha\hbeta}+\rD (\cK^{\rm cov}_{(d+1)})_{\halpha\hbeta}+Y_{[\hbeta}(\cK^{\rm cov}_{(d+1)})_{\halpha]}\right]
-\int_{\partial\Sigma} u^{\halpha\hbeta}(\cK^{\rm cov}_{(d+1)})_{\halpha\hbeta}\,,
\end{eqnarray}
Fixing $M$, the arbitrariness of $u$ implies the following off-shell bulk conservation law
\begin{eqnarray}
\rD(\cJ^{\rm cov}_{(d+1)})_{\halpha\hbeta}+ Y_{[\hbeta}(\cJ^{\rm cov}_{(d+1)})_{\halpha]}\!\!\!\!&=&\!\!\!\!0\,,
\end{eqnarray}
as well as the boundary anomalous conservation law at the off-shell level
\begin{eqnarray}\label{P.gauge.bo.cov}
\mD(\cJ^{\rm cov}_{(d)})_{\halpha\hbeta}+\mY_{[\hbeta}(\cJ^{\rm cov}_{(d)})_{\halpha]}=(\cA^{\rm cov}_{(d)})_{\halpha\hbeta}\,,
\end{eqnarray}
where $(\cA^{\rm cov}_{(d)})_{\halpha\hbeta}$ is the boundary covariant anomaly on $\Sigma$
\begin{eqnarray}
(\cA^{\rm cov}_{(d)})_{\halpha\hbeta}(\Sigma)=f_{\Sigma}^{*}[(\cJ^{\rm cov}_{(d+1)})_{\halpha\hbeta}]
=\frac{(-1)^{d+1}\ell}{4\kappa^2(d-2)!}\epsilon_{\halpha\hbeta\halpha_1\cdots \halpha_{d}} \mF^{\halpha_1\halpha_2}\wedge\mD \mY^{\halpha_{3}}\wedge\cdots \wedge  \mD \mY^{\halpha_{d}}\,.
\end{eqnarray}
Since (\ref{P.gauge.bo.cov}) works for any $M$ and $\Sigma$, the boundary anomalous conservation law must also hold without the pullback
\begin{eqnarray}\label{gauge.bo.cov}
\rD(\cK^{\rm cov}_{(d+1)})_{\halpha\hbeta}+Y_{[\hbeta}(\cK^{\rm cov}_{(d+1)})_{\halpha]}=(\cJ^{\rm cov}_{(d+1)})_{\halpha\hbeta}\,.
\end{eqnarray}
By using the resummations
\begin{eqnarray}
\!\!\!\!&&\!\!\!\!Y_{[\hbeta}\epsilon_{\halpha]\halpha_0\halpha_1\cdots\halpha_{d}} Y^{\halpha_0} \rD(F^{\halpha_1\halpha_{2}}\wedge \rD Y^{\halpha_{3}}\wedge\cdots\wedge\rD Y^{\halpha_{d}})
\cr\!\!\!\!&=&\!\!\!\!(d-2)Y_{\hbeta_0}\delta_{[\hbeta}^{\hbeta_0}\delta_{\halpha]}^{\halpha_D}\epsilon_{\halpha_D\halpha_0\halpha_1\cdots\halpha_{d}} Y^{\halpha_0} F^{\halpha_1\halpha_{2}}\wedge (FY)^{\halpha_{3}}\wedge \rD Y^{\halpha_{4}}\wedge\cdots\wedge\rD Y^{\halpha_{d}}
\cr\!\!\!\!&=&\!\!\!\!-Y_{[\hbeta}\epsilon_{\halpha]\halpha_0\halpha_1\cdots\halpha_{d}} Y^{\halpha_0} \rD(F^{\halpha_1\halpha_{2}}\wedge \rD Y^{\halpha_{3}}\wedge\cdots\wedge\rD Y^{\halpha_{d}})
\cr\!\!\!\!&&\!\!\!\!-\ell^2 \epsilon_{\halpha\hbeta\halpha_1\cdots\halpha_{d}}  \rD(F^{\halpha_1\halpha_{2}}\wedge \rD Y^{\halpha_{3}}\wedge\cdots\wedge\rD Y^{\halpha_{d}})\,,
\\\cr
\!\!\!\!&&\!\!\!\!Y_{[\hbeta}\epsilon_{\halpha]\halpha_0\halpha_1\cdots\halpha_{d}} Y^{\halpha_0} F^{\halpha_1\halpha_{2}}\wedge \rD Y^{\halpha_{3}}\wedge\cdots\wedge\rD Y^{\halpha_{d}}
\cr\!\!\!\!&=&\!\!\!\!Y_{\hbeta_0}\delta_{[\hbeta}^{\hbeta_0}\delta_{\halpha]}^{\halpha_D}\epsilon_{\halpha_D\halpha_0\halpha_1\cdots\halpha_{d}} Y^{\halpha_0} F^{\halpha_1\halpha_{2}}\wedge \rD Y^{\halpha_3}\wedge\cdots\wedge\rD Y^{\halpha_{d}}
\cr\!\!\!\!&=&\!\!\!\!-Y_{[\hbeta} \epsilon_{\halpha]\halpha_0\halpha_1\cdots\halpha_{d}} Y^{\halpha_0} F^{\halpha_1\halpha_{2}}\wedge \rD Y^{\halpha_3}\wedge\cdots\wedge\rD Y^{\halpha_{d}}
-\ell^2 \epsilon_{\halpha\hbeta\halpha_1\cdots\halpha_{d}}  F^{\halpha_1\halpha_{2}}\wedge \rD Y^{\halpha_3}\wedge\cdots\wedge\rD Y^{\halpha_{d}}
\cr\!\!\!\!&&\!\!\!\!+2\epsilon_{\halpha\hbeta\halpha_1\halpha_2\halpha_3\cdots\halpha_{d}} Y^{\halpha_1} (FY)^{\halpha_2}\wedge \rD Y^{\halpha_3}\wedge\cdots\wedge\rD Y^{\halpha_{d}}\,,
\\\cr
\!\!\!\!&&\!\!\!\!Y_{[\hbeta}\epsilon_{\halpha]\halpha_0\halpha_1\cdots\halpha_{d}} Y^{\halpha_0} \rD Y^{\halpha_1}\wedge\cdots\wedge\rD Y^{\halpha_{d}}
=Y_{\hbeta_0}\delta_{[\hbeta}^{\hbeta_0}\delta_{\halpha]}^{\halpha_D}\epsilon_{\halpha_D\halpha_0\halpha_1\cdots\halpha_{d}} Y^{\halpha_0} \rD Y^{\halpha_1}\wedge\cdots\wedge\rD Y^{\halpha_{d}}
\cr\!\!\!\!&=&\!\!\!\!-Y_{[\hbeta} \epsilon_{\halpha]\halpha_0\halpha_1\cdots\halpha_{d}} Y^{\halpha_0} \rD Y^{\halpha_1}\wedge\cdots\wedge\rD Y^{\halpha_{d}}
-\ell^2 \epsilon_{\halpha\hbeta\halpha_1\cdots\halpha_{d}}  \rD Y^{\halpha_1}\wedge\cdots\wedge\rD Y^{\halpha_{d}}\,.
\end{eqnarray}
one can verify that the off-shell conservation laws are indeed identities
\begin{eqnarray}
\!\!\!\!&&\!\!\!\!\rD(\cJ^{\rm cov}_{(d+1)})_{\halpha\hbeta}+ Y_{[\hbeta}(\cJ^{\rm cov}_{(d+1)})_{\halpha]}
\cr\!\!\!\!&=&\!\!\!\!\frac{(-1)^{d+1}\ell}{4\kappa^2(d-2)!}\Big[\rD(\epsilon_{\halpha\hbeta\halpha_1\cdots\halpha_{d}} F^{\halpha_1\halpha_2}\wedge\rD Y^{\halpha_3}\wedge\cdots\wedge\rD Y^{\halpha_{d}})
\cr&&~~~~~~~~~~~~~~~~~~+\frac{2}{\ell^2} Y_{[\hbeta}\epsilon_{\halpha]\halpha_0\halpha_1\cdots\halpha_{d}} Y^{\halpha_0} \rD(F^{\halpha_1\halpha_{2}}\wedge \rD Y^{\halpha_{3}}\wedge\cdots\wedge\rD Y^{\halpha_{d}})\Big]
\cr\!\!\!\!&=&\!\!\!\!\frac{(-1)^{d+1}\ell}{4\kappa^2(d-2)!}\Big[\epsilon_{\halpha\hbeta\halpha_1\cdots\halpha_{d}}\rD\left( F^{\halpha_1\halpha_2}\wedge\rD Y^{\halpha_3}\wedge\cdots\wedge\rD Y^{\halpha_{d}}\right) -\epsilon_{\halpha\hbeta\halpha_1\cdots\halpha_{d}}\rD\left( F^{\halpha_1\halpha_2}\wedge\rD Y^{\halpha_3}\wedge\cdots\wedge\rD Y^{\halpha_{d}}\right)\Big]
\cr\!\!\!\!&=&\!\!\!\!0\,,
\\\cr
\!\!\!\!&&\!\!\!\!\rD(\cK^{\rm cov}_{(d+1)})_{\halpha\hbeta}+Y_{[\hbeta}(\cK^{\rm cov}_{(d+1)})_{\halpha]}
\cr\!\!\!\!&=&\!\!\!\!\frac{(-1)^{d}}{2\kappa^2\ell\,(d-1)!}\Big\{-\rD(\epsilon_{\halpha\hbeta\halpha_1\cdots\halpha_{d}} Y^{\halpha_1}\rD Y^{\halpha_2}\wedge\cdots\wedge\rD Y^{\halpha_{d}})
\cr&&~~~~~~~~~~~~~~~~~~+Y_{[\hbeta}\epsilon_{\halpha]\halpha_0\halpha_1\cdots \halpha_{d}}
Y^{\halpha_0} \left[(d-1)F^{\halpha_1\halpha_2}\wedge\rD Y^{\halpha_3}\wedge\cdots \wedge  \rD Y^{\halpha_{d}}
-\frac{2}{\ell^2} \rD Y^{\halpha_1}\wedge\cdots \wedge  \rD Y^{\halpha_{d}}\right]\Big\}
\cr\!\!\!\!&=&\!\!\!\!\frac{(-1)^{d}}{2\kappa^2\ell\,(d-1)!}\Big\{-\epsilon_{\halpha\hbeta\halpha_1\cdots\halpha_{d}} \left[\rD Y^{\halpha_1}\wedge\cdots\wedge\rD Y^{\halpha_{d}}+(d-1)Y^{\halpha_1}(FY)^{\halpha_2}\wedge\rD Y^{\halpha_3}\wedge\cdots\wedge\rD Y^{\halpha_{d}}\right]
\cr&&~~~~~~~~~~~~~~~~~~+\epsilon_{\halpha\hbeta\halpha_1\cdots \halpha_{d}}
\Big[(d-1)Y^{\halpha_1}(FY)^{\halpha_2}\wedge\rD Y^{\halpha_3}\wedge\cdots\wedge\rD Y^{\halpha_{d}}
\cr&&~~~~~~~~~~~~~~~~~~~~~~~~~~~~~~~~~~~~~~~~~-\frac{\ell^2(d-1)}2 F^{\halpha_1\halpha_{2}}\wedge \rD Y^{\halpha_3}\wedge\cdots\wedge\rD Y^{\halpha_{d}}
+\rD Y^{\halpha_1}\wedge\cdots \wedge  \rD Y^{\halpha_{d}}\Big]\Big\}
\cr\!\!\!\!&=&\!\!\!\!\frac{(-1)^{d+1}\ell}{4\kappa^2\,(d-2)!} \epsilon_{\halpha\hbeta\halpha_1\cdots \halpha_{d}}F^{\halpha_1\halpha_{2}}\wedge \rD Y^{\halpha_3}\wedge\cdots\wedge\rD Y^{\halpha_{d}}=(\cJ^{\rm cov}_{(d+1)})_{\halpha\hbeta}\,.
\end{eqnarray}

\subsection{The bulk consistent action}
To construct the bulk consistent action, let us start from the following bulk CS action
\begin{eqnarray}
\!\!\!\!&&\!\!\!\!\frac{4\kappa^2(d-2)!}{(-1)^{d+1}\ell}S^{\rm CS}_{(d+1)}[A,Y]=\frac{4\kappa^2(d-2)!}{(-1)^{d+1}\ell}\int_{M}\cL_{(d+1)}^{\rm CS}
\cr\!\!\!\!&=&\!\!\!\!\frac{4\kappa^2(d-2)!}{(-1)^{d+1}\ell}\int_{M}\int_0^1\rd s\,A^{\halpha\hbeta}\wedge (\cA^{\rm cov}_{(d)})_{\halpha\hbeta}(A(s),Y)
\cr\!\!\!\!&=&\!\!\!\!\int_{M}\int_0^1\rd s\,\epsilon_{\halpha\hbeta\halpha_1\cdots\halpha_{d}}A^{\halpha\hbeta}\wedge F^{\halpha_1\halpha_2}(s)\wedge\rD(s)Y^{\halpha_3}\wedge\cdots\wedge\rD(s)Y^{\halpha_{d}}
\,,
\end{eqnarray}
where the homotopy notations are
\begin{eqnarray}
A(s)\!\!\!\!&=&\!\!\!\!sA\,,~~~~~~\rD(s)Y=\rd Y+ sA Y\,,
\cr F(s)\!\!\!\!&=&\!\!\!\!\rd A(s)+A(s)\wedge A(s)=s\rd A+s^2A\wedge A
=sF-s(1-s)A\wedge A\,.
\end{eqnarray}
and
\begin{eqnarray}
(\cA^{\rm cov}_{(d)})_{\halpha\hbeta}=(\cJ^{\rm cov}_{(d+1)})_{\halpha\hbeta}
=\frac{(-1)^{d+1}\ell}{4\kappa^2(d-2)!}\epsilon_{\halpha\hbeta\halpha_1\cdots \halpha_{d}} F^{\halpha_1\halpha_2}\wedge\rD Y^{\halpha_{3}}\wedge\cdots \wedge  \rD Y^{\halpha_{d}}\,.
\end{eqnarray}
\subsubsection{$\rd\cL_{(d+1)}^{\rm CS}=I_{(d+2)}$}
We notice that
\begin{eqnarray}
\!\!\!\!&&\!\!\!\!\rd [\epsilon_{\halpha\hbeta\halpha_1\cdots\halpha_{d}}A^{\halpha\hbeta}(s)\wedge F^{\halpha_1\halpha_2}(s)\wedge\rD(s)Y^{\halpha_3}\wedge\cdots\wedge\rD(s)Y^{\halpha_{d}}]
\cr\!\!\!\!&=&\!\!\!\!\epsilon_{\halpha\hbeta\halpha_1\cdots\halpha_{d}}\rD(s) [A^{\halpha\hbeta}(s)\wedge F^{\halpha_1\halpha_2}(s)\wedge\rD(s)Y^{\halpha_3}\wedge\cdots\wedge\rD(s)Y^{\halpha_{d}}]
\cr\!\!\!\!&=&\!\!\!\!\epsilon_{\halpha\hbeta\halpha_1\cdots\halpha_{d}}[F(s)+A(s)\wedge A(s)]^{\halpha\hbeta}\wedge F^{\halpha_1\halpha_2}(s)\wedge\rD(s)Y^{\halpha_3}\wedge\cdots\wedge\rD(s)Y^{\halpha_{d}}
\cr\!\!\!\!&&\!\!\!\!-(d-2)\epsilon_{\halpha\hbeta\halpha_1\cdots\halpha_{d}}A^{\halpha\hbeta}(s)\wedge F^{\halpha_1\halpha_2}(s)\wedge [F(s)Y]^{\halpha_3}\wedge\cdots\wedge\rD(s)Y^{\halpha_{d}}
\cr\!\!\!\!&=&\!\!\!\!\epsilon_{\halpha\hbeta\halpha_1\cdots\halpha_{d}}[F(s)+A(s)\wedge A(s)]^{\halpha\hbeta}\wedge F^{\halpha_1\halpha_2}(s)\wedge\rD(s)Y^{\halpha_3}\wedge\cdots\wedge\rD(s)Y^{\halpha_{d}}
\cr\!\!\!\!&&\!\!\!\!+\frac{d-2}2\epsilon_{\halpha\hbeta\halpha_1\cdots\halpha_{d}}F^{\halpha\hbeta}(s)\wedge F^{\halpha_1\halpha_2}(s)\wedge [A(s)Y]^{\halpha_3}\wedge\rD(s)Y^{\halpha_4}\wedge\cdots\wedge\rD(s)Y^{\halpha_{d}}
\cr\!\!\!\!&=&\!\!\!\!\epsilon_{\halpha\hbeta\halpha_1\cdots\halpha_{d}}s\partial_{s}F^{\halpha\hbeta}(s)\wedge F^{\halpha_1\halpha_2}(s)\wedge\rD(s)Y^{\halpha_3}\wedge\cdots\wedge\rD(s)Y^{\halpha_{d}}
\cr\!\!\!\!&&\!\!\!\!+\frac{d-2}2\epsilon_{\halpha\hbeta\halpha_1\cdots\halpha_{d}}F^{\halpha\hbeta}(s)\wedge F^{\halpha_1\halpha_2}(s)\wedge s\partial_s[\rD(s)Y]^{\halpha_3}\wedge\rD(s)Y^{\halpha_4}\wedge\cdots\wedge\rD(s)Y^{\halpha_{d}}
\cr\!\!\!\!&=&\!\!\!\!\frac12s\partial_{s}[\epsilon_{\halpha\hbeta\halpha_1\cdots\halpha_{d}}F^{\halpha\hbeta}(s)\wedge F^{\halpha_1\halpha_2}(s)\wedge\rD(s)Y^{\halpha_3}\wedge\cdots\wedge\rD(s)Y^{\halpha_{d}}]\,,
\end{eqnarray}
where we have used that
\begin{eqnarray}
\!\!\!\!&&\!\!\!\!\rD(s) A^{\halpha_1\halpha_2}(s)=[F(s)+A(s)\wedge A(s)]^{\halpha_1\halpha_2}=s\partial_{s}F^{\halpha_1\halpha_2}(s)\,,
~~~~s\partial_s A(s)=A(s)\,,~~~~
\end{eqnarray}
as well as the resummation
\begin{eqnarray}
\!\!\!\!&&\!\!\!\!\epsilon_{\halpha\hbeta\halpha_1\cdots\halpha_{d}}A^{\halpha\hbeta}(s)\wedge F^{\halpha_1\halpha_2}(s)\wedge F^{\halpha_3\hgamma}(s)Y_{\hgamma}\wedge\rD(s)Y^{\halpha_4}\wedge\cdots\wedge\rD(s)Y^{\halpha_{d}}
\cr\!\!\!\!&=&\!\!\!\!-\frac12\epsilon_{\halpha\hbeta\halpha_1\cdots\halpha_{d}} F^{\halpha\hbeta}(s)\wedge F^{\halpha_1\halpha_2}(s)\wedge [A(s)Y]^{\halpha_3}\wedge\rD(s)Y^{\halpha_4}\wedge\cdots\wedge\rD(s)Y^{\halpha_{d}}\,.
\end{eqnarray}
Thus the exterior derivative $\rd\cL_{(d+1)}^{\rm CS}$ is the same as $\rd\cL_{(d+1)}^{\rm cov}$
\begin{eqnarray}
\frac{4\kappa^2(d-2)!}{(-1)^{d+1}\ell}\rd \cL_{(d+1)}^{\rm con}
\!\!\!\!&=&\!\!\!\!\rd\left[\int_0^1\frac{\rd s}{s}\,\epsilon_{\halpha\hbeta\halpha_1\cdots\halpha_{d}}A^{\halpha\hbeta}(s)\wedge F^{\halpha_1\halpha_2}(s)\wedge\rD(s)Y^{\halpha_3}\wedge\cdots\wedge\rD(s)Y^{\halpha_{d}}\right]
\cr\!\!\!\!&=&\!\!\!\!\frac12\int_0^1\rd s\,\partial_{s}[\epsilon_{\halpha\hbeta\halpha_1\cdots\halpha_{d}}F^{\halpha\hbeta}(s)\wedge F^{\halpha_1\halpha_2}(s)\wedge\rD(s)Y^{\halpha_3}\wedge\cdots\wedge\rD(s)Y^{\halpha_{d}}]
\cr\!\!\!\!&=&\!\!\!\!\frac12\epsilon_{\halpha\hbeta\halpha_1\cdots\halpha_{d}}F^{\halpha\hbeta}(s)\wedge F^{\halpha_1\halpha_2}(s)\wedge\rD(s)Y^{\halpha_3}\wedge\cdots\wedge\rD(s)Y^{\halpha_{d}}\Big|^{s=1}_{s=0}
\cr\!\!\!\!&=&\!\!\!\!\frac12\epsilon_{\halpha\hbeta\halpha_1\cdots\halpha_{d}}F^{\halpha\hbeta}\wedge F^{\halpha_1\halpha_2}\wedge\rD Y^{\halpha_3}\wedge\cdots\wedge\rD Y^{\halpha_{d}}
\cr\!\!\!\!&=&\!\!\!\!\frac{4\kappa^2(d-2)!}{(-1)^{d+1}\ell}I_{(d+2)}=\frac{4\kappa^2(d-2)!}{(-1)^{d+1}\ell}\rd \cL^{\rm cov}_{(d+1)}\,.
\end{eqnarray}
It means that the two Lagrangians $\cL_{(d+1)}^{\rm con}$ and $\cL_{(d+1)}^{\rm cov}$ differ only by a closed from.
\subsubsection{$SO(2,d)$ gauge transformation of $\cL^{\rm CS}_{(d+1)}$}
Denoting 
\begin{eqnarray}
\hA=U^{-1}\rd U\,,~~~~~~\hA(s)=s\hA=sU^{-1}\rd U\,,
\end{eqnarray}
the finite gauge transformation can be expressed as
\begin{eqnarray}
Y^{\halpha}\to \tilde Y^{\halpha}=U^{\halpha}{}_{\hbeta}Y^{\hbeta}\,,~~~~~
A\to \tilde A=UAU^{-1}-\rd U U^{-1}=U(A-\hA)U^{-1}\,.
\end{eqnarray}
Thus we have
\begin{eqnarray}
A(s)\to\tilde A(s)=s\tilde A=sUAU^{-1}-sU^{-1} \rd U=U(A(s)-sU^{-1}\rd U)U^{-1}=U[A(s)-\hA(s)]U^{-1}\,.
\end{eqnarray}
It follows that
\begin{eqnarray}
\tilde F(s)\!\!\!\!&=&\!\!\!\!\rd\tilde A(s)+\tilde A(s)\wedge\tilde A(s)=s\rd\tilde A+s^2\tilde A\wedge\tilde A
=s\tilde F-s(1-s)\tilde A\wedge\tilde A
\cr\!\!\!\!&=&\!\!\!\!sU [F-(1-s)(A \wedge A-\hA\wedge A-A\wedge \hA+\hA \wedge\hA)]U^{-1}
\cr\!\!\!\!&=&\!\!\!\!U [F(s)+s(1-s)(\hA\wedge A+A\wedge \hA)+\hF(1-s)]U^{-1}
\,,
\cr\tilde\rD(s)\tilde Y\!\!\!\!&=&\!\!\!\!
U\rd Y+\rd UY +UA(s)-s\rd UY
=U[\rd Y+sA Y+(1-s)\hat AY]
\cr\!\!\!\!&=&\!\!\!\!U[\rD(s)Y+(1-s)\hA Y]=U[\hrD(1-s)Y+sA Y]\,,
\end{eqnarray}
where we have used that
\begin{eqnarray}
\hrD(s)Y\!\!\!\!&=&\!\!\!\!\rd Y+s\hat AY=\rd Y+sU^{-1}\rd UY\,,
\cr \hF(s)\!\!\!\!&=&\!\!\!\!\rd \hA(s)+\hA(s)\wedge \hA(s)
=\rd(sU^{-1}\rd U) +s^2U^{-1}\rd U\wedge U^{-1}\rd U
\cr\!\!\!\!&=&\!\!\!\!s(1-s)\rd (U^{-1}\rd U)=-s(1-s)\hA\wedge \hA=\hF(1-s)\,.
\end{eqnarray}
Thus
\begin{eqnarray}
\!\!\!\!&&\!\!\!\!\frac{4\kappa^2(d-2)!}{(-1)^{d+1}\ell}\cL^{\rm CS}_{(d+1)}[\tilde A,\tilde Y]
\cr\!\!\!\!&=&\!\!\!\!\int_0^1\rd s\,\epsilon_{\halpha\hbeta\halpha_1\cdots\halpha_{d}}\tilde A^{\halpha\hbeta}\wedge \tilde F^{\halpha_1\halpha_2}(s)\wedge\tilde\rD(s)\tilde Y^{\halpha_3}\wedge\cdots\wedge\tilde\rD(s)\tilde Y^{\halpha_{d}}
\cr\!\!\!\!&=&\!\!\!\!\int_0^1\rd s\,\epsilon_{\halpha\hbeta\halpha_1\cdots\halpha_{d}} (A-\hA)^{\halpha\hbeta}\wedge [F(s)+s(1-s)(\hA\wedge A+A\wedge \hA)+\hF(1-s)]^{\halpha_1\halpha_2}
\cr&&~~~~~~~~~~~~~~~~~~~~~\wedge[\rD(s)Y+(1-s)\hA Y]^{\halpha_3}\wedge\cdots\wedge[\rD(s)Y+(1-s)\hA Y]^{\halpha_{d}}
\cr\!\!\!\!&=&\!\!\!\!\int_0^1\rd s\,\epsilon_{\halpha\hbeta\halpha_1\cdots\halpha_{d}} (A-\hA)^{\halpha\hbeta}\wedge F(s,1-s)^{\halpha_1\halpha_2}
\wedge\rD(s,1-s)Y^{\halpha_3}\wedge\cdots\wedge \rD(s,1-s)Y^{\halpha_{d}}
\cr\!\!\!\!&=&\!\!\!\!\int_0^1\rd s\,\epsilon_{\halpha\hbeta\halpha_1\cdots\halpha_{d}} A^{\halpha\hbeta}\wedge F(s,1-s)^{\halpha_1\halpha_2}
\wedge\rD(s,1-s)Y^{\halpha_3}\wedge\cdots\wedge \rD(s,1-s)Y^{\halpha_{d}}
\cr\!\!\!\!&&\!\!\!\!-\int_0^1\rd \hs\,\epsilon_{\halpha\hbeta\halpha_1\cdots\halpha_{d}} \hA^{\halpha\hbeta}\wedge F(1-\hs,\hs)^{\halpha_1\halpha_2}
\wedge\rD(1-\hs,\hs)Y^{\halpha_3}\wedge\cdots\wedge \rD(1-\hs,\hs)Y^{\halpha_{d}}
\cr\!\!\!\!&=&\!\!\!\!\int_0^1\rd s\,\epsilon_{\halpha\hbeta\halpha_1\cdots\halpha_{d}}A^{\halpha\hbeta}\wedge F^{\halpha_1\halpha_2}(s,0)\wedge\rD(s,0)Y^{\halpha_3}\wedge\cdots\wedge\rD(s,0)Y^{\halpha_{d}}
\cr\!\!\!\!&&\!\!\!\!+\int_0^1\rd s\int_0^{1-s}\rd \hs\,\partial_{\hs}\left[\epsilon_{\halpha\hbeta\halpha_1\cdots\halpha_{d}} A^{\halpha\hbeta}\wedge F^{\halpha_1\halpha_2}(s,\hs)
\wedge\rD(s,\hs)Y^{\halpha_3}\wedge\cdots\wedge \rD(s,\hs)Y^{\halpha_{d}}\right]
\cr\!\!\!\!&&\!\!\!\!-\int_0^1\rd \hs\,\epsilon_{\halpha\hbeta\halpha_1\cdots\halpha_{d}}\hA^{\halpha\hbeta}\wedge F^{\halpha_1\halpha_2}(0,\hs)\wedge\rD(0,\hs)Y^{\halpha_3}\wedge\cdots\wedge\rD(0,\hs)Y^{\halpha_{d}}
\cr\!\!\!\!&&\!\!\!\!-\int_0^1\rd \hs\int_0^{1-\hs}\rd s\,\partial_{s}\left[\epsilon_{\halpha\hbeta\halpha_1\cdots\halpha_{d}} \hA^{\halpha\hbeta}\wedge F^{\halpha_1\halpha_2}(s,\hs)
\wedge\rD(s,\hs)Y^{\halpha_3}\wedge\cdots\wedge \rD(s,\hs)Y^{\halpha_{d}}\right]
\cr\!\!\!\!&=&\!\!\!\!\int_0^1\rd s\,\epsilon_{\halpha\hbeta\halpha_1\cdots\halpha_{d}}A^{\halpha\hbeta}\wedge F^{\halpha_1\halpha_2}(s)\wedge\rD(s)Y^{\halpha_3}\wedge\cdots\wedge\rD(s)Y^{\halpha_{d}}
\cr\!\!\!\!&&\!\!\!\!-\int_0^1\rd \hs\,\epsilon_{\halpha\hbeta\halpha_1\cdots\halpha_{d}}\hA^{\halpha\hbeta}\wedge \hF^{\halpha_1\halpha_2}(\hs)\wedge\hrD(\hs)Y^{\halpha_3}\wedge\cdots\wedge\hrD(\hs)Y^{\halpha_{d}}
\cr\!\!\!\!&&\!\!\!\!+\int_0^1\rd s\int_0^{1-s}\rd \hs\,\epsilon_{\halpha\hbeta\halpha_1\cdots\halpha_{d}}\Big\{\partial_{\hs}\left[ A^{\halpha\hbeta}\wedge F^{\halpha_1\halpha_2}(s,\hs)
\wedge\rD(s,\hs)Y^{\halpha_3}\wedge\cdots\wedge \rD(s,\hs)Y^{\halpha_{d}}\right]
\cr&&~~~~~~~~~~~~~~~~~~~~~~~~~~~~~~~~~~-\partial_{s}\left[ \hA^{\halpha\hbeta}\wedge F^{\halpha_1\halpha_2}(s,\hs)
\wedge\rD(s,\hs)Y^{\halpha_3}\wedge\cdots\wedge \rD(s,\hs)Y^{\halpha_{d}}\right]\Big\}
\cr\!\!\!\!&=&\!\!\!\!\int_0^1\rd s\,\epsilon_{\halpha\hbeta\halpha_1\cdots\halpha_{d}}A^{\halpha\hbeta}\wedge F^{\halpha_1\halpha_2}(s)\wedge\rD(s)Y^{\halpha_3}\wedge\cdots\wedge\rD(s)Y^{\halpha_{d}}
\cr\!\!\!\!&&\!\!\!\!-\int_0^1\rd \hs\,\epsilon_{\halpha\hbeta\halpha_1\cdots\halpha_{d}}\hA^{\halpha\hbeta}\wedge \hF^{\halpha_1\halpha_2}(\hs)\wedge\hrD(\hs)Y^{\halpha_3}\wedge\cdots\wedge\hrD(\hs)Y^{\halpha_{d}}
\cr\!\!\!\!&&\!\!\!\!+\int_0^1\rd s\int_0^{1-s}\rd \hs\,\epsilon_{\halpha\hbeta\halpha_1\cdots\halpha_{d}}\Big\{A^{\halpha\hbeta}\wedge \rD(s,\hs)\hA^{\halpha_1\halpha_2}
\wedge\rD(s,\hs)Y^{\halpha_3}\wedge\cdots\wedge \rD(s,\hs)Y^{\halpha_{d}}
\cr&&~~~~~~~~~~~~~~~~~~~~~~~~~~~~~~~~~~+(d-2)A^{\halpha\hbeta}\wedge F^{\halpha_1\halpha_2}(s,\hs)
\wedge (\hA Y)^{\halpha_3}\wedge\rD(s,\hs)Y^{\halpha_4}\wedge\cdots\wedge \rD(s,\hs)Y^{\halpha_{d}}
\cr&&~~~~~~~~~~~~~~~~~~~~~~~~~~~~~~~~~~-\hA^{\halpha\hbeta}\wedge \rD(s,\hs)A^{\halpha_1\halpha_2}
\wedge\rD(s,\hs)Y^{\halpha_3}\wedge\cdots\wedge \rD(s,\hs)Y^{\halpha_{d}}
\cr&&~~~~~~~~~~~~~~~~~~~~~~~~~~~~~~~~~~-(d-2)\hA^{\halpha\hbeta}\wedge F^{\halpha_1\halpha_2}(s,\hs)
\wedge (A Y)^{\halpha_3}\wedge\rD(s,\hs)Y^{\halpha_4}\wedge\cdots\wedge \rD(s,\hs)Y^{\halpha_{d}}\Big\}
\cr\!\!\!\!&=&\!\!\!\!\int_0^1\rd s\,\epsilon_{\halpha\hbeta\halpha_1\cdots\halpha_{d}}A^{\halpha\hbeta}\wedge F^{\halpha_1\halpha_2}(s)\wedge\rD(s)Y^{\halpha_3}\wedge\cdots\wedge\rD(s)Y^{\halpha_{d}}
\cr\!\!\!\!&&\!\!\!\!-\int_0^1\rd \hs\,\epsilon_{\halpha\hbeta\halpha_1\cdots\halpha_{d}}\hA^{\halpha\hbeta}\wedge \hF^{\halpha_1\halpha_2}(\hs)\wedge\hrD(\hs)Y^{\halpha_3}\wedge\cdots\wedge\hrD(\hs)Y^{\halpha_{d}}
\cr\!\!\!\!&&\!\!\!\!+\int_0^1\rd s\int_0^{1-s}\rd \hs\,\epsilon_{\halpha\hbeta\halpha_1\cdots\halpha_{d}}\Big\{ \rD(s,\hs)\hA^{\halpha\hbeta}\wedge A^{\halpha_1\halpha_2}
\wedge\rD(s,\hs)Y^{\halpha_3}\wedge\cdots\wedge \rD(s,\hs)Y^{\halpha_{d}}
\cr&&~~~~~~~~~~~~~~~~~~~~~~~~~~~~~~~~~~-\hA^{\halpha\hbeta}
\wedge\rD(s,\hs)A^{\halpha_1\halpha_2}\wedge \rD(s,\hs)Y^{\halpha_3}\wedge\cdots\wedge \rD(s,\hs)Y^{\halpha_{d}}
\cr&&~~~~~~~~~~~~~~~~~~~~~~~~~~~~~~~~~~+(d-2)\hA^{\halpha\hbeta}\wedge A^{\halpha_1\halpha_2}\wedge F^{\halpha_3\hgamma}(s,\hs) Y_{\hgamma} \wedge\rD(s,\hs)Y^{\halpha_4}\wedge\cdots\wedge \rD(s,\hs)Y^{\halpha_{d}}\Big\}
\cr\!\!\!\!&=&\!\!\!\!\int_0^1\rd s\,\epsilon_{\halpha\hbeta\halpha_1\cdots\halpha_{d}}A^{\halpha\hbeta}\wedge F^{\halpha_1\halpha_2}(s)\wedge\rD(s)Y^{\halpha_3}\wedge\cdots\wedge\rD(s)Y^{\halpha_{d}}
\cr\!\!\!\!&&\!\!\!\!-\int_0^1\rd \hs\,\epsilon_{\halpha\hbeta\halpha_1\cdots\halpha_{d}}\hA^{\halpha\hbeta}\wedge \hF^{\halpha_1\halpha_2}(\hs)\wedge\hrD(\hs)Y^{\halpha_3}\wedge\cdots\wedge\hrD(\hs)Y^{\halpha_{d}}
\cr\!\!\!\!&&\!\!\!\!+\int_0^1\rd s\int_0^{1-s}\rd \hs\,\epsilon_{\halpha\hbeta\halpha_1\cdots\halpha_{d}} \rD(s,\hs)[\hA^{\halpha\hbeta}\wedge A^{\halpha_1\halpha_2}
\wedge\rD(s,\hs)Y^{\halpha_3}\wedge\cdots\wedge \rD(s,\hs)Y^{\halpha_{d}}]
\cr\!\!\!\!&=&\!\!\!\!\frac{4\kappa^2(d-2)!}{(-1)^{d+1}\ell}\cL^{\rm CS}_{(d+1)}[A,Y]
-\int_0^1\rd \hs\,\epsilon_{\halpha\hbeta\halpha_1\cdots\halpha_{d}}\hA^{\halpha\hbeta}\wedge \hF^{\halpha_1\halpha_2}(\hs)\wedge\hrD(\hs)Y^{\halpha_3}\wedge\cdots\wedge\hrD(\hs)Y^{\halpha_{d}}
\cr\!\!\!\!&&\!\!\!\!+\int_0^1\rd s\int_0^{1-s}\rd \hs\,\rd[\epsilon_{\halpha\hbeta\halpha_1\cdots\halpha_{d}} \hA^{\halpha\hbeta}\wedge A^{\halpha_1\halpha_2}
\wedge\rD(s,\hs)Y^{\halpha_3}\wedge\cdots\wedge \rD(s,\hs)Y^{\halpha_{d}}]\,,
\end{eqnarray}
where we have used the double homotopic notation
\begin{eqnarray}
A(s,\hs)\!\!\!\!&=&\!\!\!\!sA +\hs\hA\,,
\cr\rD(s,\hs)Y\!\!\!\!&=&\!\!\!\!\rd Y+sA Y+\hs\hat AY=\rD(s)Y+\hs\hA Y=\hrD(\hs)Y+s AY\,,
\cr
F(s,\hs)\!\!\!\!&=&\!\!\!\!\rd A(s,\hs)+A(s,\hs)\wedge A(s,\hs)
=s\rd A +\hs\rd\hA+(sA +\hs\hA)\wedge(sA +\hs\hA)
\cr\!\!\!\!&=&\!\!\!\!F(s) +\hF(\hs)+s\hs(A \wedge\hA+\hA\wedge A)\,,
\cr
\rD(s,\hs)(tA+\Ht\hA)\!\!\!\!&=&\!\!\!\!\rd(tA+\Ht\hA)+(sA +\hs\hA)\wedge(tA+\Ht\hA)+(tA+\Ht\hA)\wedge(sA +\hs\hA)
\cr\!\!\!\!&=&\!\!\!\!(t\partial_s+\Ht\partial_{\hs})F(s,\hs)\,,
\cr A(t,\Ht)\!\!\!\!&=&\!\!\!\!tA+\Ht\hA=(t\partial_s+\Ht\partial_{\hs})A(s,\hs)\,,
\end{eqnarray}
as well as the resummation
\begin{eqnarray}
\!\!\!\!&&\!\!\!\!\epsilon_{\halpha\hbeta\halpha_1\cdots\halpha_{d}} \hA^{\halpha\hbeta}\wedge A^{\halpha_1\halpha_2}\wedge F^{\halpha_3\hgamma}(s,\hs) Y_{\hgamma} \wedge\rD(s,\hs)Y^{\halpha_4}\wedge\cdots\wedge \rD(s,\hs)Y^{\halpha_{d}}
\cr\!\!\!\!&=&\!\!\!\!\epsilon_{\halpha\hbeta\halpha_1\cdots\halpha_{d}} A^{\halpha\hbeta}\wedge F^{\halpha_1\halpha_2}(s,\hs) \wedge(\hA Y)^{\halpha_3}\wedge \rD(s,\hs)Y^{\halpha_4}\wedge\cdots\wedge \rD(s,\hs)Y^{\halpha_{d}}
\cr\!\!\!\!&&\!\!\!\!-\epsilon_{\halpha\hbeta\halpha_1\cdots\halpha_{d}} \hA^{\halpha\hbeta}\wedge F^{\halpha_1\halpha_2}(s,\hs)\wedge (A Y)^{\halpha_1} \wedge\rD(s,\hs)Y^{\halpha_4}\wedge\cdots\wedge \rD(s,\hs)Y^{\halpha_{d}}
\,.~~~~~~~
\end{eqnarray}
\subsubsection{$SO(2,d)$ gauge transformation of $\cL^{\rm top}_{(d+1)}$}
Under the finite gauge transformation
\begin{eqnarray}
Y^{\halpha}\to \tilde Y^{\halpha}=U^{\halpha}{}_{\hbeta}Y^{\hbeta}\,,~~~~~~~
\end{eqnarray}
the topological term transforms as
{\fontsize{10 pt}{\baselineskip}\begin{eqnarray}
\!\!\!\!&&\!\!\!\!\frac{\kappa^2\ell^3(d+1)!}{d}\cL^{\rm top}_{(d+1)}[\tilde Y]
=\epsilon_{\halpha_0\cdots\halpha_{d+1}}\rd \tilde Y^{\halpha_0}\wedge\cdots\wedge\rd \tilde Y^{\halpha_{d}}\tilde Y^{\halpha_{d+1}}
\cr\!\!\!\!&=&\!\!\!\!\epsilon_{\halpha_0\cdots\halpha_{d+1}}\rd (U^{\halpha_0}{}_{\hbeta_0}Y^{\hbeta_0})\wedge\cdots\wedge\rd (U^{\halpha_{d}}{}_{\hbeta_d}Y^{\hbeta_d}) U^{\halpha_{d+1}}{}_{\hbeta_{d+1}}Y^{\hbeta_{d+1}}
\cr\!\!\!\!&=&\!\!\!\!\epsilon_{\halpha_0\cdots\halpha_{d+1}}(\rd U^{\halpha_0}{}_{\hbeta_0}Y^{\hbeta_0}+U^{\halpha_0}{}_{\hbeta_0}\rd Y^{\hbeta_0})\wedge\cdots\wedge (\rd U^{\halpha_d}{}_{\hbeta_d}Y^{\hbeta_d}+U^{\halpha_d}{}_{\hbeta_d}\rd Y^{\hbeta_d}) U^{\halpha_{d+1}}{}_{\hbeta_{d+1}}Y^{\hbeta_{d+1}}
\cr\!\!\!\!&=&\!\!\!\!\sum_{n=0}^{d+1}\frac{(d+1)!}{n!(d+1-n)!} \epsilon_{\halpha_0\cdots\halpha_{d+1}}\rd U^{\halpha_0}{}_{\hbeta_0}Y^{\hbeta_0}\wedge\cdots\wedge\rd U^{\halpha_{n-1}}{}_{\hbeta_{n-1}}Y^{\hbeta_{n-1}}
\wedge U^{\halpha_n}{}_{\hbeta_n}\rd Y^{\hbeta_n}\wedge\cdots\wedge U^{\halpha_d}{}_{\hbeta_d}\rd Y^{\hbeta_d} U^{\halpha_{d+1}}{}_{\hbeta_{d+1}}Y^{\hbeta_{d+1}}
\cr\!\!\!\!&=&\!\!\!\!\sum_{n=0}^{d+1}\frac{(d+1)!}{n!(d+1-n)!} \epsilon_{\hbeta_0\cdots\hbeta_{d+1}}(U^{-1})^{\hbeta_0}{}_{\halpha_0}\rd U^{\halpha_0}{}_{\hgamma_0}Y^{\hgamma_0}\wedge\cdots\wedge(U^{-1})^{\hbeta_{n-1}}{}_{\halpha_{n-1}}\rd U^{\halpha_{n-1}}{}_{\hgamma_{n-1}}Y^{\hgamma_{n-1}}
\wedge \rd Y^{\hbeta_n}\wedge\cdots\wedge \rd Y^{\hbeta_d} Y^{\hbeta_{d+1}}
\cr\!\!\!\!&=&\!\!\!\!\sum_{n=0}^{d+1}\frac{(d+1)!}{n!(d+1-n)!} \epsilon_{\halpha_0\cdots\halpha_{d+1}}(U^{-1}\rd UY)^{\halpha_0}\wedge\cdots\wedge(U^{-1}\rd UY)^{\halpha_{n-1}}
\wedge \rd Y^{\halpha_n}\wedge\cdots\wedge \rd Y^{\halpha_d} Y^{\halpha_{d+1}}
\cr\!\!\!\!&=&\!\!\!\! \epsilon_{\halpha_0\cdots\halpha_{d+1}}(\rd Y+U^{-1}\rd UY)^{\halpha_0}\wedge\cdots\wedge(\rd Y+U^{-1}\rd UY)^{\halpha_d} Y^{\halpha_{d+1}}
\cr\!\!\!\!&=&\!\!\!\! \epsilon_{\halpha_0\cdots\halpha_{d+1}}\hrD Y^{\halpha_0}\wedge\cdots\wedge\hrD Y^{\halpha_d} Y^{\halpha_{d+1}}
\cr\!\!\!\!&=&\!\!\!\!\epsilon_{\halpha_0\cdots\halpha_{d+1}}\rd  Y^{\halpha_0}\wedge\cdots\wedge\rd  Y^{\halpha_{d}} Y^{\halpha_{d+1}}
+\int_{0}^{1}\rd s\,\partial_s[\epsilon_{\halpha_0\cdots\halpha_{d+1}}\hrD(s) Y^{\halpha_0}\wedge\cdots\wedge\hrD(s) Y^{\halpha_d} Y^{\halpha_{d+1}}]
\cr\!\!\!\!&=&\!\!\!\!\epsilon_{\halpha_0\cdots\halpha_{d+1}}\rd  Y^{\halpha_0}\wedge\cdots\wedge\rd  Y^{\halpha_{d}} Y^{\halpha_{d+1}}
+(d+1)\int_{0}^{1}\rd s\,\epsilon_{\halpha_0\cdots\halpha_{d+1}}(\hat AY)^{\halpha_0}\wedge\hrD(s) Y^{\halpha_1}\wedge\cdots\wedge\hrD(s) Y^{\halpha_d} Y^{\halpha_{d+1}}
\cr\!\!\!\!&=&\!\!\!\!\epsilon_{\halpha_0\cdots\halpha_{d+1}}\rd Y^{\halpha_0}\wedge\cdots\wedge\rd  Y^{\halpha_{d}}Y^{\halpha_{d+1}}
-\frac{(-1)^{d}(d+1)\ell^2}2\int_0^{1}\rd s\,\epsilon_{\halpha_0\cdots\halpha_{d+1}}\hA^{\halpha_0\halpha_{1}}\wedge \hrD(s) Y^{\halpha_{2}}\wedge\cdots\wedge\hrD(s) Y^{\halpha_{d+1}}
\cr\!\!\!\!&=&\!\!\!\!\epsilon_{\halpha_0\cdots\halpha_{d+1}}\rd Y^{\halpha_0}\wedge\cdots\wedge\rd  Y^{\halpha_{d}}Y^{\halpha_{d+1}}
-\frac{(d+1)\ell^2}2\rd\Big[\int_0^{1}\rd s\,\epsilon_{\halpha_0\cdots\halpha_{d+1}}\hA^{\halpha_0\halpha_{1}}\wedge \hrD(s) Y^{\halpha_{2}}\wedge\cdots\wedge\hrD(s)Y^{\halpha_{d}} Y^{\halpha_{d+1}}\Big]
\cr\!\!\!\!&&\!\!\!\!+\frac{(d+1)\ell^2}2\int_0^{1}\rd s\,\epsilon_{\halpha_0\cdots\halpha_{d+1}}\partial_s\hF(s)^{\halpha_0\halpha_{1}}\wedge \hrD(s) Y^{\halpha_{2}}\wedge\cdots\wedge\hrD(s)Y^{\halpha_{d}} Y^{\halpha_{d+1}}
\cr\!\!\!\!&&\!\!\!\!-\frac{(d^2-1)\ell^2}2\int_0^{1}\rd s\,\epsilon_{\halpha_0\cdots\halpha_{d+1}}\hA^{\halpha_0\halpha_{1}}\wedge [\hF(s) Y]^{\halpha_{2}}\wedge\hrD(s) Y^{\halpha_{3}}\wedge\cdots\wedge\hrD(s)Y^{\halpha_{d}} Y^{\halpha_{d+1}}
\cr\!\!\!\!&=&\!\!\!\!\epsilon_{\halpha_0\cdots\halpha_{d+1}}\rd Y^{\halpha_0}\wedge\cdots\wedge\rd  Y^{\halpha_{d}}Y^{\halpha_{d+1}}
-\frac{(d+1)\ell^2}2\rd\Big[\int_0^{1}\rd s\,\epsilon_{\halpha_0\cdots\halpha_{d+1}}\hA^{\halpha_0\halpha_{1}}\wedge \hrD(s) Y^{\halpha_{2}}\wedge\cdots\wedge\hrD(s)Y^{\halpha_{d}} Y^{\halpha_{d+1}}\Big]
\cr\!\!\!\!&&\!\!\!\!+\frac{(d+1)\ell^2}2\int_0^{1}\rd s\,\epsilon_{\halpha_0\cdots\halpha_{d+1}}\partial_s\Big[\hF(s)^{\halpha_0\halpha_{1}}\wedge \hrD(s) Y^{\halpha_{2}}\wedge\cdots\wedge\hrD(s)Y^{\halpha_{d}} Y^{\halpha_{d+1}}\Big]
\cr\!\!\!\!&&\!\!\!\!-\frac{(d^2-1)\ell^2}2\int_0^{1}\rd s\,\epsilon_{\halpha_0\cdots\halpha_{d+1}}\hF(s)^{\halpha_0\halpha_{1}}\wedge (\hA Y)^{\halpha_{2}}\wedge\hrD(s) Y^{\halpha_3}\wedge\cdots\wedge\hrD(s)Y^{\halpha_{d}} Y^{\halpha_{d+1}}
\cr\!\!\!\!&&\!\!\!\!-\frac{(d^2-1)\ell^2}2\int_0^{1}\rd s\,\epsilon_{\halpha_0\cdots\halpha_{d+1}}\hA^{\halpha_0\halpha_{1}}\wedge [\hF(s) Y]^{\halpha_{2}}\wedge\hrD(s) Y^{\halpha_{3}}\wedge\cdots\wedge\hrD(s)Y^{\halpha_{d}} Y^{\halpha_{d+1}}
\cr\!\!\!\!&=&\!\!\!\!\epsilon_{\halpha_0\cdots\halpha_{d+1}}\rd  Y^{\halpha_0}\wedge\cdots\wedge\rd  Y^{\halpha_{d}} Y^{\halpha_{d+1}}
\cr\!\!\!\!&&\!\!\!\!
-\frac{(d+1)\ell^2}2\rd\Big[\int_0^{1}\rd s\,\epsilon_{\halpha_0\cdots\halpha_{d+1}}\hA^{\halpha_0\halpha_{1}}\wedge \hrD(s) Y^{\halpha_{2}}\wedge\cdots\wedge\hrD(s)Y^{\halpha_{d}} Y^{\halpha_{d+1}}\Big]
\cr\!\!\!\!&&\!\!\!\!+\frac{(-1)^d(d^2-1)\ell^4}4\int_0^{1}\rd s\,\epsilon_{\halpha_0\cdots\halpha_{d+1}}\hA^{\halpha_0\halpha_{1}}\wedge \hF^{\halpha_{2}\halpha_{3}}(s)\wedge\hrD(s) Y^{\halpha_{4}}\wedge\cdots\wedge\hrD(s)Y^{\halpha_{d+1}}
\,,~~~~~~~
\end{eqnarray}}
where we have used that
\begin{eqnarray}
\!\!\!\!&&\!\!\!\!\hat F(s)=\rd\hat A(s)+\hat A(s)\wedge\hat A(s)=\rd(sU^{-1}\rd U)+(sU^{-1}\rd U)\wedge(sU^{-1}\rd U)
\cr\!\!\!\!&=&\!\!\!\!-sU^{-1}\rd UU^{-1}\wedge\rd U+s^2U^{-1}\rd U\wedge U^{-1}\rd U
=s(s-1)U^{-1}\rd U\wedge U^{-1}\rd U
\cr\!\!\!\!&=&\!\!\!\!s(1-s)\rd(U^{-1}\rd U)\,,
\\\cr \!\!\!\!&&\!\!\!\!
\epsilon_{\hbeta_0\cdots\hbeta_{d+1}}(U^{-1})^{\hbeta_0}{}_{\halpha_0}\cdots (U^{-1})^{\hbeta_{n}}{}_{\halpha_{n}}
=
\epsilon_{\halpha_0\cdots\halpha_{d+1}}U^{\halpha_{n+1}}{}_{\hbeta_{n+1}}\cdots U^{\halpha_{d+1}}{}_{\hbeta_{d+1}}\,,
\end{eqnarray}
as well as the resummations
\begin{eqnarray}
\!\!\!\!&&\!\!\!\!\epsilon_{\halpha_0\cdots\halpha_{d+1}}\hat A^{\halpha_0\hbeta}Y_{\hbeta}\wedge\hrD(s) Y^{\halpha_1}\wedge\cdots\wedge\hrD(s) Y^{\halpha_d} Y^{\halpha_{d+1}}
\cr\!\!\!\!&=&\!\!\!\!-\frac{(-1)^d\ell^2}2\epsilon_{\halpha_0\cdots\halpha_{d+1}}\hat A^{\halpha_0\halpha_{1}}\wedge\hrD(s) Y^{\halpha_2}\wedge\cdots\wedge\hrD(s) Y^{\halpha_{d+1}}\,,
\\\cr
\!\!\!\!&&\!\!\!\!
-\ell^2\epsilon_{\halpha_0\cdots\halpha_{d+1}}\hA^{\halpha_0\halpha_{1}}\wedge \hF^{\halpha_{2}\halpha_{3}}(s)\wedge\hrD(s) Y^{\halpha_{4}}\wedge\cdots\wedge\hrD(s)Y^{\halpha_{d+1}}
\cr\!\!\!\!&=&\!\!\!\!2(-1)^d\epsilon_{\halpha_0\cdots\halpha_{d+1}} \hF^{\halpha_{0}\halpha_{1}}(s)\wedge(\hA Y)^{\halpha_2}\wedge \hrD(s) Y^{\halpha_{3}}\wedge\cdots\wedge\hrD(s)Y^{\halpha_{d}}Y^{\halpha_{d+1}}
\cr\!\!\!\!&&\!\!\!\!+2(-1)^d\epsilon_{\halpha_0\cdots\halpha_{d+1}}\hA^{\halpha_0\halpha_{1}}\wedge [\hF(s) Y]^{\halpha_{2}}\wedge\hrD(s) Y^{\halpha_3}\wedge\cdots\wedge\hrD(s)Y^{\halpha_d}Y^{\halpha_{d+1}}\,.
\end{eqnarray}

Together with $\cL^{\rm top}_{(d+1)}$, the bulk consistent action
\begin{eqnarray}
\!\!\!\!&&\!\!\!\!2\kappa^2\ell(d-1)!S^{\rm con}_{(d+1)}=2\kappa^2\ell(d-1)!(S^{\rm CS}_{(d+1)}-S^{\rm top}_{(d+1)})
\cr\!\!\!\!&=&\!\!\!\!\frac{(-1)^{d+1}(d-1)\ell^2}{2}\int_{M}\epsilon_{\halpha_0\halpha_1\cdots\halpha_{d+1}} \Big[\int_0^1\rd s\,A^{\halpha_0\halpha_1}\wedge F^{\halpha_2\halpha_3}(s)\wedge\rD(s)Y^{\halpha_4}\wedge\cdots\wedge\rD(s)Y^{\halpha_{d+1}}
\cr&&~~~~~~~~~~~~~~~~~~~~~~~~~~~~~~~~~~~~~~~~~~~+\frac{4(-1)^{d}}{(d^2-1)\ell^4} \rd Y^{\halpha_0}\wedge\cdots\wedge\rd Y^{\halpha_{d}}Y^{\halpha_{d+1}}\Big]
\,,~~~~~~~
\end{eqnarray}
transforms as
\begin{eqnarray}\label{A.Con.F}
\!\!\!\!&&\!\!\!\!\frac{4\kappa^2(d-2)!}{(-1)^{d+1}\ell}S^{\rm con}_{(d+1)}[\tilde A,\tilde Y]
-\frac{4\kappa^2(d-2)!}{(-1)^{d+1}\ell}S^{\rm con}_{(d+1)}[A,Y]
\cr\!\!\!\!&=&\!\!\!\!\int_{\Sigma}\int_0^1\rd s\,\epsilon_{\halpha\hbeta\halpha_1\cdots\halpha_{d}} \hA^{\halpha\hbeta}\wedge\Big[\int_0^{1-s}\rd \hs\,A^{\halpha_1\halpha_2}
\wedge\rD(s,\hs)Y^{\halpha_3}\wedge\cdots\wedge \rD(s,\hs)Y^{\halpha_{d}}
\cr&&~~~~~~~~~~~~~~~~~~~~~~~~~~~~~~~~~~~~~
+\frac{2}{(d-1)\ell^2} Y^{\halpha_{1}}\hrD(s) Y^{\halpha_{2}}\wedge\cdots\wedge\hrD(s)Y^{\halpha_{d}} \Big]
\,.
\end{eqnarray}
The large gauge transforms in $S^{\rm CS}_{(d+1)}$ and $S^{\rm top}_{(d+1)}$ cancel each other, and thus $S^{\rm con}_{(d+1)}$ is gauge invariant up to boundary terms.  

\subsubsection{The variation of $S^{\rm con}_{(d+1)}$}
The variation of $S^{\rm CS}_{(d+1)}$ is given by
\begin{eqnarray}\label{variS_cs}
\!\!\!\!&&\!\!\!\!
\delta\int_{M}\int_0^1\rd s\,\epsilon_{\halpha\hbeta\halpha_1\cdots\halpha_{d}} A^{\halpha\hbeta}\wedge F^{\halpha_1\halpha_2}(s)\wedge\rD(s)Y^{\halpha_3}\wedge\cdots\wedge\rD(s)Y^{\halpha_{d}}
\cr\!\!\!\!&=&\!\!\!\!\int_{M}\int_0^1\rd s\,\epsilon_{\halpha\hbeta\halpha_1\cdots\halpha_{d}} \Big\{\delta A^{\halpha\hbeta}\wedge F^{\halpha_1\halpha_2}(s)\wedge\rD(s)Y^{\halpha_3}\wedge\cdots\wedge\rD(s)Y^{\halpha_{d}}
\cr&&~~~~~~~~~~~~~~~~~~~~~~~~~~+\delta A^{\halpha\hbeta}\wedge\rD(s)[A^{\halpha_1\halpha_2}(s) \wedge\rD(s)Y^{\halpha_3}\wedge\cdots\wedge\rD(s)Y^{\halpha_{d}}]
\cr&&~~~~~~~~~~~~~~~~~~~~~~~~~~-(d-2)\delta A^{\halpha}{}_{\hgamma}Y^{\hgamma}\wedge A^{\hbeta\halpha_3}(s)\wedge F^{\halpha_1\halpha_2}(s)\wedge \rD(s)Y^{\halpha_4}\wedge\cdots\wedge\rD(s)Y^{\halpha_{d}}
\cr&&~~~~~~~~~~~~~~~~~~~~~~~~~~+(d-2)\delta Y^{\halpha}\rD(s)[A^{\hbeta\halpha_1}\wedge F^{\halpha_2\halpha_3}(s)\wedge\rD(s)Y^{\halpha_4}\wedge\cdots\wedge\rD(s)Y^{\halpha_{d}}]\Big\}
\cr\!\!\!\!&&\!\!\!\!+\int_{\Sigma}\int_0^1\rd s\,\epsilon_{\halpha\hbeta\halpha_1\cdots\halpha_{d}} \Big\{\delta A^{\halpha\hbeta}\wedge A^{\halpha_1\halpha_2}(s) \wedge\rD(s)Y^{\halpha_3}\wedge\cdots\wedge\rD(s)Y^{\halpha_{d}}
\cr&&~~~~~~~~~~~~~~~~~~~~~~~~~~~~-(d-2)\delta Y^{\halpha}A^{\hbeta\halpha_1}\wedge F^{\halpha_2\halpha_3}(s)\wedge\rD(s)Y^{\halpha_4}\wedge\cdots\wedge\rD(s)Y^{\halpha_{d}}\Big\}
\cr\!\!\!\!&=&\!\!\!\!\int_{M}\int_0^1\rd s\,\epsilon_{\halpha\hbeta\halpha_1\cdots\halpha_{d}} \Big\{\delta A^{\halpha\hbeta}\wedge F^{\halpha_1\halpha_2}(s)\wedge\rD(s)Y^{\halpha_3}\wedge\cdots\wedge\rD(s)Y^{\halpha_{d}}
\cr&&~~~~~~~~~~~~~~~~~~~~~~~~~~+\delta A^{\halpha\hbeta}\wedge s\partial_sF^{\halpha_1\halpha_2}(s) \wedge\rD(s)Y^{\halpha_3}\wedge\cdots\wedge\rD(s)Y^{\halpha_{d}}
\cr&&~~~~~~~~~~~~~~~~~~~~~~~~~~-(d-2)\delta A^{\halpha\hbeta}\wedge A^{\halpha_1\halpha_2}(s) \wedge [F(s)Y]^{\halpha_3}\wedge \rD(s)Y^{\halpha_4}\wedge\cdots\wedge\rD(s)Y^{\halpha_{d}}
\cr&&~~~~~~~~~~~~~~~~~~~~~~~~~~-(d-2)\delta A^{\halpha}{}_{\hgamma}Y^{\hgamma}\wedge A^{\hbeta\halpha_3}(s)\wedge F^{\halpha_1\halpha_2}(s)\wedge \rD(s)Y^{\halpha_4}\wedge\cdots\wedge\rD(s)Y^{\halpha_{d}}
\cr&&~~~~~~~~~~~~~~~~~~~~~~~~~~+(d-2)\delta Y^{\halpha}\partial_sF^{\hbeta\halpha_1}(s)\wedge F^{\halpha_2\halpha_3}(s)\wedge\rD(s)Y^{\halpha_4}\wedge\cdots\wedge\rD(s)Y^{\halpha_{d}}
\cr&&~~~~~~~~~~~~~~~~~~~~~~~~~~-(d-2)(d-3)\delta Y^{\halpha}A^{\hbeta\halpha_1}\wedge F^{\halpha_2\halpha_3}(s)\wedge [F(s)Y]^{\halpha_4}\wedge\rD(s)Y^{\halpha_5}\wedge\cdots\wedge\rD(s)Y^{\halpha_{d}}\Big\}
\cr\!\!\!\!&&\!\!\!\!+\int_{\Sigma}\int_0^1\rd s\,\epsilon_{\halpha\hbeta\halpha_1\cdots\halpha_{d}} \Big\{\delta A^{\halpha\hbeta}\wedge A^{\halpha_1\halpha_2}(s) \wedge\rD(s)Y^{\halpha_3}\wedge\cdots\wedge\rD(s)Y^{\halpha_{d}}
\cr&&~~~~~~~~~~~~~~~~~~~~~~~~~~~~-\frac{2(d-2)}{\ell^2}\delta Y^{\halpha}Y^{\hbeta}F^{\halpha_1\halpha_2}(s)\wedge(AY)^{\halpha_3}\wedge \rD(s)Y^{\halpha_4}\wedge\cdots\wedge\rD(s)Y^{\halpha_{d}}
\cr&&~~~~~~~~~~~~~~~~~~~~~~~~~~~~-\frac{2(d-2)}{\ell^2}\delta Y^{\halpha}Y^{\hbeta}A^{\halpha_1\halpha_2}\wedge [F(s) Y]^{\halpha_3}\wedge\rD(s)Y^{\halpha_4}\wedge\cdots\wedge\rD(s)Y^{\halpha_{d}}\Big\}
\cr\!\!\!\!&=&\!\!\!\!\int_{M}\int_0^1\rd s\,\epsilon_{\halpha\hbeta\halpha_1\cdots\halpha_{d}} \Big\{\delta A^{\halpha\hbeta}\wedge F^{\halpha_1\halpha_2}(s)\wedge\rD(s)Y^{\halpha_3}\wedge\cdots\wedge\rD(s)Y^{\halpha_{d}}
\cr&&~~~~~~~~~~~~~~~~~~~~~~~~~~+\delta A^{\halpha\hbeta}\wedge s\partial_sF^{\halpha_1\halpha_2}(s) \wedge\rD(s)Y^{\halpha_3}\wedge\cdots\wedge\rD(s)Y^{\halpha_{d}}
\cr&&~~~~~~~~~~~~~~~~~~~~~~~~~~+(d-2)\delta A^{\halpha\hbeta}\wedge F^{\halpha_1\halpha_2}(s)\wedge s\partial_s[\rD(s)Y^{\halpha_3}]\wedge\rD(s)Y^{\halpha_4}\wedge\cdots\wedge\rD(s)Y^{\halpha_{d}}
\cr&&~~~~~~~~~~~~~~~~~~~~~~~~~~+(d-2)\delta Y^{\halpha}\partial_sF^{\hbeta\halpha_1}(s)\wedge F^{\halpha_2\halpha_3}(s)\wedge\rD(s)Y^{\halpha_4}\wedge\cdots\wedge\rD(s)Y^{\halpha_{d}}
\cr&&~~~~~~~~~~~~~~~~~~~~~~~~~~+\frac{(d-2)(d-3)}2\delta Y^{\halpha}F^{\hbeta\halpha_1}(s)\wedge F^{\halpha_2\halpha_3}(s)\wedge \partial_s[\rD(s)Y^{\halpha_4}]\wedge\rD(s)Y^{\halpha_5}\wedge\cdots\wedge\rD(s)Y^{\halpha_{d}}\Big\}
\cr\!\!\!\!&&\!\!\!\!+\int_{\Sigma}\int_0^1\rd s\,\epsilon_{\halpha\hbeta\halpha_1\cdots\halpha_{d}} \Big\{\delta A^{\halpha\hbeta}\wedge A^{\halpha_1\halpha_2}(s) \wedge\rD(s)Y^{\halpha_3}\wedge\cdots\wedge\rD(s)Y^{\halpha_{d}}
\cr&&~~~~~~~~~~~~~~~~~~~~~~~~~~~~-\frac{2(d-2)}{\ell^2}\delta Y^{\halpha}Y^{\hbeta}F^{\halpha_1\halpha_2}(s)\wedge \partial_s\rD(s)Y^{\halpha_3}\wedge\cdots\wedge\rD(s)Y^{\halpha_{d}}
\cr&&~~~~~~~~~~~~~~~~~~~~~~~~~~~~-\frac{2(d-2)}{\ell^2}\delta Y^{\halpha}Y^{\hbeta}A^{\halpha_1\halpha_2}\wedge \rD(s)\rD(s) Y^{\halpha_3}\wedge\rD(s)Y^{\halpha_4}\wedge\cdots\wedge\rD(s)Y^{\halpha_{d}}\Big\}
\cr\!\!\!\!&=&\!\!\!\!\int_{M}\int_0^1\rd s\,\epsilon_{\halpha\hbeta\halpha_1\cdots\halpha_{d}} \Big\{\partial_s[s\delta A^{\halpha\hbeta}\wedge F^{\halpha_1\halpha_2}(s) \wedge\rD(s)Y^{\halpha_3}\wedge\cdots\wedge\rD(s)Y^{\halpha_{d}}]
\cr&&~~~~~~~~~~~~~~~~~~~~~~~~~~+\frac{d-2}2\delta Y^{\halpha}\partial_s[F^{\hbeta\halpha_1}(s)\wedge F^{\halpha_2\halpha_3}(s)\wedge\rD(s)Y^{\halpha_4}\wedge\cdots\wedge\rD(s)Y^{\halpha_{d}}]
\Big\}
\cr\!\!\!\!&&\!\!\!\!+\int_{\Sigma}\int_0^1\rd s\,\epsilon_{\halpha\hbeta\halpha_1\cdots\halpha_{d}} \Big\{\delta A^{\halpha\hbeta}\wedge A^{\halpha_1\halpha_2}(s) \wedge\rD(s)Y^{\halpha_3}\wedge\cdots\wedge\rD(s)Y^{\halpha_{d}}
\cr&&~~~~~~~~~~~~~~~~~~~~~~~~~~~~-\frac{2}{\ell^2}\delta Y^{\halpha}Y^{\hbeta}\partial_s[F^{\halpha_1\halpha_2}(s)\wedge \rD(s)Y^{\halpha_3}\wedge\cdots\wedge\rD(s)Y^{\halpha_{d}}]
\cr&&~~~~~~~~~~~~~~~~~~~~~~~~~~~~+\frac{2}{\ell^2}\delta Y^{\halpha}Y^{\hbeta}[\partial_sF(s)-\rD(s) A]^{\halpha_1\halpha_2}\wedge \rD(s)Y^{\halpha_3}\wedge\cdots\wedge\rD(s)Y^{\halpha_{d}}
\cr&&~~~~~~~~~~~~~~~~~~~~~~~~~~~~+\frac{2}{\ell^2}\delta Y^{\halpha}Y^{\hbeta}\rD(s)[A^{\halpha_1\halpha_2}\wedge \rD(s)Y^{\halpha_3}\wedge\cdots\wedge\rD(s)Y^{\halpha_{d}}]\Big\}
\cr\!\!\!\!&=&\!\!\!\!\int_{M}\epsilon_{\halpha\hbeta\halpha_1\cdots\halpha_{d}} \Big\{\delta A^{\halpha\hbeta}\wedge F^{\halpha_1\halpha_2} \wedge\rD Y^{\halpha_3}\wedge\cdots\wedge\rD Y^{\halpha_{d}}
+\frac{2}{\ell^2}\delta Y^{\halpha}Y^{\hbeta}\rD[F^{\halpha_1\halpha_2}\wedge\rD Y^{\halpha_3}\wedge\cdots\wedge\rD Y^{\halpha_{d}}]
\Big\}
\cr\!\!\!\!&&\!\!\!\!+\int_{\Sigma}\epsilon_{\halpha\hbeta\halpha_1\cdots\halpha_{d}} \Big\{\int_0^1\!\!\rd s\,\delta A^{\halpha\hbeta}\wedge A^{\halpha_1\halpha_2}(s) \wedge\rD(s)Y^{\halpha_3}\wedge\cdots\wedge\rD(s)Y^{\halpha_{d}}
\cr&&~~~~-\frac{2}{\ell^2}\delta Y^{\halpha}Y^{\hbeta}\left[F^{\halpha_1\halpha_2}\wedge \rD Y^{\halpha_3}\wedge\cdots\wedge\rD Y^{\halpha_{d}}
-\int_0^1\!\!\!\rd s\,\rD(s)[A^{\halpha_1\halpha_2}\wedge \rD(s)Y^{\halpha_3}\wedge\cdots\wedge\rD(s)Y^{\halpha_{d}}]\right]\Big\}\,,
\end{eqnarray}
where we have used that
\begin{eqnarray}
\!\!\!\!&&\!\!\!\!\partial_s \rD(s)Y^{\halpha}=(AY)^{\halpha}\,,
\cr\!\!\!\!&&\!\!\!\!\rD(s) A^{\halpha_1\halpha_2}(s)=[F(s)+A(s)\wedge A(s)]^{\halpha_1\halpha_2}=s\partial_{s}F^{\halpha_1\halpha_2}(s)\,,
\cr\!\!\!\!&&\!\!\!\!\rD(s) A^{\halpha_1\halpha_2}=s^{-1}[F(s)+A(s)\wedge A(s)]^{\halpha_1\halpha_2}=\partial_{s}F^{\halpha_1\halpha_2}(s)\,,
\end{eqnarray}
as well as the resummations 
\begin{eqnarray}
\!\!\!\!&&\!\!\!\!\epsilon_{\halpha\hbeta\halpha_1\cdots\halpha_{d}}\delta A^{\halpha\hbeta}\wedge A^{\halpha_1\halpha_2}(s)\wedge F^{\halpha_3\hgamma}(s)Y_{\hgamma}\wedge\rD(s)Y^{\halpha_4}\wedge\cdots\wedge\rD(s)Y^{\halpha_{d}}
\cr\!\!\!\!&=&\!\!\!\!-\epsilon_{\halpha\hbeta\halpha_1\cdots\halpha_{d}}\delta A^{\halpha\hgamma}Y_{\hgamma}\wedge A^{\hbeta\halpha_1}(s)\wedge F^{\halpha_2\halpha_3}(s)\wedge\rD(s)Y^{\halpha_4}\wedge\cdots\wedge\rD(s)Y^{\halpha_{d}}
\cr\!\!\!\!&&\!\!\!\!-\epsilon_{\halpha\hbeta\halpha_1\cdots\halpha_{d}}\delta A^{\halpha\hbeta}\wedge F^{\halpha_1\halpha_2}(s)\wedge s\partial_s[\rD(s)Y^{\halpha_3}]\wedge\rD(s)Y^{\halpha_4}\wedge\cdots\wedge\rD(s)Y^{\halpha_{d}}\,,
\\\cr\!\!\!\!&&\!\!\!\!
\epsilon_{\halpha\hbeta\halpha_1\cdots\halpha_{d}}\delta Y^{\halpha}A^{\hbeta\halpha_1}\wedge F^{\halpha_2\halpha_3}(s)\wedge F^{\halpha_4\hgamma}(s)Y_{\gamma}\wedge\rD(s)Y^{\halpha_5}\wedge\cdots\wedge\rD(s)Y^{\halpha_{d}}
\cr\!\!\!\!&=&\!\!\!\!-\frac12\epsilon_{\halpha\hbeta\halpha_1\cdots\halpha_{d}}\delta Y^{\halpha}F^{\hbeta\halpha_1}(s)\wedge F^{\halpha_2\halpha_3}(s)\wedge \partial_s[\rD(s)Y^{\halpha_4}]\wedge\rD(s)Y^{\halpha_5}\wedge\cdots\wedge\rD(s)Y^{\halpha_{d}}\,,
\\\cr\!\!\!\!&&\!\!\!\!
-\ell^2\epsilon_{\halpha\hbeta\halpha_1\cdots\halpha_{d}}\delta Y^{\halpha}F^{\hbeta\halpha_1}\wedge F^{\halpha_2\halpha_3}\wedge\rD Y^{\halpha_4}\wedge\cdots\wedge\rD Y^{\halpha_{d}}
\cr\!\!\!\!&=&\!\!\!\!-\frac{4}{d-2}\epsilon_{\halpha\hbeta\halpha_1\cdots\halpha_{d}}\delta Y^{\halpha}Y^{\hbeta}\rD[F^{\halpha_1\halpha_2}\wedge \rD Y^{\halpha_3}\wedge\cdots\wedge\rD Y^{\halpha_{d}}]\,,
\\\cr\!\!\!\!&&\!\!\!\!
-\ell^2\delta Y^{\halpha}A^{\hbeta\halpha_1}\wedge F^{\halpha_2\halpha_3}(s)\wedge\rD(s)Y^{\halpha_4}\wedge\cdots\wedge\rD(s)Y^{\halpha_{d}}
\cr\!\!\!\!&=&\!\!\!\!\delta Y^{\halpha}A^{\hbeta\halpha_1}\wedge F^{\halpha_2\halpha_3}(s)\wedge\rD(s)Y^{\halpha_4}\wedge\cdots\wedge\rD(s)Y^{\halpha_{d}}Y^{\hgamma}Y_{\hgamma}
\cr\!\!\!\!&=&\!\!\!\!-\frac{2}{d-2}\delta Y^{\halpha}Y^{\hbeta} \partial_s[F^{\halpha_1\halpha_2}(s)\wedge\rD(s)Y^{\halpha_3}\wedge\cdots\wedge\rD(s)Y^{\halpha_{d}}]
\cr\!\!\!\!&&\!\!\!\!+\frac{2}{d-2}\delta Y^{\halpha}Y^{\hbeta}\rD(s)[A^{\halpha_1\halpha_2}\wedge \rD(s)Y^{\halpha_3}\wedge\cdots\wedge\rD(s)Y^{\halpha_d}]\,.
\end{eqnarray}
On the other hand, the variation of $S^{\rm top}_{(d+1)}$ only gives rise to the boundary term
\begin{eqnarray}\label{variS_top}
\!\!\!\!&&\!\!\!\!\frac{\kappa^2\ell^3(d+1)!}{d}\delta S^{\rm top}_{(d+1)}
\cr\!\!\!\!&=&\!\!\!\!\delta\int_{M}\epsilon_{\halpha_1\cdots\halpha_{d+2}}\rd Y^{\halpha_1}\wedge\cdots\wedge\rd Y^{\halpha_{d+1}}Y^{\halpha_{d+2}}
\cr\!\!\!\!&=&\!\!\!\!\int_{M}\epsilon_{\halpha_1\cdots\halpha_{d+2}}[(d+1)\rd \delta Y^{\halpha_1}\wedge\rd Y^{\halpha_2}\wedge\cdots\wedge\rd Y^{\halpha_{d+1}}Y^{\halpha_{d+2}}
+\rd Y^{\halpha_1}\wedge\rd Y^{\halpha_2}\wedge\cdots\wedge\rd Y^{\halpha_{d+1}}\delta Y^{\halpha_{d+2}}]
\cr\!\!\!\!&=&\!\!\!\!(-1)^d(d+1)\int_{\Sigma}\epsilon_{\halpha_1\cdots\halpha_{d+2}}\delta Y^{\halpha_1}Y^{\halpha_{2}}\rd Y^{\halpha_3}\wedge\cdots\wedge\rd Y^{\halpha_{d+2}}
\cr\!\!\!\!&&\!\!\!\!-(-1)^d(d+2)\int_{M}\epsilon_{\halpha_1\cdots\halpha_{d+2}}\delta Y^{\halpha_1}\rd Y^{\halpha_2}\wedge\cdots\wedge\rd Y^{\halpha_{d+1}}\wedge \rd Y^{\halpha_{d+2}}
\cr\!\!\!\!&=&\!\!\!\!(-1)^d(d+1)\int_{\Sigma}\epsilon_{\halpha_1\cdots\halpha_{d+2}}\delta  Y^{\halpha_1} Y^{\halpha_{2}}\rd  Y^{\halpha_3}\wedge\cdots\wedge\rd  Y^{\halpha_{d+2}}
\,,~~~~~~~
\end{eqnarray}
where we have used the resummation
\begin{eqnarray}
\!\!\!\!&&\!\!\!\!-\ell^2\epsilon_{\halpha_1\cdots\halpha_{d+2}}\delta  Y^{\halpha_1} \rd  Y^{\halpha_2}\wedge\cdots\wedge\rd  Y^{\halpha_{d+2}}
\cr\!\!\!\!&=&\!\!\!\!\epsilon_{\halpha_1\cdots\halpha_{d+2}}\delta  Y^{\halpha_1} \rd  Y^{\halpha_2}\wedge\cdots\wedge\rd  Y^{\halpha_{d+2}}Y^{\hgamma}Y_{\hgamma}
=0\,.
\end{eqnarray}

In total, we have
\begin{eqnarray}
\!\!\!\!&&\!\!\!\!\frac{4\kappa^2(d-2)!}{(-1)^{d+1}\ell}\delta S^{\rm con}_{(d+1)}=2\kappa^2\ell(d-1)!\frac{2(-1)^{d+1}}{(d-1)\ell^2}(\delta S^{\rm CS}_{(d+1)}-\delta S^{\rm top}_{(d+1)})
\cr\!\!\!\!&=&\!\!\!\!\int_{M}\epsilon_{\halpha\hbeta\halpha_1\cdots\halpha_{d}} \Big\{\delta A^{\halpha\hbeta}\wedge F^{\halpha_1\halpha_2} \wedge\rD Y^{\halpha_3}\wedge\cdots\wedge\rD Y^{\halpha_{d}}
+\frac{2}{\ell^2}\delta Y^{\halpha}Y^{\hbeta}\rD[F^{\halpha_1\halpha_2}\wedge\rD Y^{\halpha_3}\wedge\cdots\wedge\rD Y^{\halpha_{d}}]\Big\}
\cr\!\!\!\!&&\!\!\!\!+\int_{\Sigma}\epsilon_{\halpha\hbeta\halpha_1\cdots\halpha_{d}} \Big\{\int_0^1\!\!\rd s\,\delta A^{\halpha\hbeta}\wedge A^{\halpha_1\halpha_2}(s) \wedge\rD(s)Y^{\halpha_3}\wedge\cdots\wedge\rD(s)Y^{\halpha_{d}}
\cr&&~~~~~~~~~~~~~~~~~~~+\frac{2}{\ell^2}\delta Y^{\halpha}Y^{\hbeta}\Big[-F^{\halpha_1\halpha_2}\wedge \rD Y^{\halpha_3}\wedge\cdots\wedge\rD Y^{\halpha_{d}}+\frac{2}{(d-1)\ell^2}\rd  Y^{\halpha_1}\wedge\cdots\wedge\rd Y^{\halpha_{d}}
\cr&&~~~~~~~~~~~~~~~~~~~~~~~~~~~~~~~~~~~~~+\int_0^1\!\!\!\rd s\,\rD(s)[A^{\halpha_1\halpha_2}\wedge \rD(s)Y^{\halpha_3}\wedge\cdots\wedge\rD(s)Y^{\halpha_{d}}]\Big]\Big\}
\,.~~~~~~~
\end{eqnarray}
Especially, we have verified 
\begin{eqnarray}
\cJ^{\rm cov}_{(d+1)}=\cJ^{\rm CS}_{(d+1)}=\cJ^{\rm con}_{(d+1)}
\end{eqnarray}
which confirms that $S^{\rm cov}_{(d+1)},S^{\rm CS}_{(d+1)}$ and $S^{\rm con}_{(d+1)}$ gives rise to the same bulk EOM.

\subsection{The boundary effective action }
Starting from  either consistent or covariant current, we can construct the regularized effective action. Begin with the covariant current, we get
\begin{eqnarray}
W_{(d)}\!\!\!\!&=&\!\!\!\!\int_{\Sigma}\cL^{\rm eff}_{(d)}=\int_{\Sigma}\int_{0}^1\rd s\, A^{\halpha\hbeta }\wedge(\cJ_{(d)}^{\rm cov})_{\halpha\hbeta}(A(s),Y)
\cr\!\!\!\!&=&\!\!\!\!-\frac1{2\kappa^2\ell(d-1)!}\int_{\Sigma}\int_{0}^1\rd s\,\epsilon_{\halpha\hbeta\halpha_1\cdots\halpha_{d}} A^{\halpha\hbeta }\wedge \rD(s)Y^{\halpha_1}\wedge\cdots\wedge \rD(s)Y^{\halpha_{d-1}}Y^{\halpha_{d}}
\,.
\end{eqnarray}

\subsubsection{$\rd \cL_{(d)}^{\rm eff}=\cL^{\rm con}_{(d+1)}-\cL^{\rm cov}_{(d+1)}$}
By using the resummations
\begin{eqnarray}
\!\!\!\!&&\!\!\!\!\epsilon_{\halpha\hbeta\halpha_1\cdots\halpha_{d}} A^{\halpha\hbeta}\wedge F^{\halpha_1\hgamma}(s) Y_{\hgamma}\wedge \rD(s) Y^{\halpha_2}\wedge\cdots\wedge \rD(s) Y^{\halpha_{d-1}} Y^{\halpha_{d}}
\cr\!\!\!\!&=&\!\!\!\!-\epsilon_{\halpha\hbeta\halpha_1\cdots\halpha_{d}}  F^{\halpha\hbeta}(s)\wedge \partial_s(\rD(s) Y^{\halpha_1})\wedge\rD(s) Y^{\halpha_2}\wedge\cdots\wedge \rD(s) Y^{\halpha_{d-1}} Y^{\halpha_{d}}
\cr\!\!\!\!&&\!\!\!\!-\frac{\ell^2(-1)^d}2\epsilon_{\halpha\hbeta\halpha_1\cdots\halpha_{d}} A^{\halpha\hbeta}\wedge F^{\halpha_1\halpha_2}(s)\wedge \rD(s) Y^{\halpha_3}\wedge\cdots\wedge \rD(s) Y^{\halpha_{d}}\,,
\\\cr\!\!\!\!&&\!\!\!\!
-\ell^2\epsilon_{\halpha\hbeta\halpha_1\cdots\halpha_{d}} A^{\halpha\hbeta}\wedge \rD(s) Y^{\halpha_1}\wedge\cdots\wedge\rD(s) Y^{\halpha_{d}}
\cr\!\!\!\!&=&\!\!\!\!\epsilon_{\halpha\hbeta\halpha_1\cdots\halpha_{d}} A^{\halpha\hbeta}\wedge \rD(s) Y^{\halpha_1}\wedge\cdots \wedge\rD(s) Y^{\halpha_{d}} Y^{\hgamma} Y_{\hgamma}
\cr\!\!\!\!&=&\!\!\!\!(-1)^d2\epsilon_{\halpha\hbeta\halpha_1\cdots\halpha_{d}}\partial_s[\rD(s) Y^{\halpha}]\wedge \rD(s) Y^{\hbeta}\wedge \rD(s) Y^{\halpha_1}\wedge\cdots\wedge\rD(s) Y^{\halpha_{d-1}} Y^{\halpha_{d}}\,.
\end{eqnarray}
we find that the exterior derivative of $\cL^{\rm eff}_{(d)}$ is
\begin{eqnarray}
\!\!\!\!&&\!\!\!\!-2\kappa^2\ell(d-1)!\rd\cL_{(d)}^{\rm eff} 
\cr\!\!\!\!&=&\!\!\!\!\rd\left[\int_{0}^1\rd s\,\epsilon_{\halpha\hbeta\halpha_1\cdots\halpha_{d}} A^{\halpha\hbeta} \wedge \rD(s) Y^{\halpha_1}\wedge\cdots\wedge \rD(s) Y^{\halpha_{d-1}} Y^{\halpha_{d}}\right]
\cr\!\!\!\!&=&\!\!\!\!\int_{0}^1\rd s\,\epsilon_{\halpha\hbeta\halpha_1\cdots\halpha_{d}} \rD(s)[ A^{\halpha\hbeta}\wedge \rD(s) Y^{\halpha_1}\wedge\cdots\wedge \rD(s) Y^{\halpha_{d-1}} Y^{\halpha_{d}}]
\cr\!\!\!\!&=&\!\!\!\!\int_{0}^1\rd s\,\epsilon_{\halpha\hbeta\halpha_1\cdots\halpha_{d}} \Big\{\partial_s F^{\halpha\hbeta}(s)\wedge \rD(s) Y^{\halpha_1}\wedge\cdots\wedge \rD(s) Y^{\halpha_{d-1}} Y^{\halpha_{d}}
\cr&&~~~~~~~~~~~~~~~~-(d-1) A^{\halpha\hbeta}\wedge [ F(s) Y]^{\halpha_1}\wedge \rD(s) Y^{\halpha_2}\wedge\cdots\wedge \rD(s) Y^{\halpha_{d-1}} Y^{\halpha_{d}}
\cr&&~~~~~~~~~~~~~~~~+(-1)^d A^{\halpha\hbeta}\wedge \rD(s) Y^{\halpha_1}\wedge\cdots\wedge \rD(s) Y^{\halpha_{d-1}}\wedge\rD(s) Y^{\halpha_{d}}]\Big\}
\cr\!\!\!\!&=&\!\!\!\!\int_{0}^1\rd s\,\epsilon_{\halpha\hbeta\halpha_1\cdots\halpha_{d}} \Big\{\partial_s F^{\halpha\hbeta}(s)\wedge \rD(s) Y^{\halpha_1}\wedge\cdots\wedge \rD(s) Y^{\halpha_{d-1}} Y^{\halpha_{d}}
\cr&&~~~~~~~~~~~~~~~~+(d-1) F^{\halpha\hbeta}(s)\wedge \partial_s(\rD(s) Y^{\halpha_1})\wedge\rD(s) Y^{\halpha_2}\wedge\cdots\wedge \rD(s) Y^{\halpha_{d-1}} Y^{\halpha_{d}}
\cr&&~~~~~~~~~~~~~~~~+\frac{\ell^2(-1)^d(d-1)}2 A^{\halpha\hbeta}\wedge  F^{\halpha_1\halpha_2}(s)\wedge \rD(s) Y^{\halpha_3}\wedge\cdots\wedge \rD(s) Y^{\halpha_{d}}
\cr&&~~~~~~~~~~~~~~~~+\frac{2}{\ell^2}\partial_s[\rD(s) Y^{\halpha}] \wedge \rD(s) Y^{\hbeta}\wedge \rD(s) Y^{\halpha_1}\wedge\cdots\wedge\rD(s) Y^{\halpha_{d-1}} Y^{\halpha_{d}}]\Big\}
\cr\!\!\!\!&=&\!\!\!\!\int_{0}^1\rd s\,\epsilon_{\halpha\hbeta\halpha_1\cdots\halpha_{d}} \Big\{\partial_s[ F^{\halpha\hbeta}(s)\wedge \rD(s) Y^{\halpha_1}\wedge\cdots\wedge \rD(s) Y^{\halpha_{d-1}} Y^{\halpha_{d}}]
\cr&&~~~~~~~~~~~~~~~~+\frac{\ell^2(-1)^d(d-1)}2 A^{\halpha\hbeta}\wedge  F^{\halpha_1\halpha_2}(s)\wedge \rD(s) Y^{\halpha_3}\wedge\cdots\wedge \rD(s) Y^{\halpha_{d}}
\cr&&~~~~~~~~~~~~~~~~-\frac{2}{\ell^2(d+1)}\partial_s[\rD(s) Y^{\halpha}\wedge \rD(s) Y^{\hbeta}\wedge \rD(s) Y^{\halpha_1}\wedge\cdots\wedge\rD(s) Y^{\halpha_{d-1}} Y^{\halpha_{d}}]\Big\}
\cr\!\!\!\!&=&\!\!\!\!\epsilon_{\halpha\hbeta\halpha_1\cdots\halpha_{d}} \Big\{ \Big[F^{\halpha\hbeta}
-\tfrac{2}{(d+1)\ell^2}\rD Y^{\halpha}\wedge \rD Y^{\hbeta}\Big]\wedge \rD Y^{\halpha_1}\wedge\cdots\wedge\rD Y^{\halpha_{d-1}} Y^{\halpha_{d}}
\cr&&~~~~~~~~~~~~~~~~+\tfrac{2}{(d+1)\ell^2}\rd Y^{\halpha}\wedge\rd Y^{\hbeta} \wedge\rd Y^{\halpha_1}\wedge\cdots\wedge\rd Y^{\halpha_{d-1}}Y^{\halpha_{d}}
\cr&&~~~~~~~~~~~~~~~~-\tfrac{\ell^2(-1)^{d+1}(d-1)}2\int_{0}^1\rd s\, A^{\halpha\hbeta}\wedge  F^{\halpha_1\halpha_2}(s)\wedge \rD(s) Y^{\halpha_3}\wedge\cdots\wedge \rD(s) Y^{\halpha_{d}}\Big\}
\cr\!\!\!\!&=&\!\!\!\!2\kappa^2\ell\,(d-1)!(\cL^{\rm cov}_{(d+1)}-\cL^{\rm con}_{(d+1)})\,.
\end{eqnarray}
It implies that
\begin{eqnarray}
S^{\rm cov}_{(d+1)}\!\!\!\!&=&\!\!\!\!S^{\rm con}_{(d+1)}-W_{(d)}\,,
\end{eqnarray}

\subsubsection{$SO(2,d)$ gauge transformation of $\cL^{\rm eff}_{(d)}$}
By using the resummation
\begin{eqnarray}
\!\!\!\!&&\!\!\!\!-\ell^2\epsilon_{\halpha\hbeta\halpha_1\cdots\halpha_{d}}\hA^{\halpha\hbeta}\wedge A^{\halpha_1\halpha_2}\wedge\rD(s,\hs)Y^{\halpha_3}\wedge\cdots\wedge \rD(s,\hs)Y^{\halpha_{d}}
\cr\!\!\!\!&=&\!\!\!\!\epsilon_{\halpha\hbeta\halpha_1\cdots\halpha_{d}}\hA^{\halpha\hbeta}\wedge A^{\halpha_1\halpha_2}\wedge\rD(s,\hs)Y^{\halpha_3}\wedge\cdots\wedge \rD(s,\hs)Y^{\halpha_{d}}Y^{\hgamma}Y_{\hgamma}
\cr\!\!\!\!&=&\!\!\!\!-2(-1)^d\epsilon_{\halpha\hbeta\halpha_1\cdots\halpha_{d}} A^{\halpha\hbeta}\wedge(\hA Y)^{\halpha_1}\wedge \rD(s,\hs)Y^{\halpha_2}\wedge\cdots\wedge \rD(s,\hs)Y^{\halpha_{d-1}}Y^{\halpha_d}
\cr\!\!\!\!&&\!\!\!\!+2(-1)^d\epsilon_{\halpha\hbeta\halpha_1\cdots\halpha_{d}}\hA^{\halpha\hbeta}\wedge (AY)^{\halpha_1}\wedge\rD(s,\hs)Y^{\halpha_2}\wedge\cdots\wedge \rD(s,\hs)Y^{\halpha_{d-1}}Y^{\halpha_d}\,,
\end{eqnarray}
we find the finite gauge transformation of $\cL_{(d)}^{\rm eff}$ is 
\begin{eqnarray}
\!\!\!\!&&\!\!\!\!-2\kappa^2\ell\,(d-1)!\cL_{(d)}^{\rm eff}[\tilde A,\tilde Y]
\cr\!\!\!\!&=&\!\!\!\!\int_0^1\rd s\,\epsilon_{\halpha\hbeta\halpha_1\cdots\halpha_{d}}\tilde A^{\halpha\hbeta}\wedge\tilde\rD(s)\tilde Y^{\halpha_1}\wedge\cdots\wedge\tilde\rD(s)\tilde Y^{\halpha_{d-1}}Y^{\halpha_{d}}
\cr\!\!\!\!&=&\!\!\!\!\int_0^1\rd s\,\epsilon_{\halpha\hbeta\halpha_1\cdots\halpha_{d}}(A-\hA)^{\halpha\hbeta}\wedge [\rD(s)Y+(1-s)\hA Y]^{\halpha_1}\wedge\cdots\wedge[\rD(s)Y+(1-s)\hA Y]^{\halpha_{d-1}}Y^{\halpha_{d}}
\cr\!\!\!\!&=&\!\!\!\!\int_0^1\rd s\,\epsilon_{\halpha\hbeta\halpha_1\cdots\halpha_{d}}A^{\halpha\hbeta}\wedge \rD(s,1-s)Y^{\halpha_1}\wedge\cdots\wedge\rD(s,1-s)Y^{\halpha_{d-1}}Y^{\halpha_{d}}
\cr\!\!\!\!&&\!\!\!\!-\int_0^1\rd \hs\,\epsilon_{\halpha\hbeta\halpha_1\cdots\halpha_{d}}\hA^{\halpha\hbeta}\wedge \rD(1-\hs,\hs) Y^{\halpha_1}\wedge\cdots\wedge\rD(1-\hs,\hs)Y^{\halpha_{d-1}}Y^{\halpha_{d}}
\cr\!\!\!\!&=&\!\!\!\!\int_0^1\rd s\,\epsilon_{\halpha\hbeta\halpha_1\cdots\halpha_{d}}  A^{\halpha\hbeta}\wedge \rD(s,0)Y^{\halpha_1}\wedge\cdots\wedge \rD(s,0)Y^{\halpha_{d-1}}Y^{\halpha_{d}}
\cr\!\!\!\!&&\!\!\!\!+\int_0^1\rd s\int_0^{1-s}\rd\hs\,\epsilon_{\halpha\hbeta\halpha_1\cdots\halpha_{d}}  \partial_{\hs}[A^{\halpha\hbeta}\wedge \rD(s,\hs)Y^{\halpha_1}\wedge\cdots\wedge \rD(s,\hs)Y^{\halpha_{d-1}}Y^{\halpha_{d}}]
\cr\!\!\!\!&&\!\!\!\!-\int_0^1\rd \hs\,\epsilon_{\halpha\hbeta\halpha_1\cdots\halpha_{d}}  \hA^{\halpha\hbeta}\wedge \rD(0,\hs)Y^{\halpha_1}\wedge\cdots\wedge \rD(0,\hs)Y^{\halpha_{d-1}}Y^{\halpha_{d}}
\cr\!\!\!\!&&\!\!\!\!-\int_0^1\rd \hs\int_{0}^{1-\hs}\rd s\,\epsilon_{\halpha\hbeta\halpha_1\cdots\halpha_{d}}  \partial_s[\hA^{\halpha\hbeta}\wedge \rD(s,\hs)Y^{\halpha_1}\wedge\cdots\wedge \rD(s,\hs)Y^{\halpha_{d-1}}Y^{\halpha_{d}}]
\cr\!\!\!\!&=&\!\!\!\!\int_0^1\rd s\,\epsilon_{\halpha\hbeta\halpha_1\cdots\halpha_{d}}A^{\halpha\hbeta}\wedge \rD(s,1-s)Y^{\halpha_1}\wedge\cdots\wedge\rD(s,1-s)Y^{\halpha_{d-1}}Y^{\halpha_{d}}
\cr\!\!\!\!&&\!\!\!\!-\int_0^1\rd \hs\,\epsilon_{\halpha\hbeta\halpha_1\cdots\halpha_{d}}\hA^{\halpha\hbeta}\wedge \rD(1-\hs,\hs) Y^{\halpha_1}\wedge\cdots\wedge\rD(1-\hs,\hs)Y^{\halpha_{d-1}}Y^{\halpha_{d}}
\cr\!\!\!\!&=&\!\!\!\!\int_0^1\rd s\,\epsilon_{\halpha\hbeta\halpha_1\cdots\halpha_{d}}  A^{\halpha\hbeta}\wedge \rD(s)Y^{\halpha_1}\wedge\cdots\wedge \rD(s)Y^{\halpha_{d-1}}Y^{\halpha_{d}}
\cr\!\!\!\!&&\!\!\!\!-\int_0^1\rd \hs\,\epsilon_{\halpha\hbeta\halpha_1\cdots\halpha_{d}}  \hA^{\halpha\hbeta}\wedge \hat\rD(\hs)Y^{\halpha_1}\wedge\cdots\wedge \hat\rD(\hs)Y^{\halpha_{d-1}}Y^{\halpha_{d}}
\cr\!\!\!\!&&\!\!\!\!+\int_0^1\rd s\int_0^{1-s}\rd\hs\,\epsilon_{\halpha\hbeta\halpha_1\cdots\halpha_{d}}  \Big\{\partial_{\hs}[A^{\halpha\hbeta}\wedge \rD(s,\hs)Y^{\halpha_1}\wedge\cdots\wedge \rD(s,\hs)Y^{\halpha_{d-1}}Y^{\halpha_{d}}]
\cr&&~~~~~~~~~~~~~~~~~~~~~~~~~~~~~~~~~~~ -\partial_s[\hA^{\halpha\hbeta}\wedge \rD(s,\hs)Y^{\halpha_1}\wedge\cdots\wedge \rD(s,\hs)Y^{\halpha_{d-1}}Y^{\halpha_{d}}]\Big\}
\cr\!\!\!\!&=&\!\!\!\!\int_0^1\rd s\,\epsilon_{\halpha\hbeta\halpha_1\cdots\halpha_{d}}  A^{\halpha\hbeta}\wedge \rD(s)Y^{\halpha_1}\wedge\cdots\wedge \rD(s)Y^{\halpha_{d-1}}Y^{\halpha_{d}}
\cr\!\!\!\!&&\!\!\!\!-\int_0^1\rd \hs\,\epsilon_{\halpha\hbeta\halpha_1\cdots\halpha_{d}}  \hA^{\halpha\hbeta}\wedge \hat\rD(\hs)Y^{\halpha_1}\wedge\cdots\wedge \hat\rD(\hs)Y^{\halpha_{d-1}}Y^{\halpha_{d}}
\cr\!\!\!\!&&\!\!\!\!+\int_0^1\rd s\int_0^{1-s}\rd\hs\,\epsilon_{\halpha\hbeta\halpha_1\cdots\halpha_{d}}  \Big\{(d-1)A^{\halpha\hbeta}\wedge (\hA Y)^{\halpha_1}\wedge\rD(s,\hs)Y^{\halpha_2}\wedge\cdots\wedge \rD(s,\hs)Y^{\halpha_{d-1}}Y^{\halpha_{d}}
\cr&&~~~~~~~~~~~~~~~~~~~~~~~~~~~~ -(d-1)\hA^{\halpha\hbeta}\wedge (A Y)^{\halpha_1}\wedge\rD(s,\hs)Y^{\halpha_2}\wedge\cdots\wedge \rD(s,\hs)Y^{\halpha_{d-1}}Y^{\halpha_{d}}\Big\}
\cr\!\!\!\!&=&\!\!\!\!-2\kappa^2\ell\,(d-1)!\cL_{(d)}^{\rm eff}[A,Y]
\cr\!\!\!\!&&\!\!\!\!-\int_0^1\rd \hs\,\epsilon_{\halpha\hbeta\halpha_1\cdots\halpha_{d}}  \hA^{\halpha\hbeta}\wedge \hat\rD(\hs)Y^{\halpha_1}\wedge\cdots\wedge \hat\rD(\hs)Y^{\halpha_{d-1}}Y^{\halpha_{d}}
\cr\!\!\!\!&&\!\!\!\!+\frac{\ell^2(d-1)}{2(-1)^d}\int_0^1\rd s\int_0^{1-s}\rd\hs\,\epsilon_{\halpha\hbeta\halpha_1\cdots\halpha_{d}}\hA^{\halpha\hbeta}\wedge A^{\halpha_1\halpha_2}\wedge\rD(s,\hs)Y^{\halpha_3}\wedge\cdots\wedge \rD(s,\hs)Y^{\halpha_{d}}
\,.~~~~~~~
\end{eqnarray}
It correctly reproduces the boundary term appeared in the gauge transformation (\ref{A.Con.F}) of $\cL^{\rm con}_{(d+1)}$. That is 
\begin{eqnarray}
\Delta_{U}\cL_{(d+1)}^{\rm con}\!\!\!\!&=&\!\!\!\!\rd(\Delta_{U}\cL_{(d)}^{\rm eff})\,.
\end{eqnarray}
This result is consistent with the facts that $\rd \cL_{(d)}^{\rm eff}=\cL^{\rm con}_{(d+1)}-\cL^{\rm cov}_{(d+1)}$ and $\cL^{\rm cov}_{(d+1)}$ is manifestly gauge invariant.

At the infinitesimal level, we have
\begin{eqnarray}
\!\!\!\!&&\!\!\!\!2\kappa^2\ell\,(d-1)!\delta_{u}\cL_{(d)}^{\rm eff}
\cr\!\!\!\!&=&\!\!\!\!\int_0^1\rd s\,\epsilon_{\halpha\hbeta\halpha_1\cdots\halpha_{d}}  \Big[\rd u^{\halpha\hbeta}\wedge \rd Y^{\halpha_1}\wedge\cdots\wedge \rd Y^{\halpha_{d-1}}Y^{\halpha_{d}}
\cr&&~~~~~~~~~~~~~~~~~~~~~ -\frac{\ell^2(d-1)}{2(-1)^d}\int_0^{1-s}\rd\hs\,\rd u^{\halpha\hbeta}\wedge A^{\halpha_1\halpha_2}\wedge\rD(s)Y^{\halpha_3}\wedge\cdots\wedge \rD(s)Y^{\halpha_{d}}\Big]
\cr\!\!\!\!&=&\!\!\!\!\rd\Big\{\epsilon_{\halpha\hbeta\halpha_1\cdots\halpha_{d}}u^{\halpha\hbeta}\Big[\rd Y^{\halpha_1}\wedge\cdots\wedge \rd Y^{\halpha_{d-1}}Y^{\halpha_{d}}
-\int_0^1\rd s\frac{\ell^2(d-1)(1-s)}{2(-1)^d} A^{\halpha_1\halpha_2}\wedge\rD(s)Y^{\halpha_3}\wedge\cdots\wedge \rD(s)Y^{\halpha_{d}}\Big]\Big\}
\cr\!\!\!\!&&\!\!\!\!+(-1)^{d}\epsilon_{\halpha\hbeta\halpha_1\cdots\halpha_{d}}u^{\halpha\hbeta}\Big[\rd Y^{\halpha_1}\wedge\cdots\wedge \rd Y^{\halpha_{d}}
\cr&&~~~~~~~~~~~~~~~~~~~~~~~+\int_0^1\rd s\frac{\ell^2(d-1)(1-s)}{2} \rd\left(A^{\halpha_1\halpha_2}\wedge\rD(s)Y^{\halpha_3}\wedge\cdots\wedge \rD(s)Y^{\halpha_{d}}\right)\Big]
\,,
\end{eqnarray}
thus the boundary consistent anomaly is
\begin{eqnarray}
\!\!\!\!&&\!\!\!\!2\kappa^2\ell\,(d-1)!(\cA^{\rm con}_{(d)})_{\halpha\hbeta}
\cr\!\!\!\!&=&\!\!\!\!(-1)^{d}\epsilon_{\halpha\hbeta\halpha_1\cdots\halpha_{d}}\Big[\rd Y^{\halpha_1}\wedge\cdots\wedge \rd Y^{\halpha_{d}}
\cr&&~~~~~~~~~~~~~~~~~~~~+\int_0^1\rd s\frac{\ell^2(d-1)(1-s)}{2}\rd\left( A^{\halpha_1\halpha_2}\wedge\rD(s)Y^{\halpha_3}\wedge\cdots\wedge \rD(s)Y^{\halpha_{d}}\right)\Big]\,.~~~~
\end{eqnarray}
It satisfies that
\begin{eqnarray}
\!\!\!\!&&\!\!\!\!\rd[u^{\halpha\hbeta}(\cA^{\rm con}_{(d)})_{\halpha\hbeta}]
=\delta_{u} I^{\rm con}_{(d+1)}=\delta_{u} I^{\rm CS}_{(d+1)}-\delta_{u} \cL^{\rm top}_{(d+1)}\,.~~~~~
\end{eqnarray}

\end{document}